\documentclass[apj]{emulateapj}

\begin{document}

\title{Numerical Simulation of an EUV Coronal Wave Based on the February 13, 2009 \\CME Event Observed by STEREO}

\author{Ofer Cohen\altaffilmark{1}, Gemma D.~R. Attrill\altaffilmark{1}, Ward B. Manchester {\sc{IV}}\altaffilmark{2} \\ 
and Meredith J. Wills-Davey\altaffilmark{1}}

\altaffiltext{1}{Harvard-Smithsonian Center for Astrophysics, 60 Garden St. Cambridge, MA 02138}
\altaffiltext{2}{Center for Space Environment Modeling, University of Michigan, 2455 Hayward St., 
Ann Arbor, MI 48109}

\begin{abstract}  

On 13 February 2009, a coronal wave -- CME -- dimming event was observed 
in quadrature by the STEREO spacecraft.  We analyze this event 
using a three-dimensional, global magnetohydrodynamic (MHD) model for the solar corona. 
The numerical simulation is driven and constrained by the observations, and indicates where
magnetic reconnection occurs between the expanding CME core and surrounding environment.  
We focus primarily on the lower corona, extending out to $3R_{\odot}$; 
this range allows simultaneous comparison with both EUVI and COR1 data.
Our simulation produces a diffuse coronal bright front
remarkably similar to that observed by STEREO/EUVI at 195 \AA\/.
It is made up of \emph{two} 
components, and is the result of a combination of both wave and non-wave mechanisms.  

The CME becomes large-scale quite low ($<$ 200 Mm) in the corona.
It is not, however, an inherently large-scale event; rather, the expansion 
is facilitated by magnetic reconnection between the expanding CME core 
and the surrounding magnetic environment.  
In support of this, we also find numerous secondary dimmings, 
many far from the initial CME source region.  Relating such dimmings to reconnecting
field lines within the simulation provides further evidence that CME expansion
leads to the ``opening'' of coronal field lines on a global scale. 
Throughout the CME expansion, the coronal wave maps directly to the CME footprint.

Our results suggest that the ongoing debate over the ``true'' nature of diffuse
coronal waves may be mischaracterized. It appears that \emph{both} wave and 
non-wave models are required to explain the observations and 
understand the complex nature of these events.

\end{abstract}



\section{INTRODUCTION}
\label{sec:Intro}


The physical nature of EUV coronal waves has been ambiguous ever since their discovery in 1996 
\citep{Dere97, Moses97, Thompson98}, with the Extreme Imaging Telescope \citep[EIT;][]{Delaboudiniere95} 
on board the Solar and Heliospheric Observatory \citep[SOHO;][]{Domingo95}. Following the launch of the 
Solar Terrestrial Relations Observatory \citep[STEREO;][]{Kaiser08}, coronal waves are now observed with 
increased temporal and spatial resolution, and have now been seen from two perspectives using the Extreme Ultra-Violet Imagers 
\citep[EUVI;][]{Wuelser04}.  Using SOHO data, \cite{Biesecker02} showed (after correcting for observational biases),
that every EIT wave in their study could be credibly associated with a coronal mass ejection (CME).  

In the past decade, there has been ongoing debate about the physical nature of the coronal wave bright front. 
Many models have been developed and they can be split into two main groups. The first group of models 
describes coronal waves as a pure MHD wave, either a: fast-mode 
\citep[e.g.][]{Thompson99,Klassen00,Vrsnak02,Cliver04,Warmuth04b}, 
slow-mode \citep[][]{Krasnoselskikh07,Wang09}, or a slow-mode solitary wave \citep[][]{WillsDavey07}. 
In many respects, a wave model is appropriate. 
Fast-mode solutions are consistent with 
propagation across the magnetic field; slow-modes can account
for the large intensity enhancements and some of the lower measured velocities; and solitary waves 
are consistent with single-pulse structure and coherence over large distances.
In these models, the wave front may be freely propagating (generated by an initial pressure pulse), or piston-driven where the wave is 
constantly supplied by energy from the expanding CME \citep[e.g.][]{Vrsnak08}.  

The second group of models describe the 
bright front as being related to the actual expansion of the CME, in a non-wave context. 
Work by \cite{Delannee99, Delannee00b, Delannee08} attribute the regions of enhanced emission at low altitude
to compression of the plasma between stable flux domains.  \cite{Chen02, Chen05a, Chen09} suggest that the expansion of the magnetic 
field during the CME lift-off gradually ``opens''  the overlying magnetic field, compressing the plasma in the legs of 
the CME.  And \cite{Attrill07a} propose magnetic reconnections between the expanding CME and favorably 
oriented surrounding quiet Sun magnetic field as an explanation for the patchy bright front.  
Although the connection with CMEs factors strongly in many of these models, because the exact relationship 
between diffuse coronal waves \citep{Biesecker02, Vrsnak05b} and CMEs is not well understood, the physical nature of the bright 
front has remained inconclusive. 

Space-based white-light coronagraph observations of the low corona are available from STEREO/COR1.  However, 
the field of view only starts at 1.3 R$_{\odot}$, and the domain where coronal waves are observed, below 200 Mm 
\citep[e.g.][]{Warmuth04a, Warmuth05, Vrsnak05a, WillsDavey07, Patsourakos09a} is blocked by the coronagraph occulting disk. Thus, a 
combination of EUVI and COR1 observations, 
coupled with numerical simulations, is required to develop a coherent picture of the early stages of CME evolution 
in the low corona. 

In this paper, we present an analysis of observations combined with a global numerical simulation of a coronal wave event 
observed by STEREO on 13 February 2009. This is the first numerical simulation of an EUV coronal wave based on this 
observation in quadrature. We use a global MHD model for the solar corona that is driven by 
real magnetogram data, and we drive the CME in a realistic way that matches the observations.  Numerical studies 
of coronal waves have been carried out previously. However, these models were either two-dimensional \citep[e.g.][]{Pomoell08}, 
considered only the expanding flux tube, neglecting interaction with the surroundings \citep{Delannee08}, simulated 
the interaction of coronal waves with only a local active region \citep{Uchida74,Ofman02}, or they drove the coronal wave by a 
pressure pulse \citep{Wang00, Wu01}. 

By constraining the simulation as much as possible with the observations, we obtain a full picture of the three-dimensional 
evolution of the coronal magnetic field during the eruption, including those regions that are not observed by white-light 
coronagraphs.  In this way, we hope to shed some light on the nature of coronal waves, their relationship to CMEs, and 
their theoretical description. 

We describe the observations of the CME event in Section~\ref{sec:Obs}, and the numerical simulation in 
Section~\ref{sec:Model}. We present the results in Section~\ref{sec:Results} and discuss the implications 
for the various descriptions of coronal waves in Section~\ref{sec:Discussion}. We conclude our findings in 
Section~\ref{sec:Conclusions}.


\section{OBSERVATIONS}
\label{sec:Obs}

The coronal wave -- CME -- dimmings event on 13 February 2009 occurred when the STEREO spacecraft 
had a separation angle of 91$^{\circ}$. The Sun was in the very quietest part of its cycle, just at the start of the rise phase of 
solar cycle 24.  NOAA active region 1012 was the only active region on the solar disk. Thus the majority of the 
surrounding environment was quiet Sun, with a low-latitude coronal hole to the East of the active region (see top panels 
Figure \ref{fig:f1}).  Pre-eruption, the active region hosted a forward ``S'' sigmoid.
The Sun produced a CME that was seen by spacecraft A right on the East limb, and by spacecraft 
B expanding from the center of the disk.

\subsection{STEREO/COR1}

The COR1 instruments are internally occulted coronagraphs and observe the inner solar corona in 
white light from 1.3 - 4 R$_{\odot}$  \citep{Thompson03}.  Base difference 
images (where a pre-event image at 05:45 UT, is subtracted from all subsequent images) of COR1-A data, are shown in the left panels 
of Figure \ref{fig:f2}. The COR1 images were taken with a temporal cadence of 10 minutes, and have a pixel size of 
7.5$^{\prime\prime}$.  COR1-B also observed the CME as a halo event, first becoming 
apparent beyond the occulting disk at 06:55 UT.  The non-differenced COR1-A data (not shown) show a helmet 
streamer located north of AR 1012, and open streamers to the south.  

\subsection{STEREO/EUVI}

The EUVI imagers observe the Sun out to 1.7 R$_{\odot}$, and produce 2048$\times$2048 images with 
a pixel size of 1.6$^{\prime\prime}$ \citep{Wuelser04}. We analyze the 195 \AA\ EUVI images, which have a temporal 
cadence of 10 minutes. 
Base difference images from EUVI-B are shown in Figure \ref{fig:f3}. The base difference images are produced by 
first compensating for the solar rotation using {\tt SolarSoft}'s {\tt drot\_map} routine 
(http://www.lmsal.com/solarsoft), so that all images are de-rotated to the pre-event image time at 05:15 UT. Then, 
the pre-event image is subtracted from all subsequent images. Base difference images highlight real intensity 
changes, with bright areas showing an increase in emission with respect to the pre-event data.  

\subsubsection{Coronal dimmings}
As well as showing the coronal wave bright front, base difference images also show depletions in intensity, 
known as ``coronal dimmings''.  These are regions where 
the plasma density has dramatically decreased due to plasma evacuation along the ``opened'' magnetic field,
usually occurring during an eruption \citep[e.g.][]{Hudson96, Harra01, Harra07}. Various works 
have shown that the core coronal dimming regions (i.e. black 
regions, bottom panels of Figure \ref{fig:f1}), located on either side of the bright post-eruptive arcade (PEA) mark the 
footpoints of the erupting flux rope \citep[e.g.][]{Webb00, Mandrini05, Crooker06, Attrill06, McIntosh07}.  
The intensity drop in core coronal dimmings is substantial \citep[typically $\sim$ 40 - 60\%; e.g.][]{Chertok05}.

We process the EUVI base difference data using the automatic dimmings detection algorithm described in 
\cite{Attrill09b}. Figure \ref{fig:f4} shows the output from this algorithm. In addition to the core dimmings 
near to the PEA, secondary dimmings are also detected which develop remote from the active region and are 
spread across the solar disk.  These secondary dimmings are more subtle than the deep, core dimmings 
and are not easy to identify by eye in the base difference data.  We discuss the secondary dimmings further 
in \textsection \ref{subsec:sec_dims}. 

\subsubsection{Persistent brightenings}
The base difference images also show relatively concentrated, persistent brightenings at the edge of the deep, core dimmings. 
Examination of the location of coronal holes with respect to these brightenings shows that two brightenings 
(marked ``A'' and ``B'' in Figure \ref{fig:f1}) are situated along the boundary of a low-latitude coronal hole extending to the East of 
the active region.  Additionally, brightenings ``B'' and ``C'' are located at the edge of the core dimmings.


\section{NUMERICAL MODEL}
\label{sec:Model}

In order to simulate the CME event, we use the Solar Corona (SC) model developed at the University of Michigan 
\citep{roussev03b,cohen07, cohen08b}. The model is based on the global MHD BATSRUS code \citep{powell99} and is part of 
the Space Weather Modeling Framework (SWMF) \citep{toth05}. The model solves the set of MHD equations on a non-uniform 
Cartesian grid and is designed with highly efficient parallel architecture. Here, we describe the model briefly. We 
refer the reader to the references above for a more detailed description.

\begin{figure*}[h!]
\centering
\includegraphics[width=3.0in]{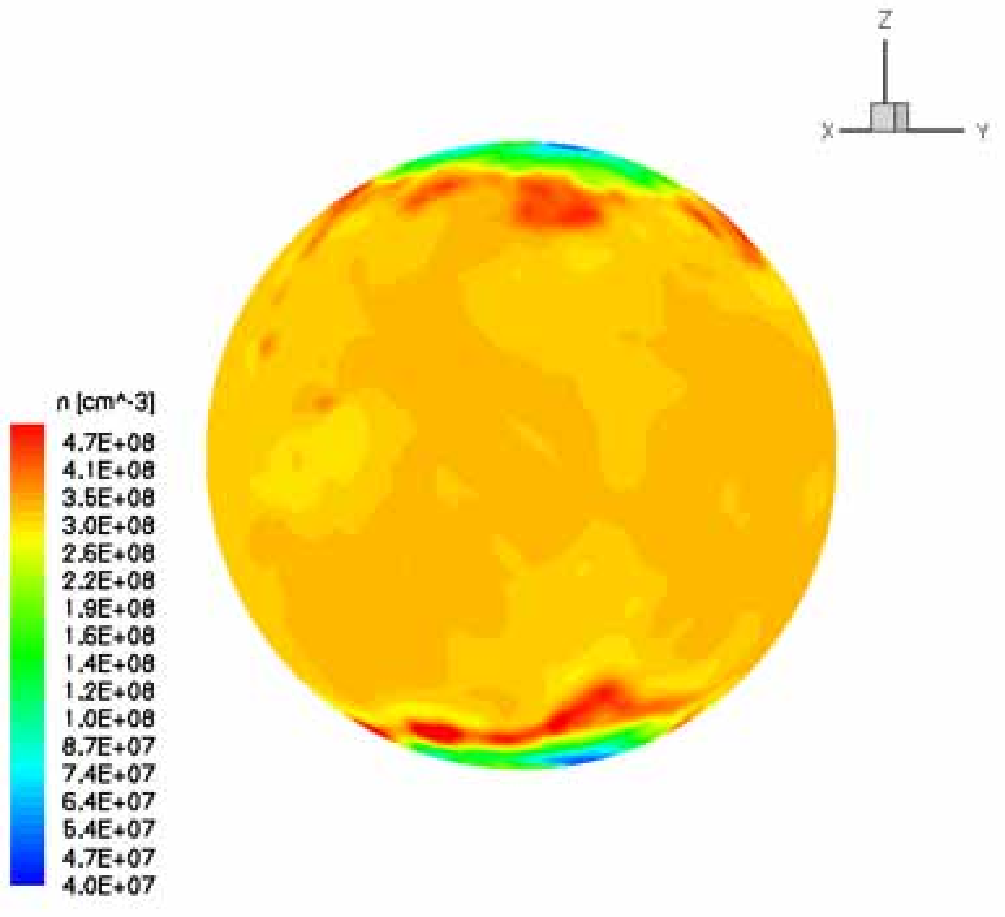} 
\includegraphics[width=3.0in]{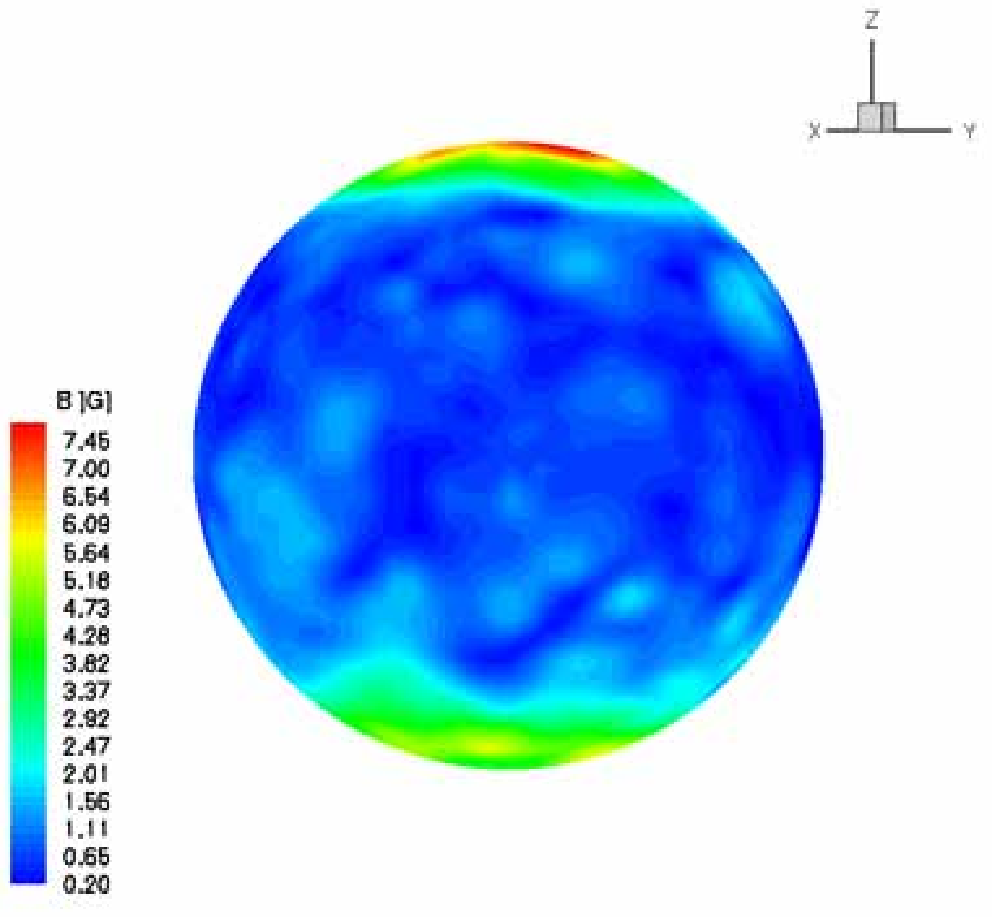}  \\
\includegraphics[width=3.0in]{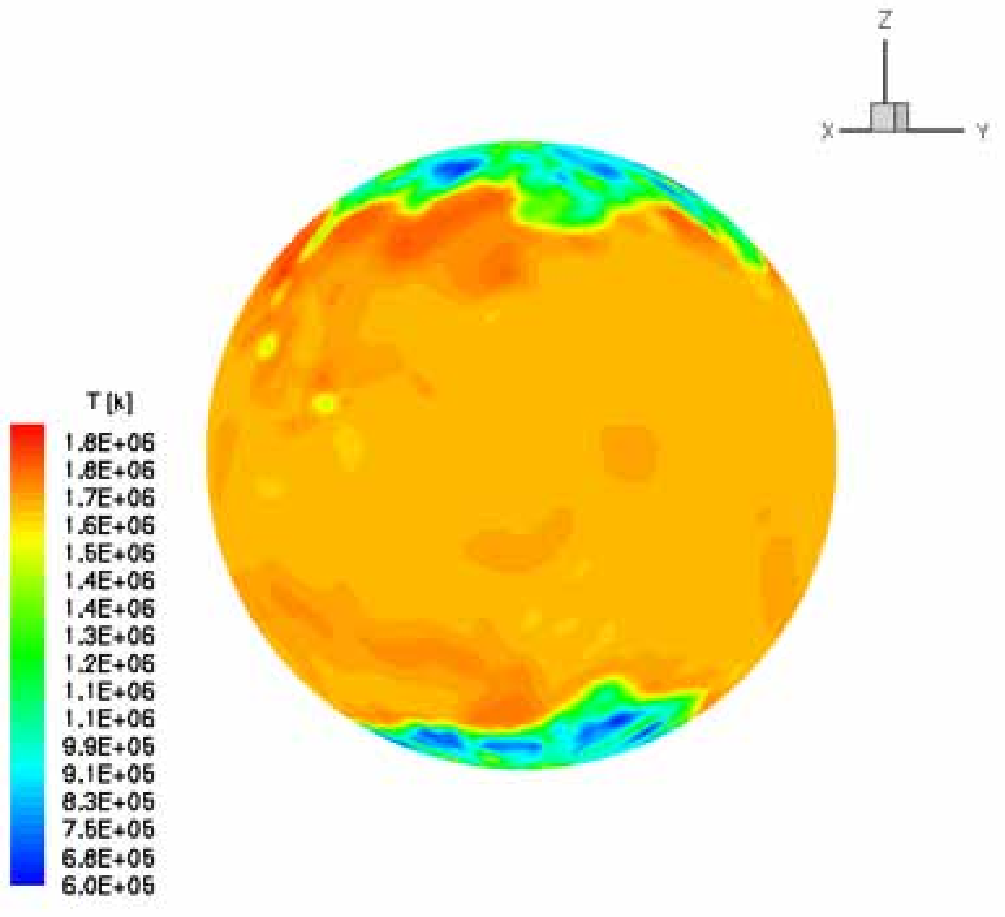} 
\includegraphics[width=3.0in]{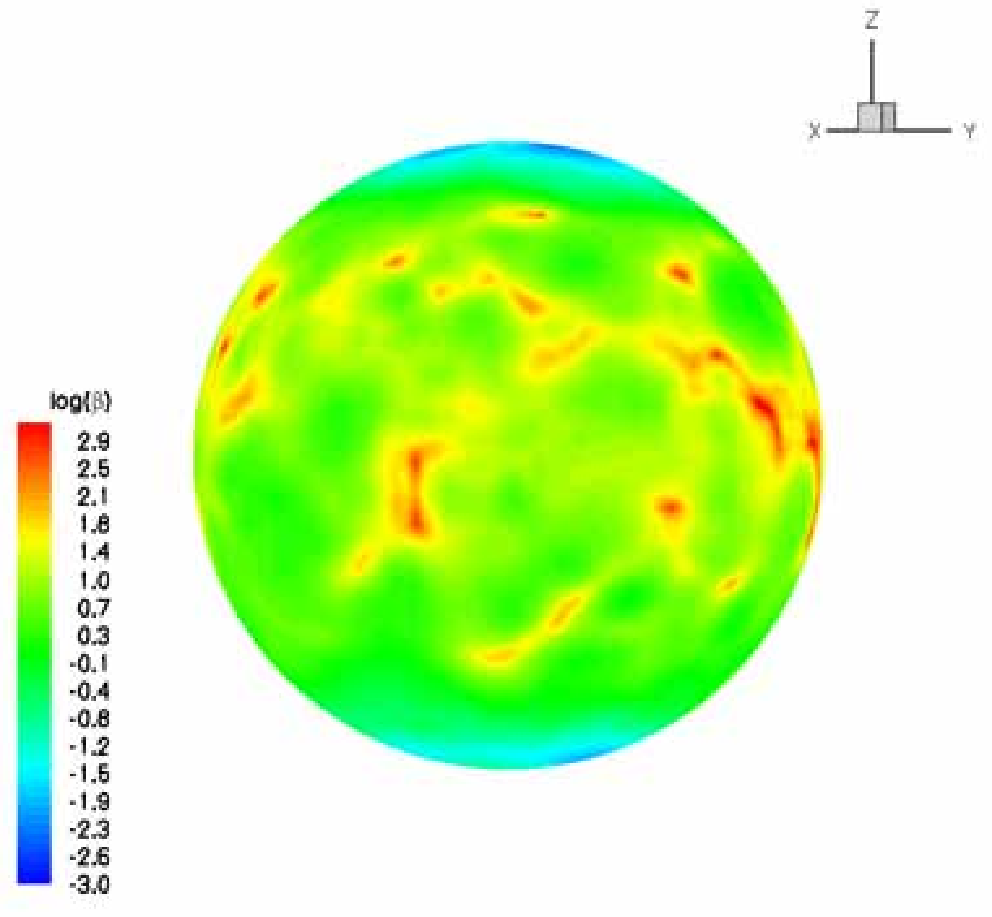} \\
\includegraphics[width=3.0in]{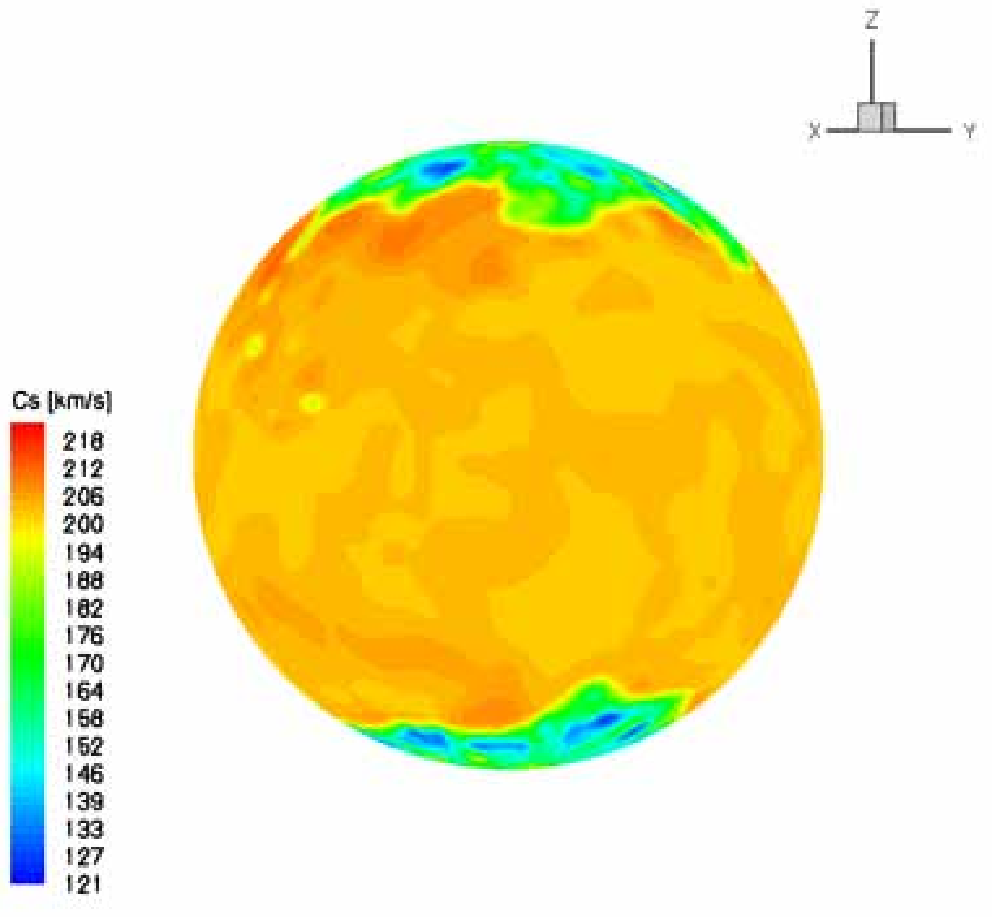} 
\includegraphics[width=3.0in]{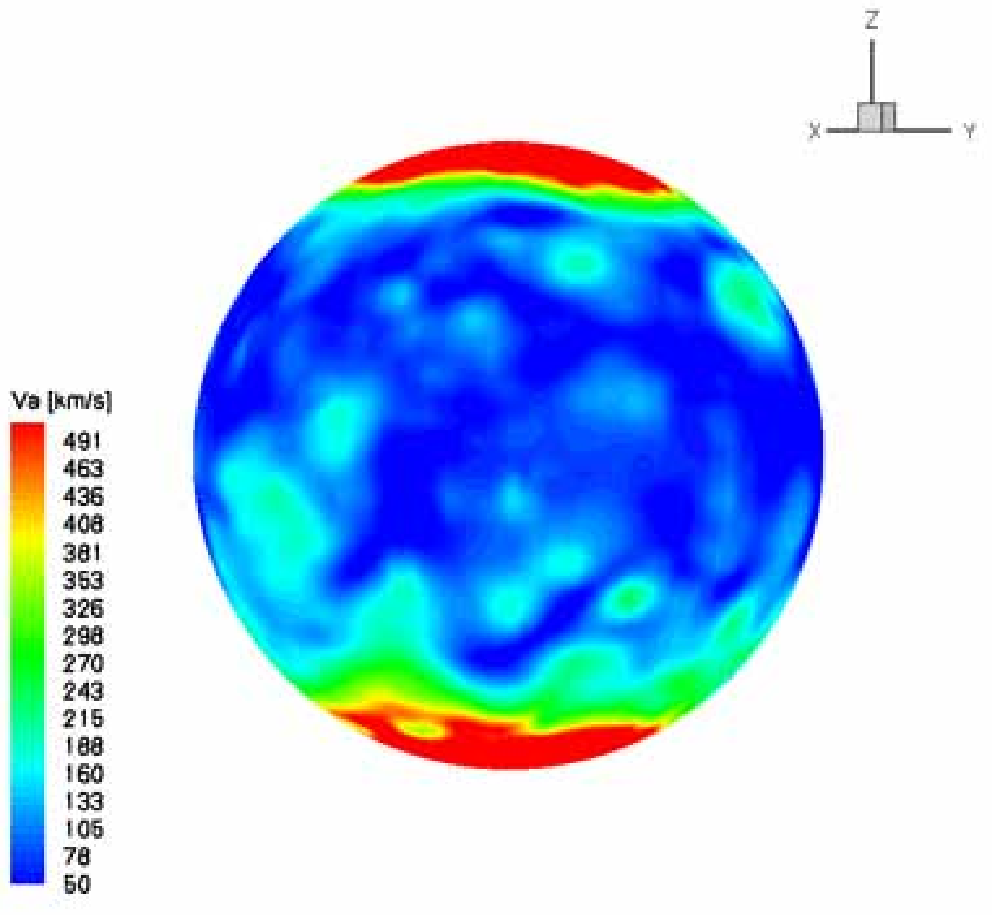}
\caption{The steady-state values of number density (top left), magnetic field strength (top right), temperature (middle left), 
plasma $\beta$ (middle right), sound speed (bottom left), and Alfv\'en speed (bottom right) at height of $r=1.1R_\odot$.}
\label{fig:f5}
\end{figure*}

The ambient solar wind conditions in the model are obtained under the assumption that the source of energy required to 
power the solar wind 
is the change in the polytropic index, $\gamma$ in a non-polytropic medium. The numerical procedure is as follows. 
First, the potential magnetic field is calculated 
using high resolution SOHO/Michelson Doppler Imager \citep[MDI;][]{Scherrer95} magnetogram synoptic maps 
(http://soi.stanford.edu). Second, this potential field distribution 
is used to calculate the distribution of the terminal solar wind speed, $u_{wsa}$ as a function of the flux tube expansion 
factor, $f_s$, based on the Wang-Sheeley-Arge (WSA) model \citep{wangy90,argepizzo00}. Third, the photospheric boundary 
conditions for $\gamma$, and the terminal speed, $u_{wsa}$, are related by tracing the total energy (Bernoulli Integral) 
along the flux tubes. The spatial distribution of $\gamma$ is then specified as a radial function of the photospheric values, 
and the MHD equations are solved self-consistently until a steady-state with a wind solution is obtained. Figure~\ref{fig:f5} 
shows the steady-state values of number density, $n$, magnetic field strength, $B$, temperature, $T$, plasma $\beta$, 
sound speed, $C_s$, and Alfv\'en speed, $v_A$ respectively, at height of $r=1.1R_\odot$. 

In order to drive the CME, we superimpose an unstable, semi-circular flux rope based on the \cite{titov99} model, on top of the 
ambient solution \citep{roussev03a}. We match the location and orientation of the flux rope to those of the source 
active region and its inversion line as they appear in the magnetogram data. The free energy of the CME is obtained by 
a prescribed toroidal current; we modify the magnitude of this current to match the observed CME speed. We would like 
to stress that even though the CME initiation method is not based on actual photospheric motions, it has been 
successful in mimicking the CME conditions once it is already propagating and expanding \citep{lugaz07,cohen08a,manchester08}.
Since we are interested in the development of the CME after it has already been initiated, this model is appropriate for our study.

We run the simulation in a Cartesian box of $20R_\odot\times20R_\odot\times20R_\odot$, with 9 levels of grid refinement 
around the solar surface. The grid size around the active region and up to a height of about $3R_\odot$ is of the order 
of $1/50R_\odot$. The ambient coronal conditions are driven by MDI magnetogram data for Carrington Rotation 2080.  The MHD 
simulation was performed using the Pleiades cluster at NASA's Advanced Supercomputing (NAS) center.


\section{RESULTS}
\label{sec:Results}

\subsection{CME Expansion}
\label{subsec:CMEprop}

Prior to a detailed analysis of the magnetic field three-dimensional evolution, we validate the timing of 
the simulated CME by comparing its propagation with STEREO/COR1 observations. This validation is required for any further 
assumptions about the dynamic evolution and its relation to the observed coronal waves. Fortunately, the 13 February 2009
event was observed by both STEREO-A and STEREO-B simultaneously, in quadrature. 

\begin{figure*}[h!]
\centering
\includegraphics[width=4.3in]{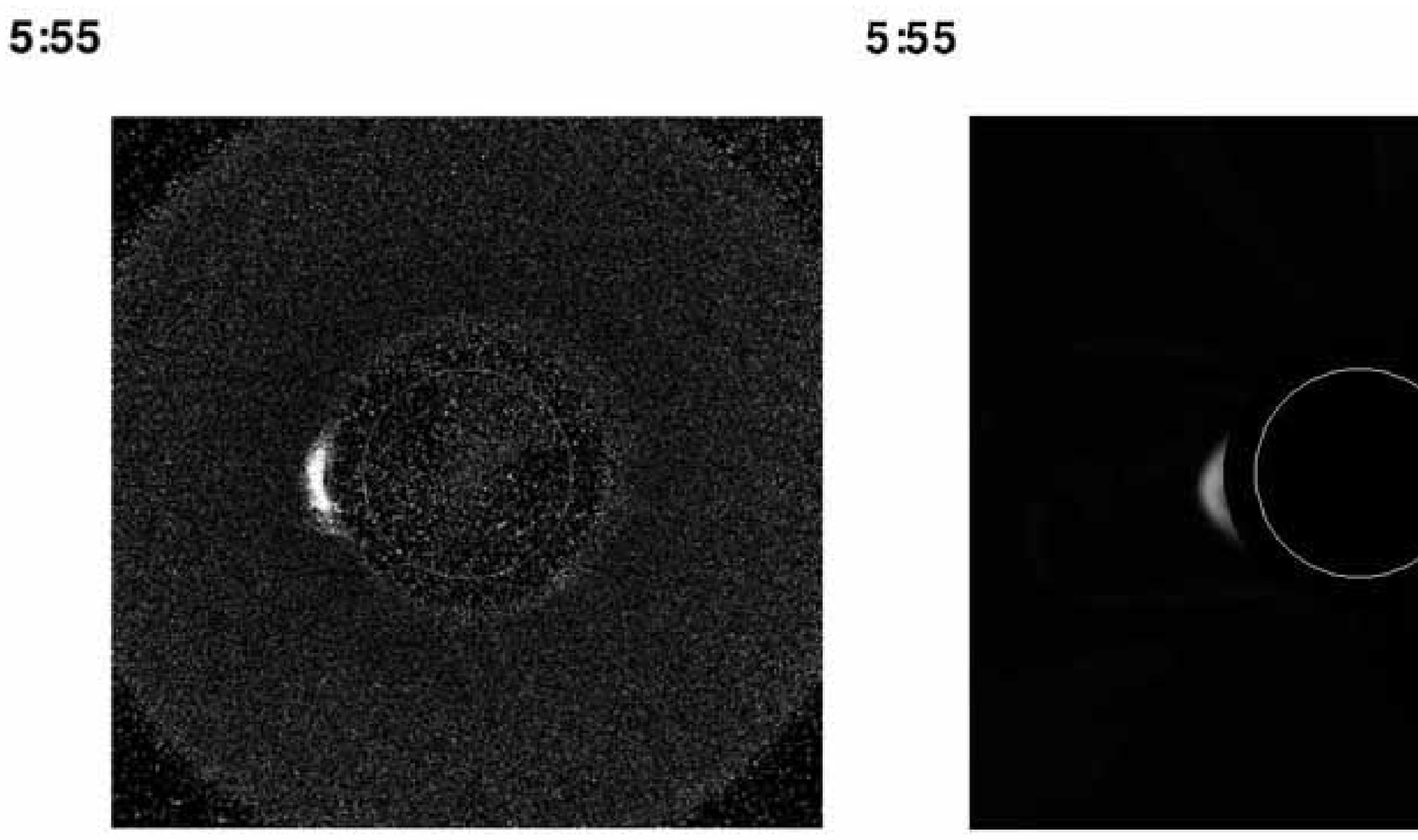}
\includegraphics[width=2.07in]{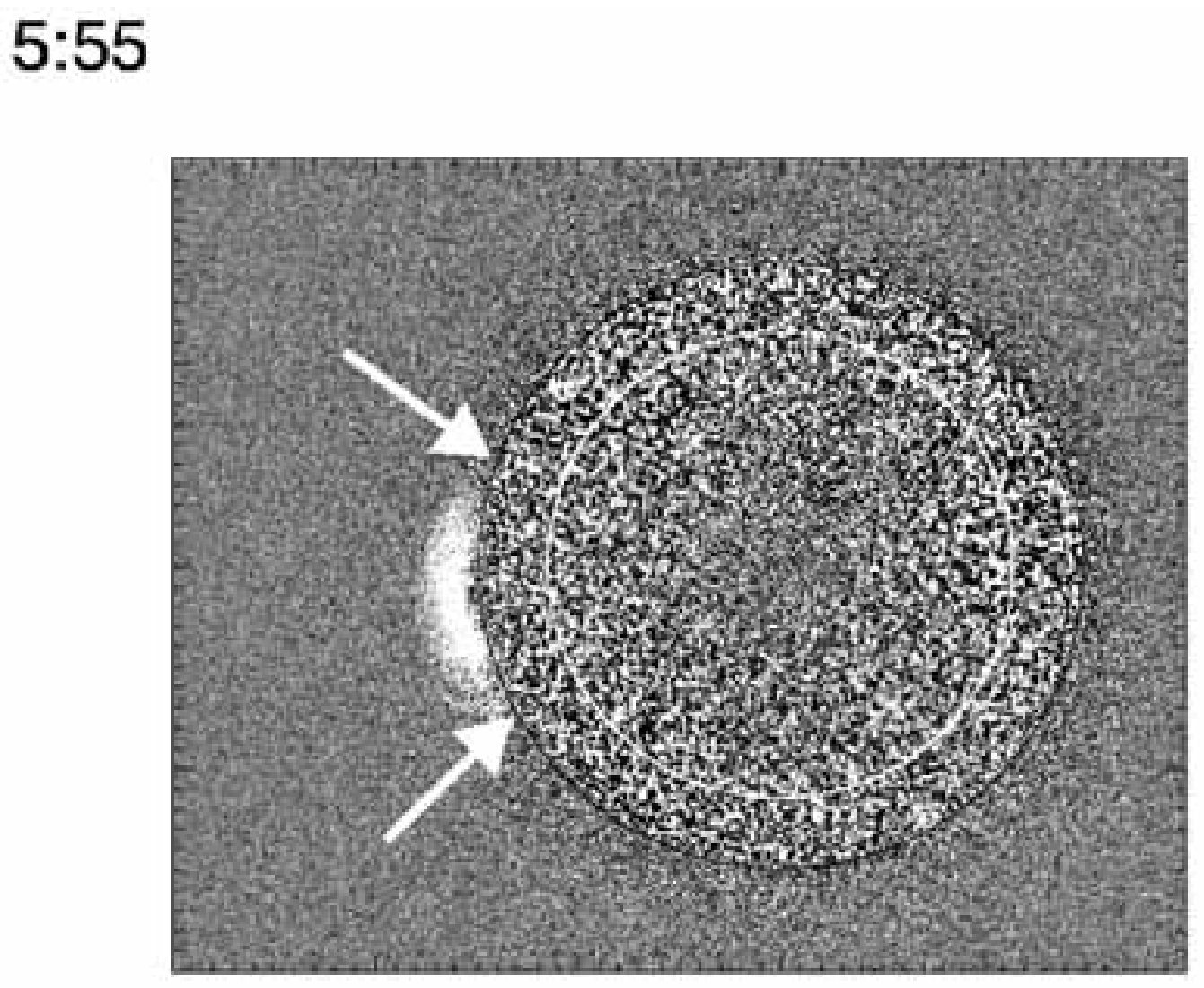}\\
\includegraphics[width=4.3in]{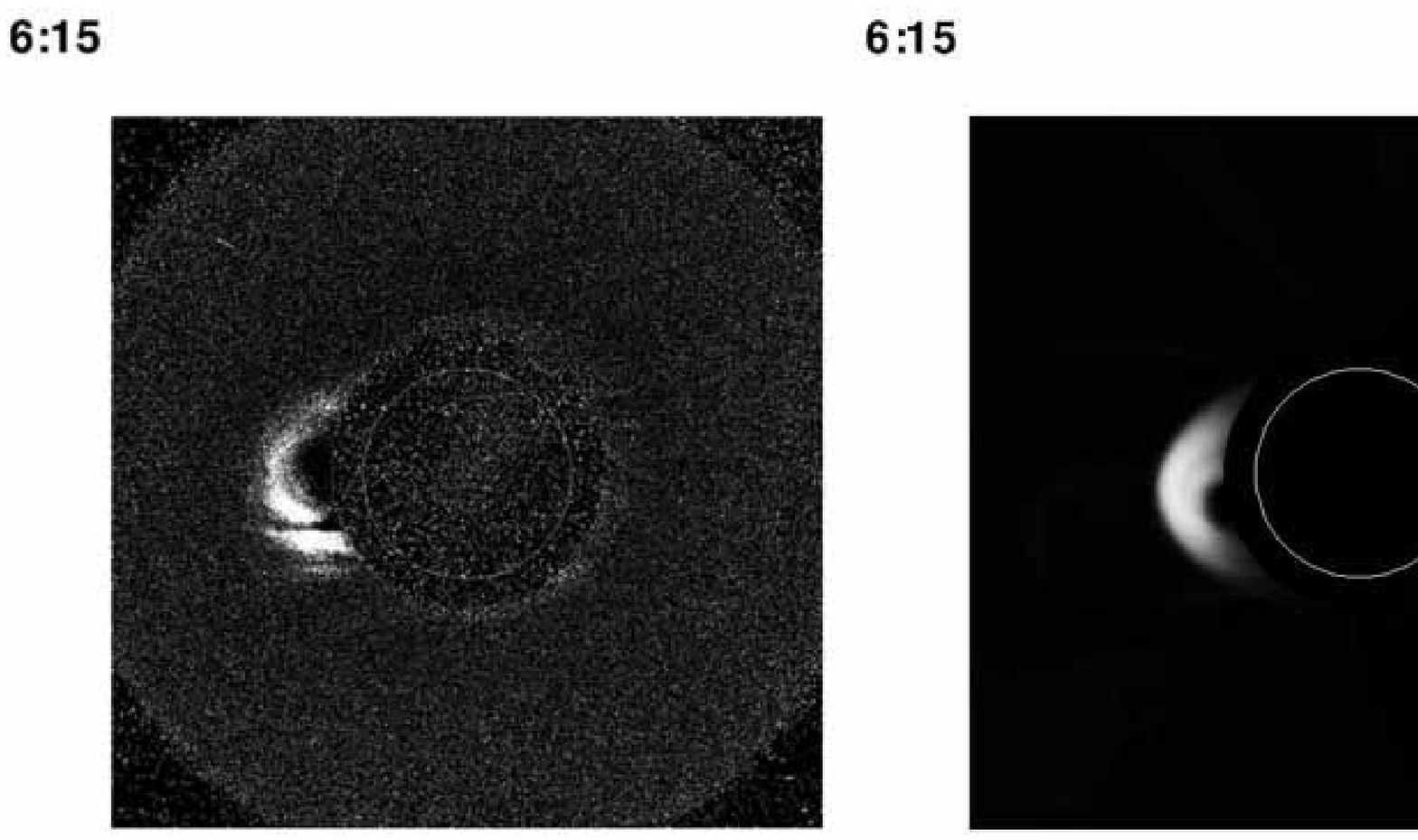}
\includegraphics[width=2.075in]{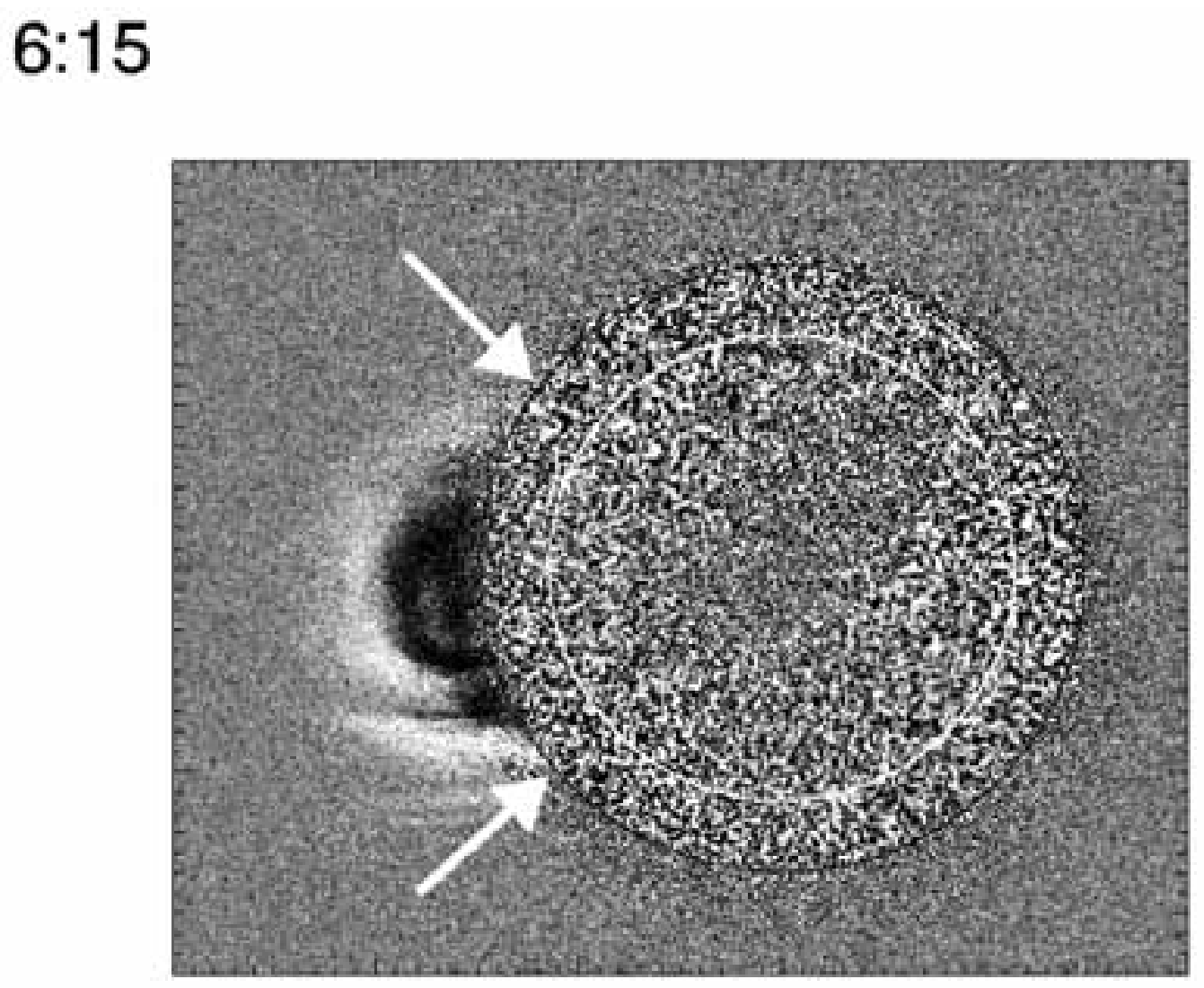} \\
\includegraphics[width=4.3in]{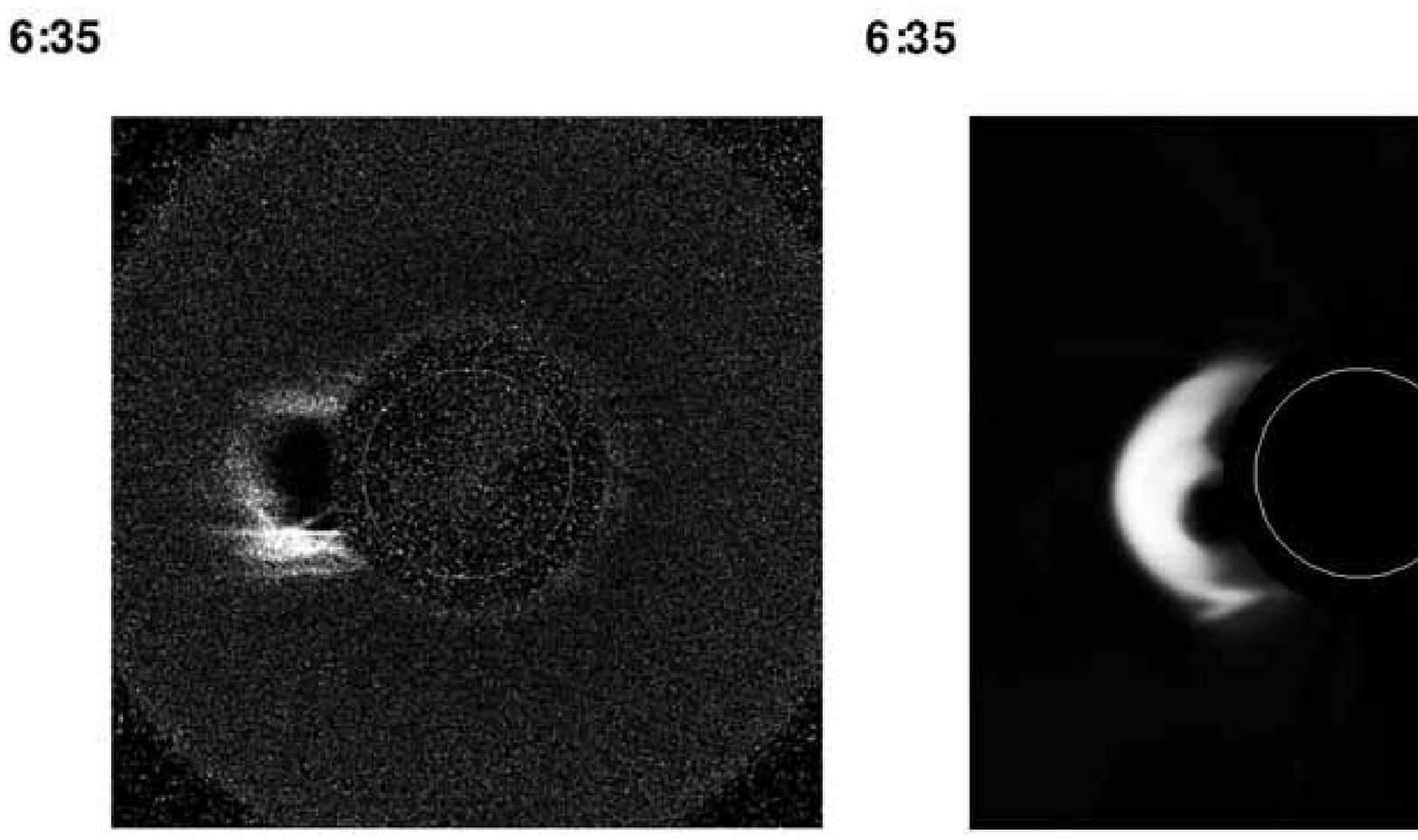}
\includegraphics[width=2.07in]{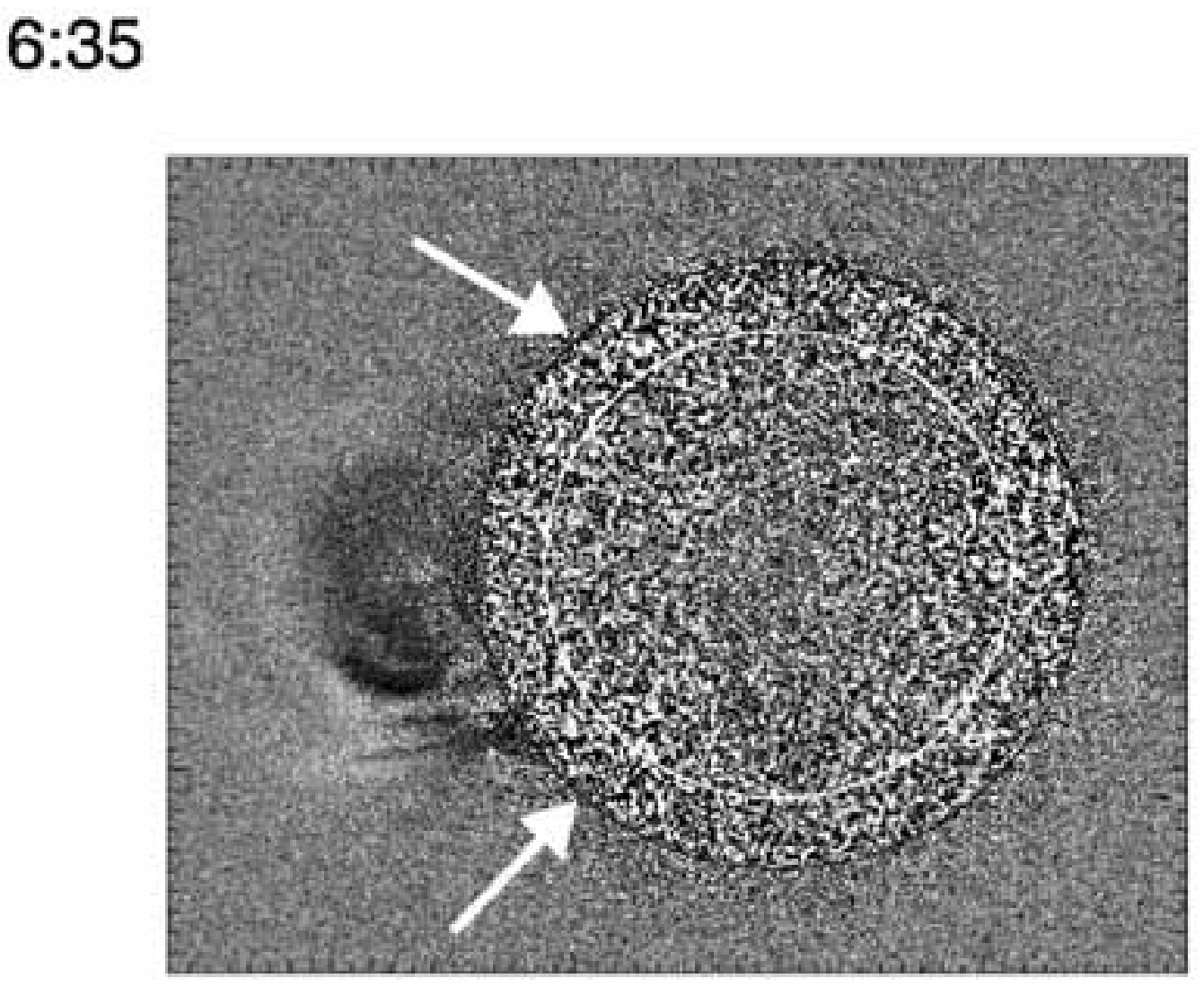}
\caption{Comparison of Stereo-A COR1 images and synthetic white-light images from the simulation at 5:55, 6:15, 
and 6:35 UT. Each row shows a COR1 base difference image (left), base difference image of the simulated 
white-light scattering (center) and a COR1 running difference image (right).  The white arrows in the right panels 
indicate the maximum lateral extent of the CME as seen in the COR1 running difference movie - the reader is 
encouraged to view the COR1 running difference movie accompanying this Figure ({\tt{COR1\_rdiff.mov}}). 
The expansion of the subtle outermost CME shell to the North is more easily discerned than the expansion to the South,
where streamers complicate the observations.  See also the corresponding movies: {\tt{COR1joint1.mov}} and {\tt{COR1\_Joint\_Zoom.mov}.}}
\label{fig:f2}
\end{figure*}

Figure~\ref{fig:f2} shows a side view of the CME propagation, comparing STEREO-A COR1 base difference white-light images 
with synthetic base difference white-light images generated from the simulation domain for 5:55, 6:15, and 6:35 UT. 
We would like to stress that both real and simulated data sets have been processed in the same manner, and the scale of the 
images has been chosen so that it provides the best display. 

The comparison of the simulation (center column) with the base difference images (left column) is
used to verify the structure of the CME.  
Comparison with the running difference images (right column) is used to verify the lateral extent.  We emphasise that the outer shell
of the CME is a very subtle feature, and is only really evident when successive frames are switched back and forth.
Thus the reader is encouraged to examine the COR1 running difference movie ({\tt COR1\_rdiff.mov}), using the white arrows in the right column of 
Figure~\ref{fig:f2} as a reference.  One can see that the simulated CME front and the global structure matches well to the observed CME.

\subsection{Coronal Wave Expansion}
\label{subsec:cw_expansion}

\begin{figure*}[h!]
\centering
\includegraphics[width=1.45in]{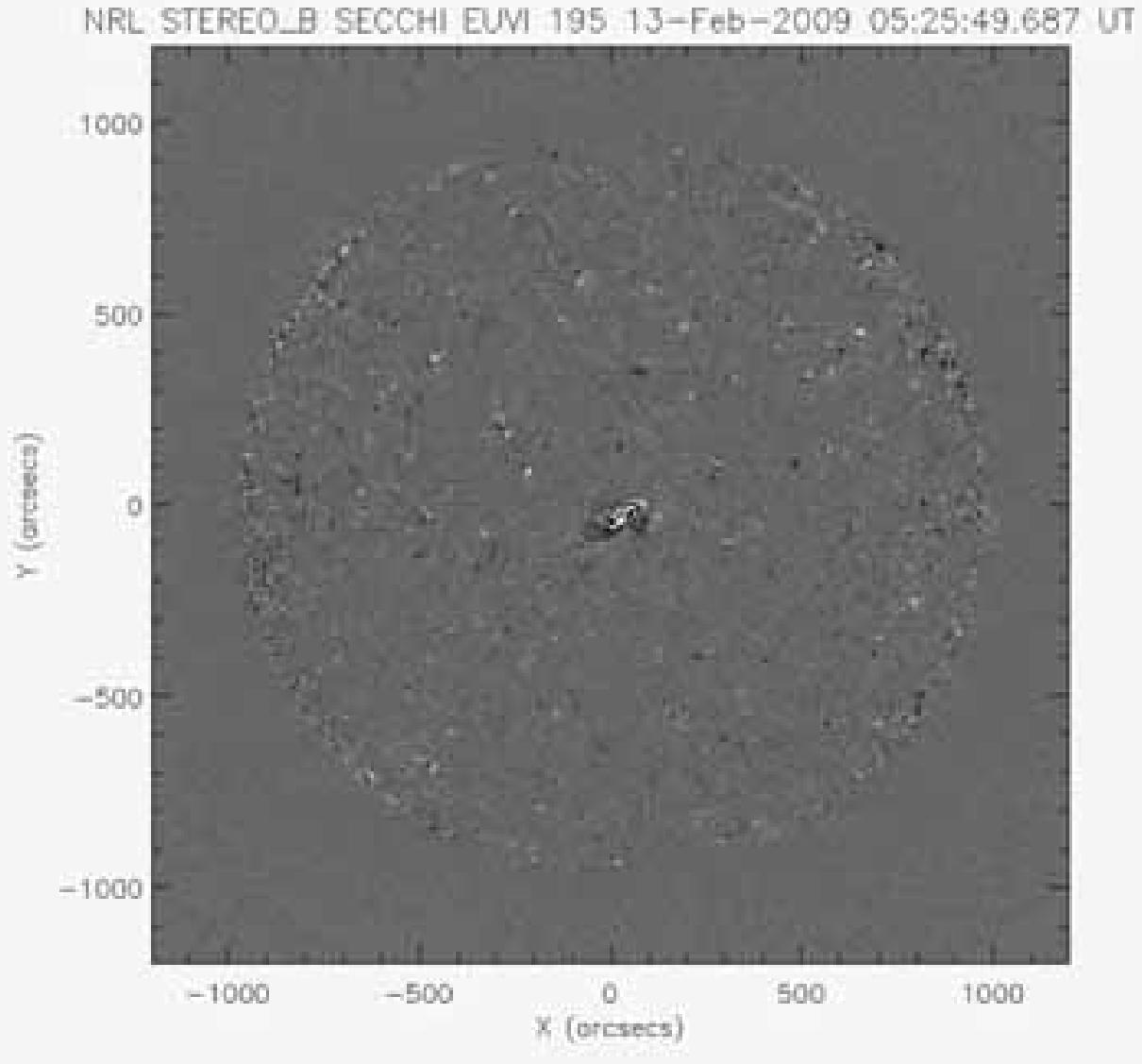}
\includegraphics[width=1.7in]{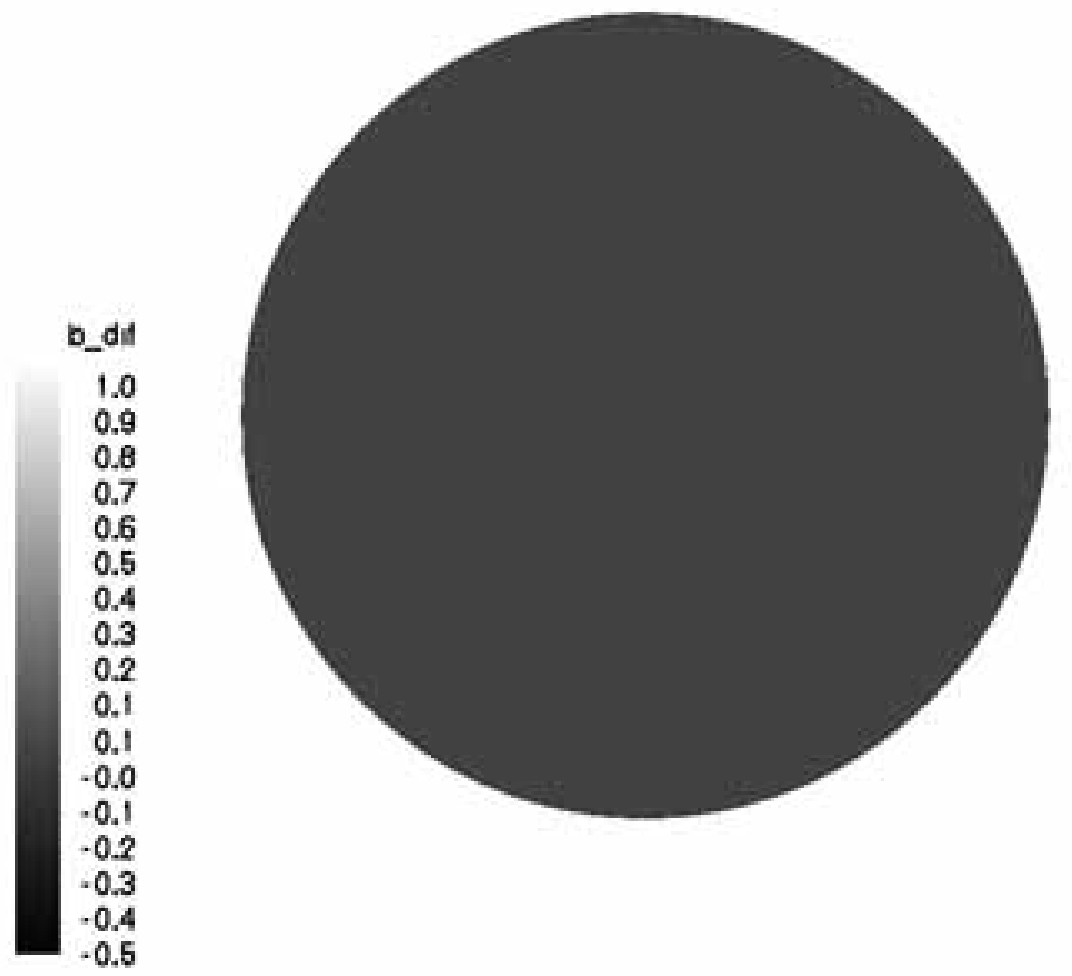}
\includegraphics[width=1.45in]{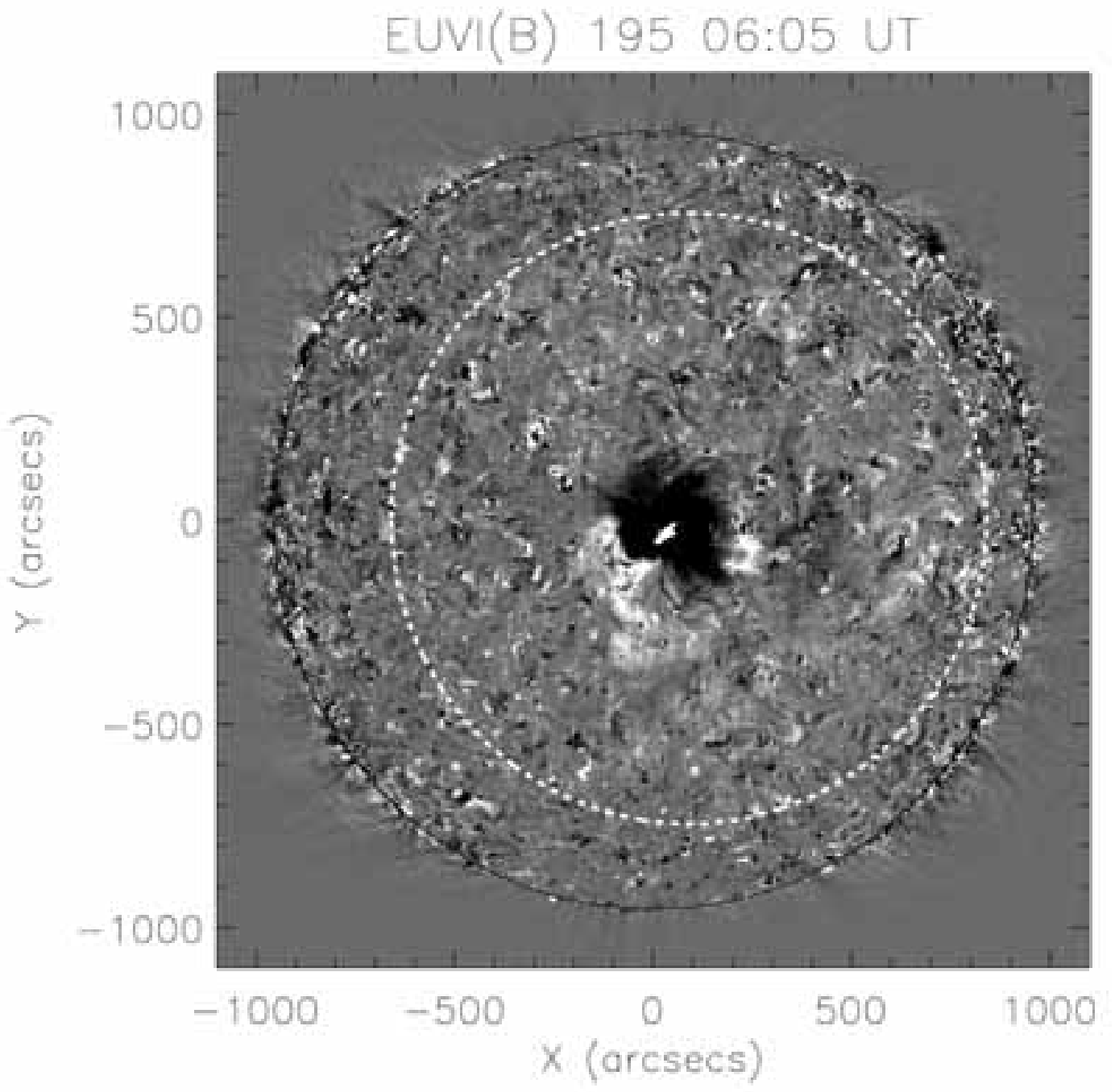}
\includegraphics[width=1.7in]{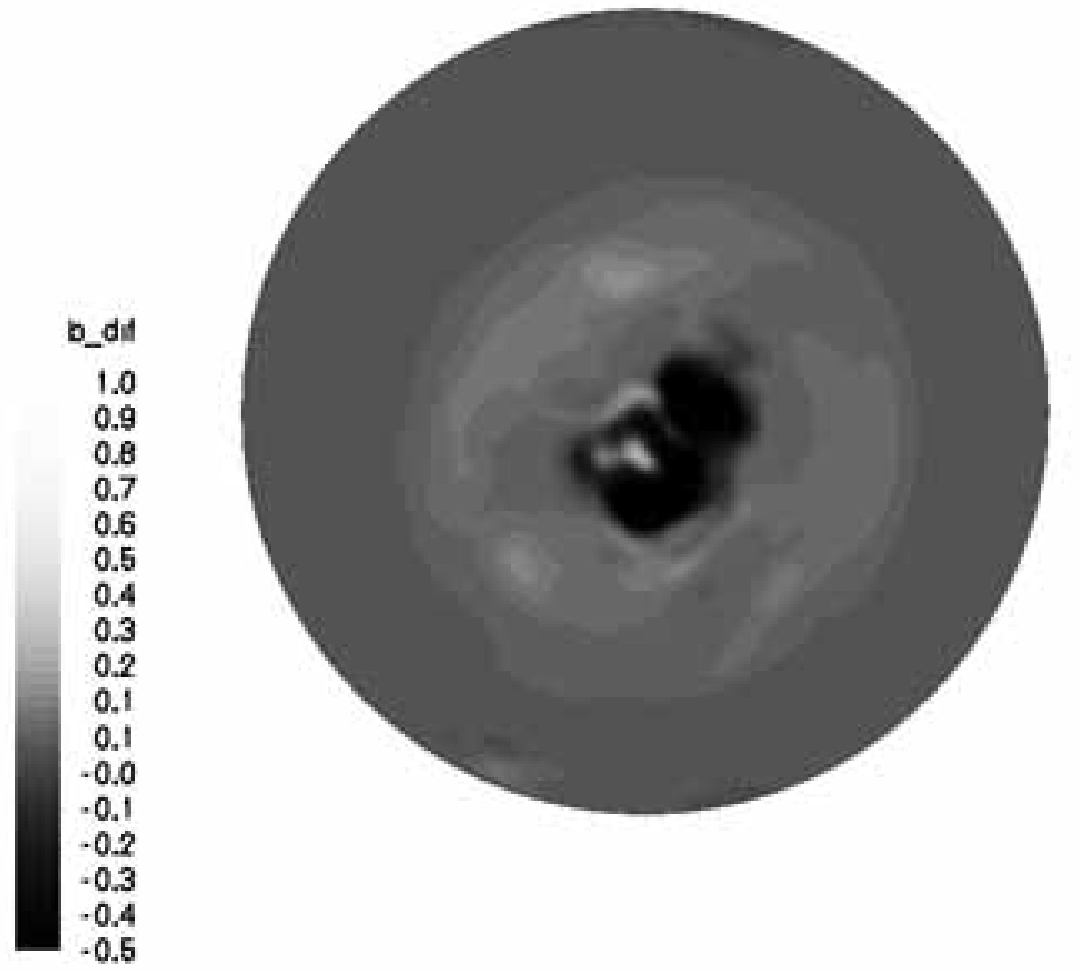}\\
\includegraphics[width=1.45in]{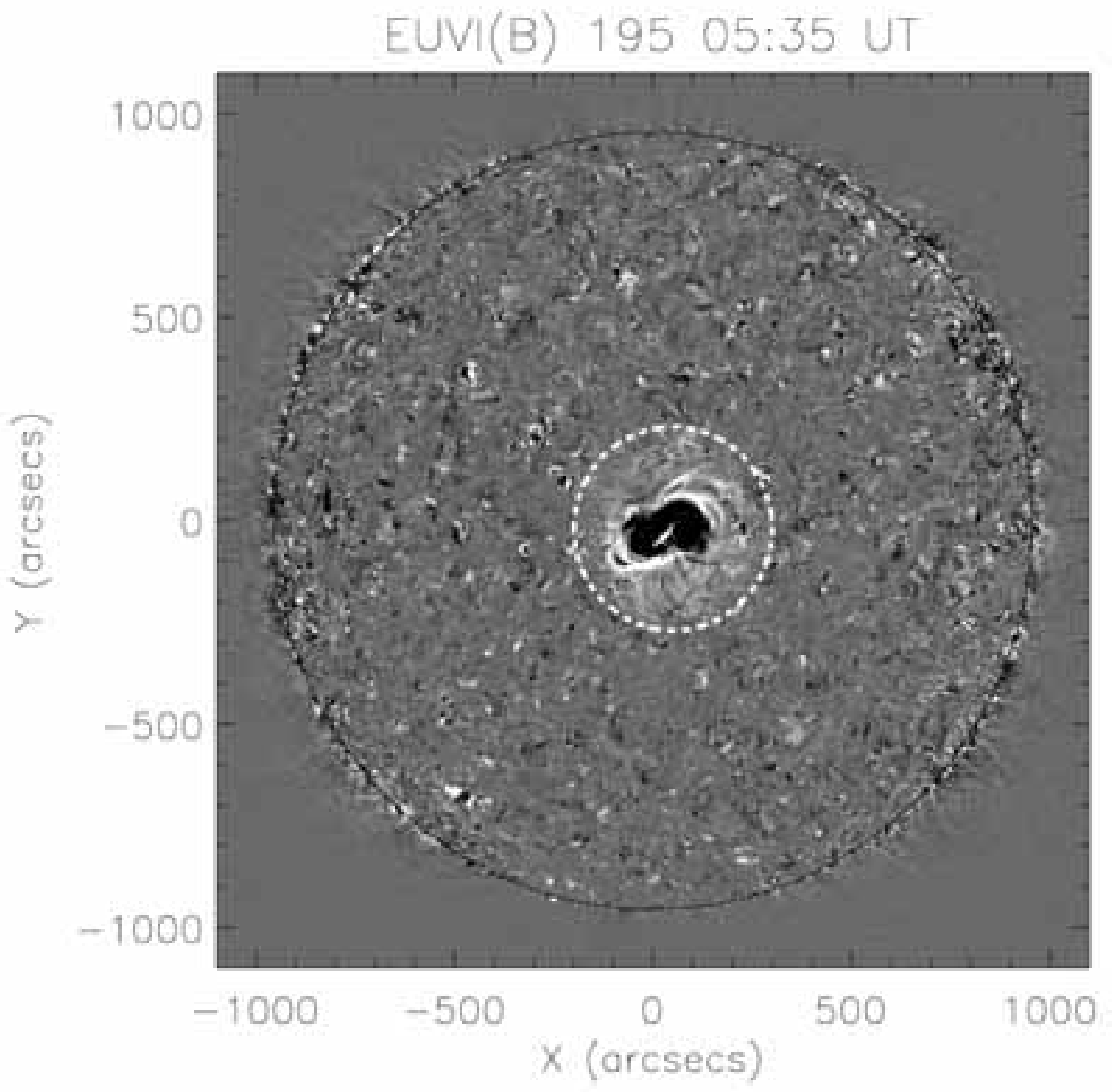}
\includegraphics[width=1.7in]{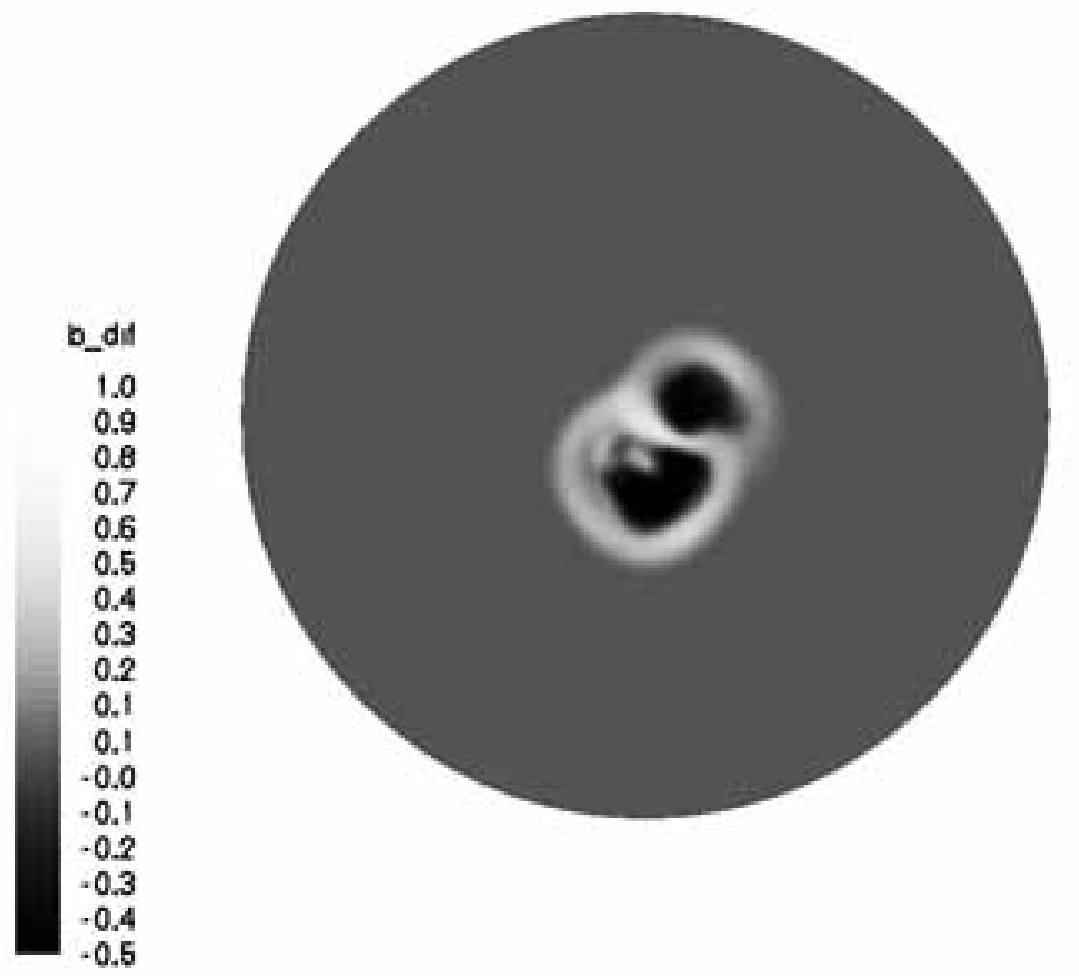} 
\includegraphics[width=1.45in]{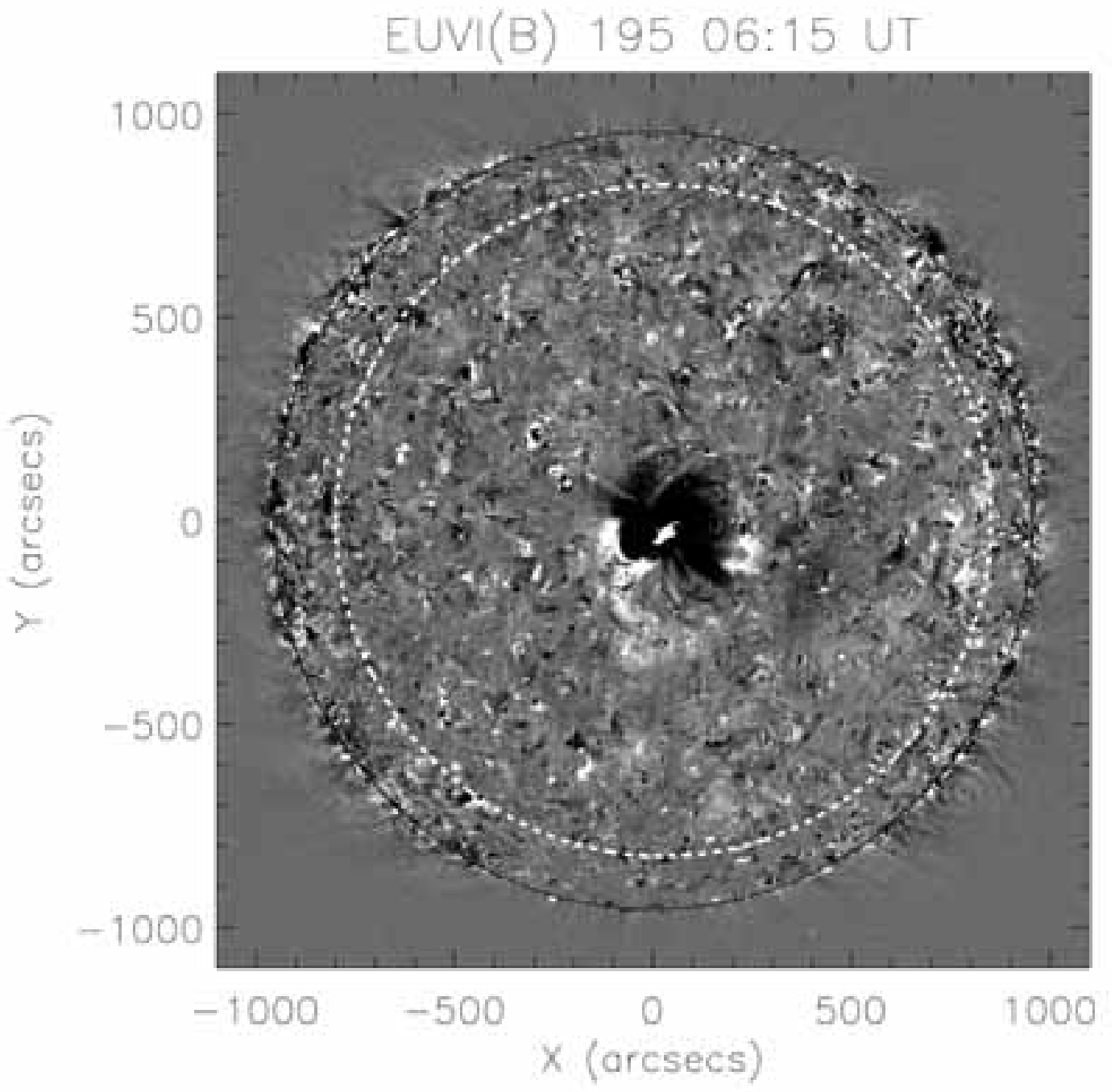}
\includegraphics[width=1.7in]{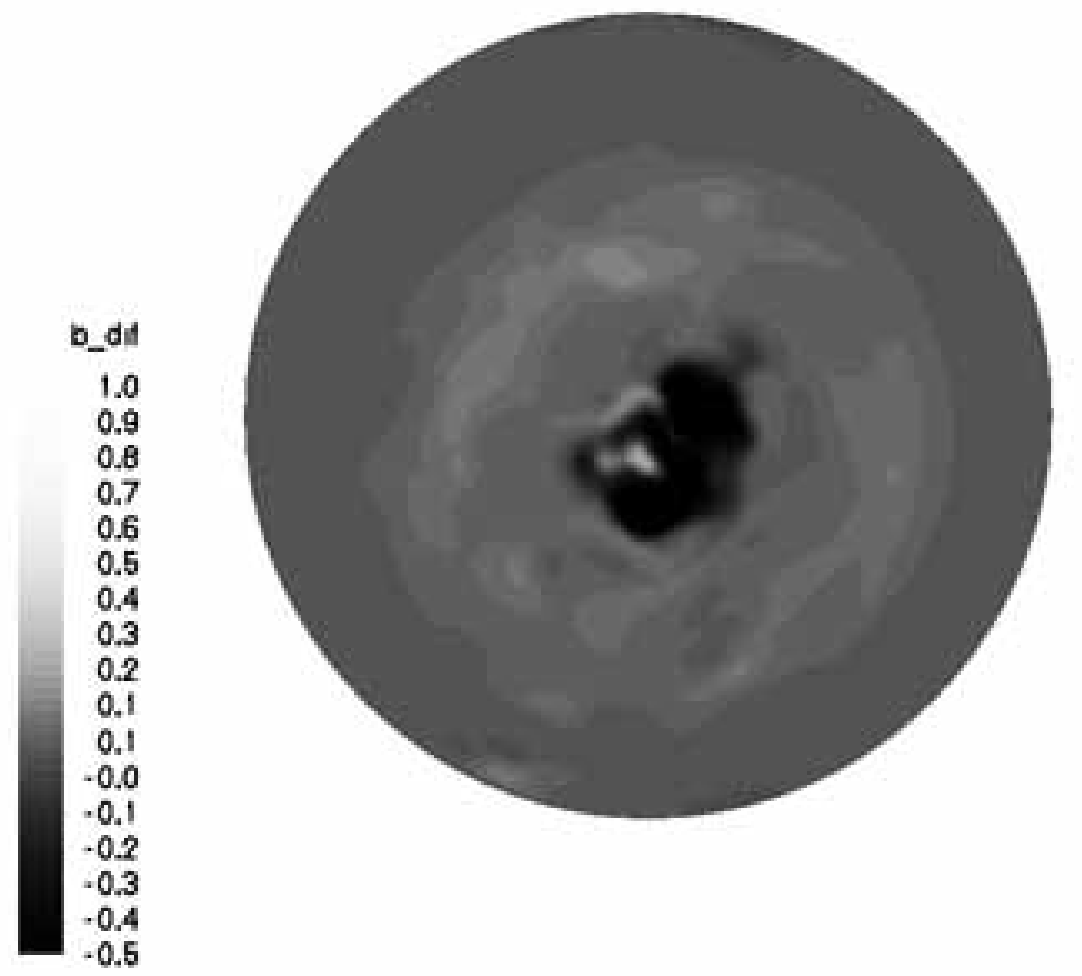} \\
\includegraphics[width=1.45in]{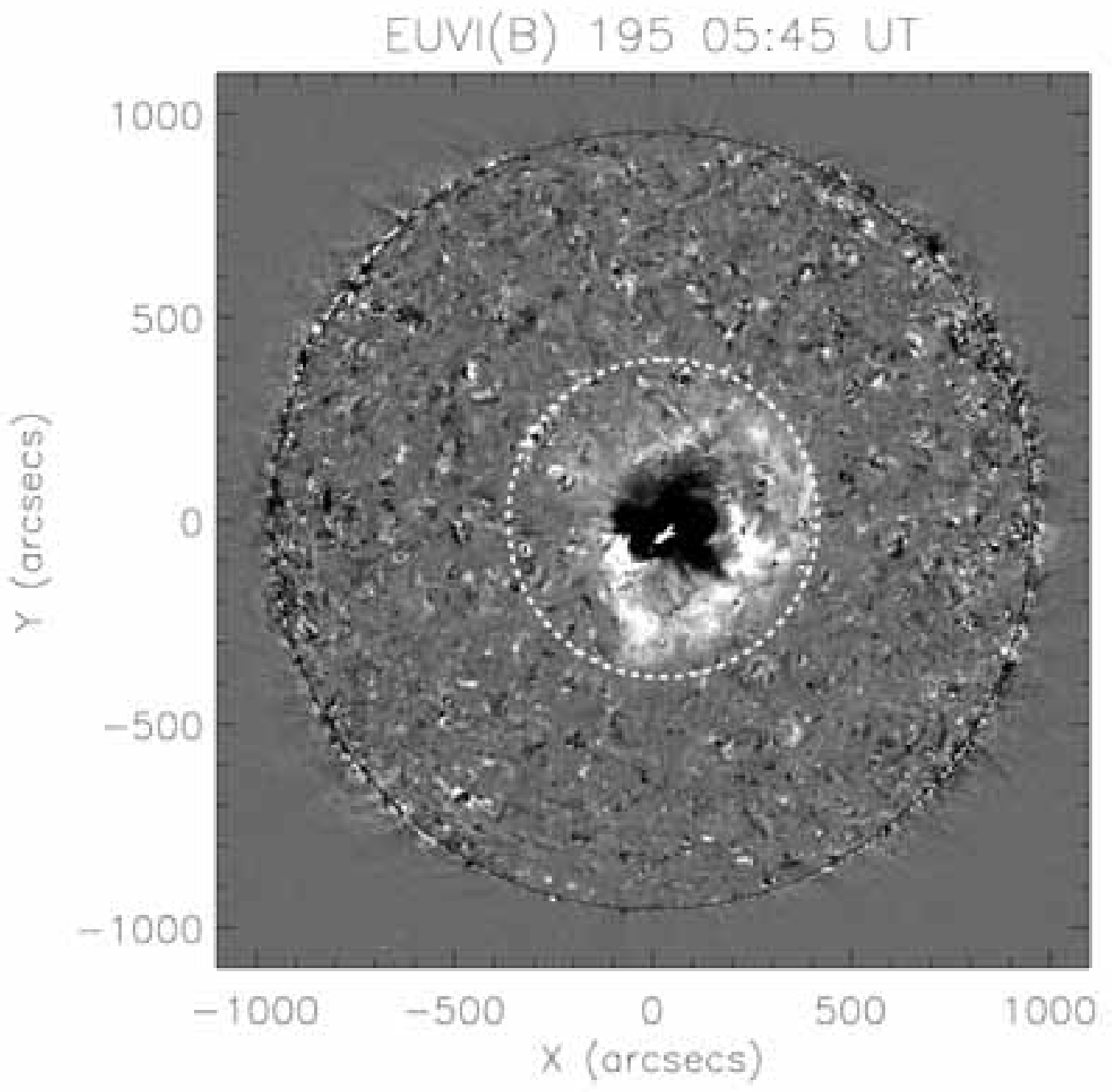} 
\includegraphics[width=1.7in]{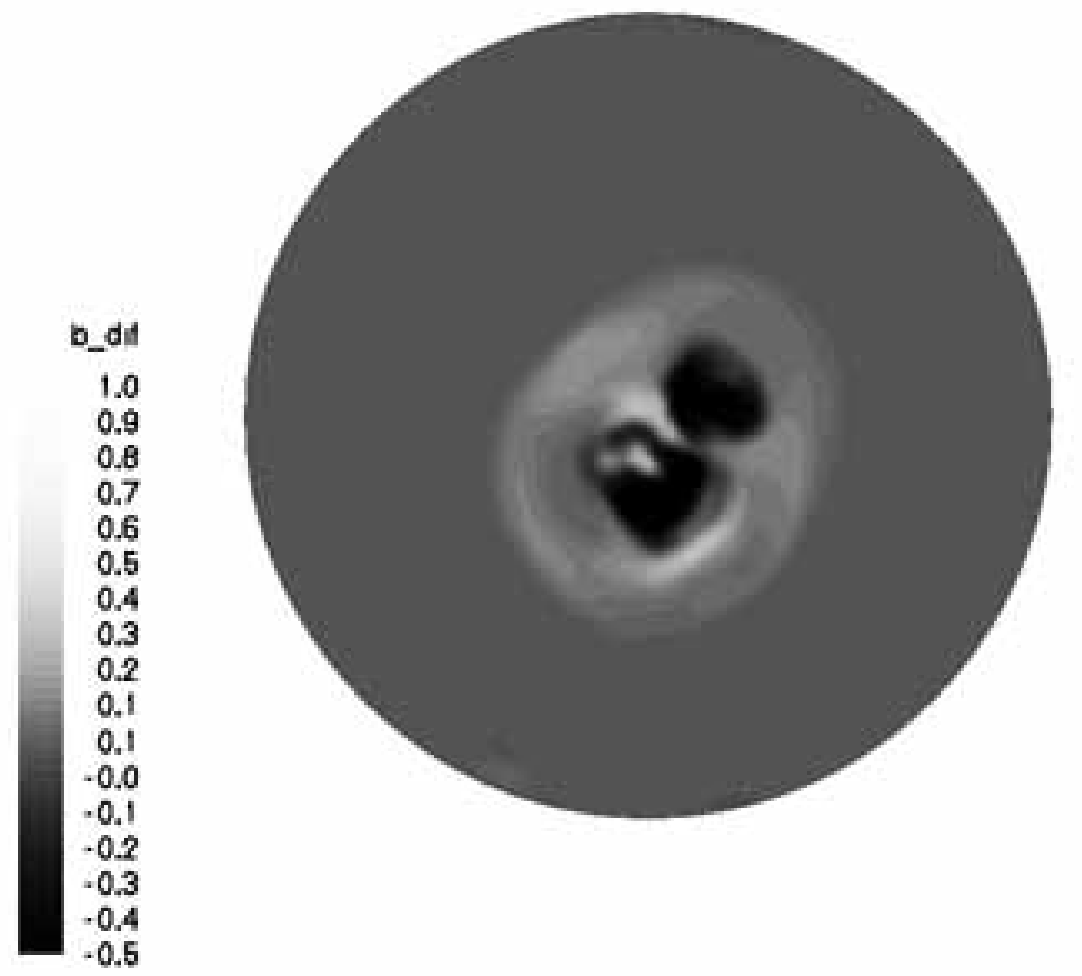}
\includegraphics[width=1.45in]{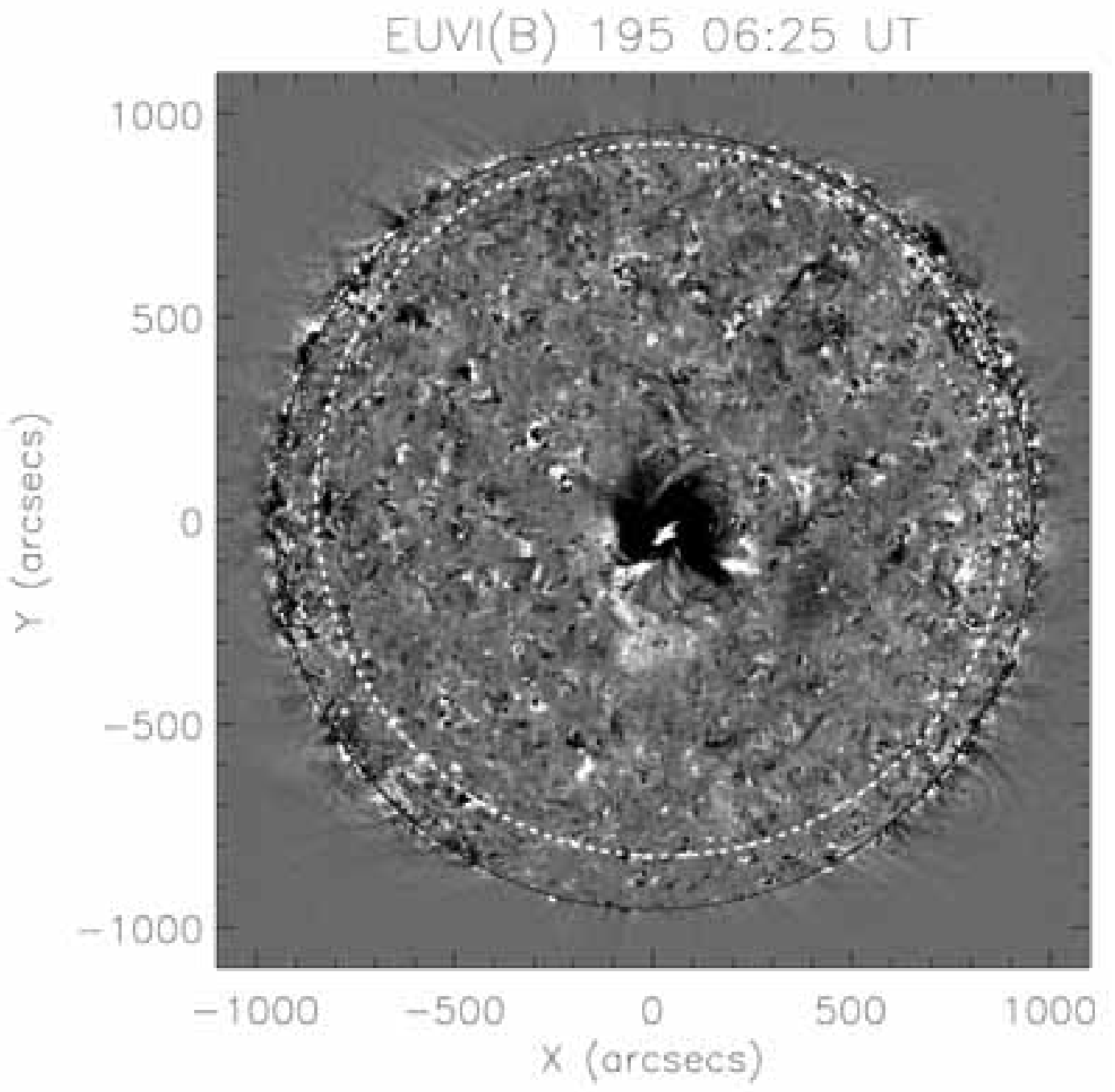}
\includegraphics[width=1.7in]{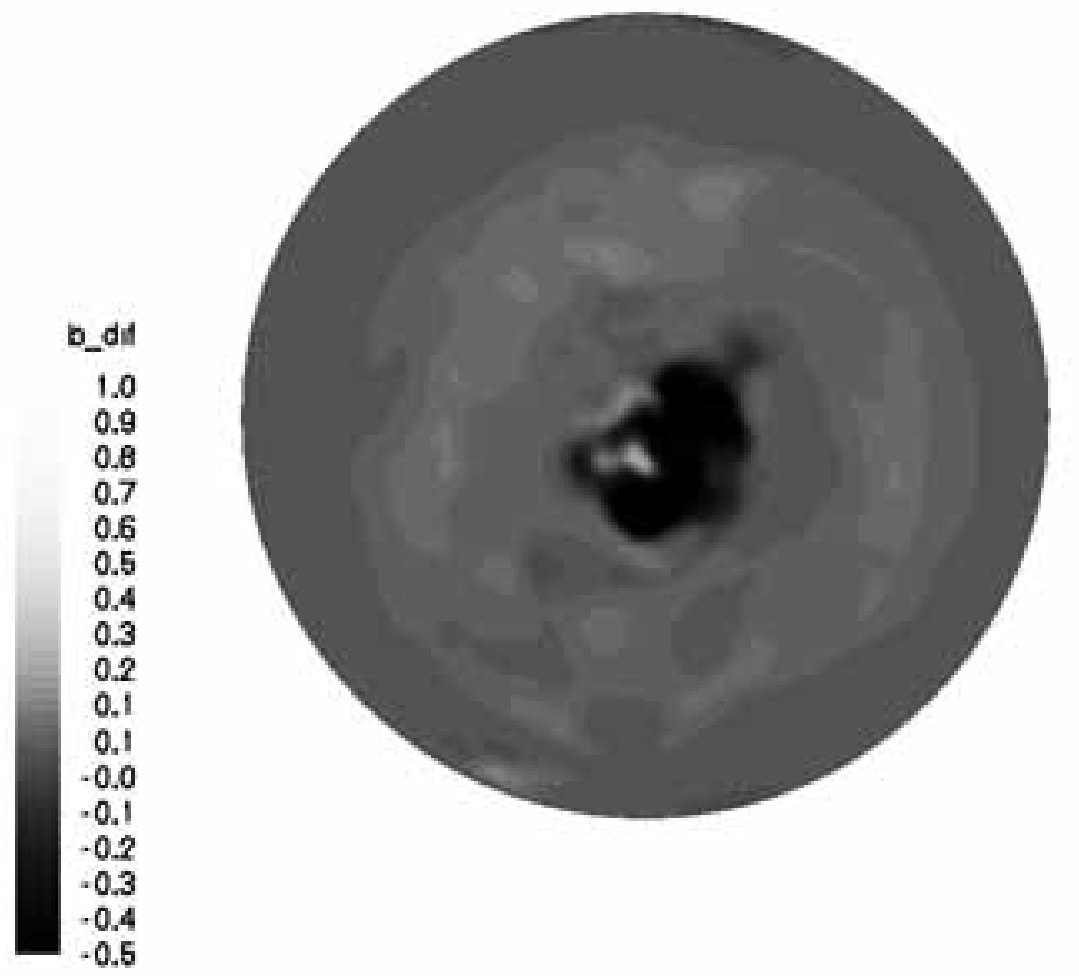}\\
\includegraphics[width=1.45in]{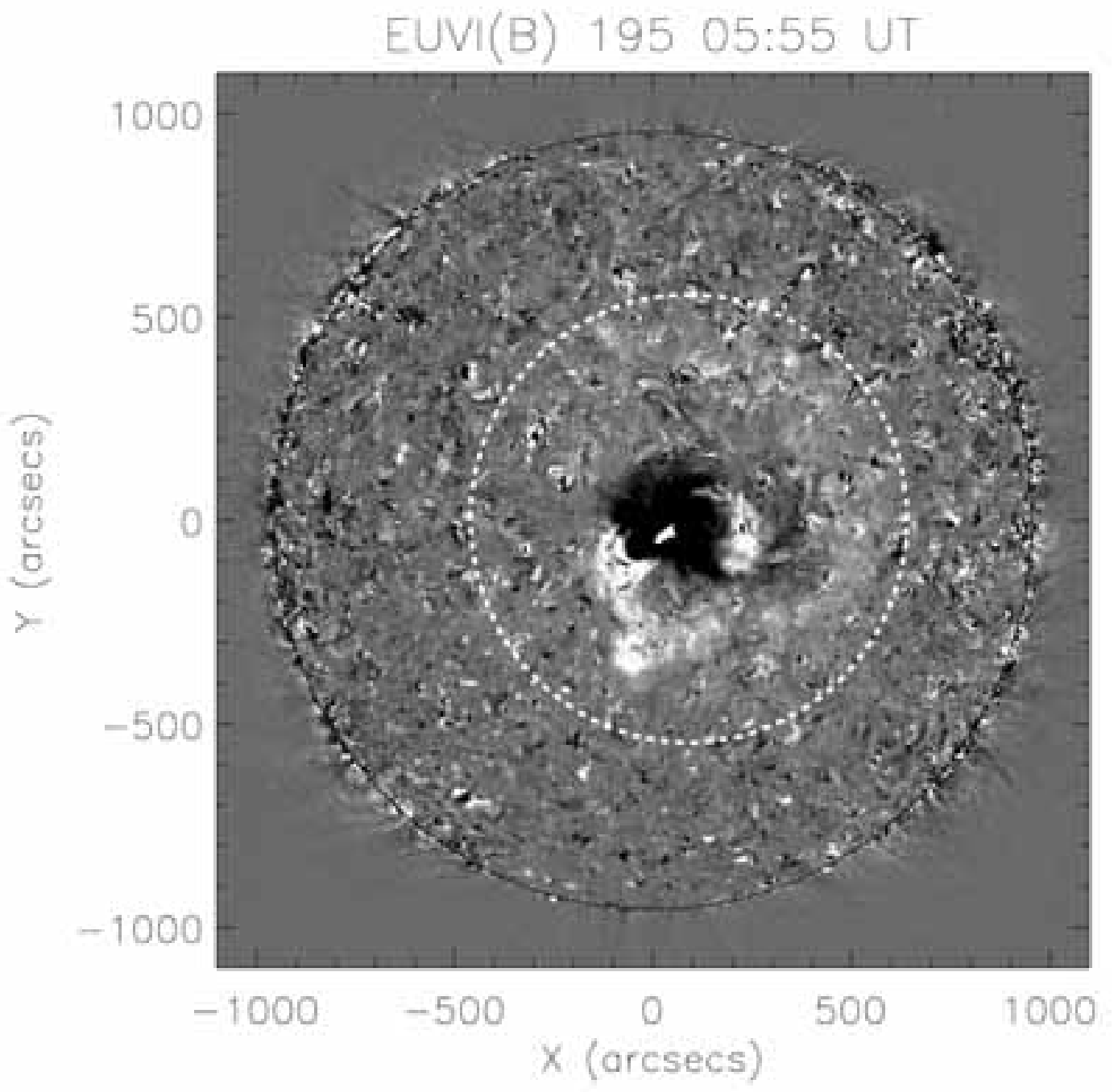}
\includegraphics[width=1.7in]{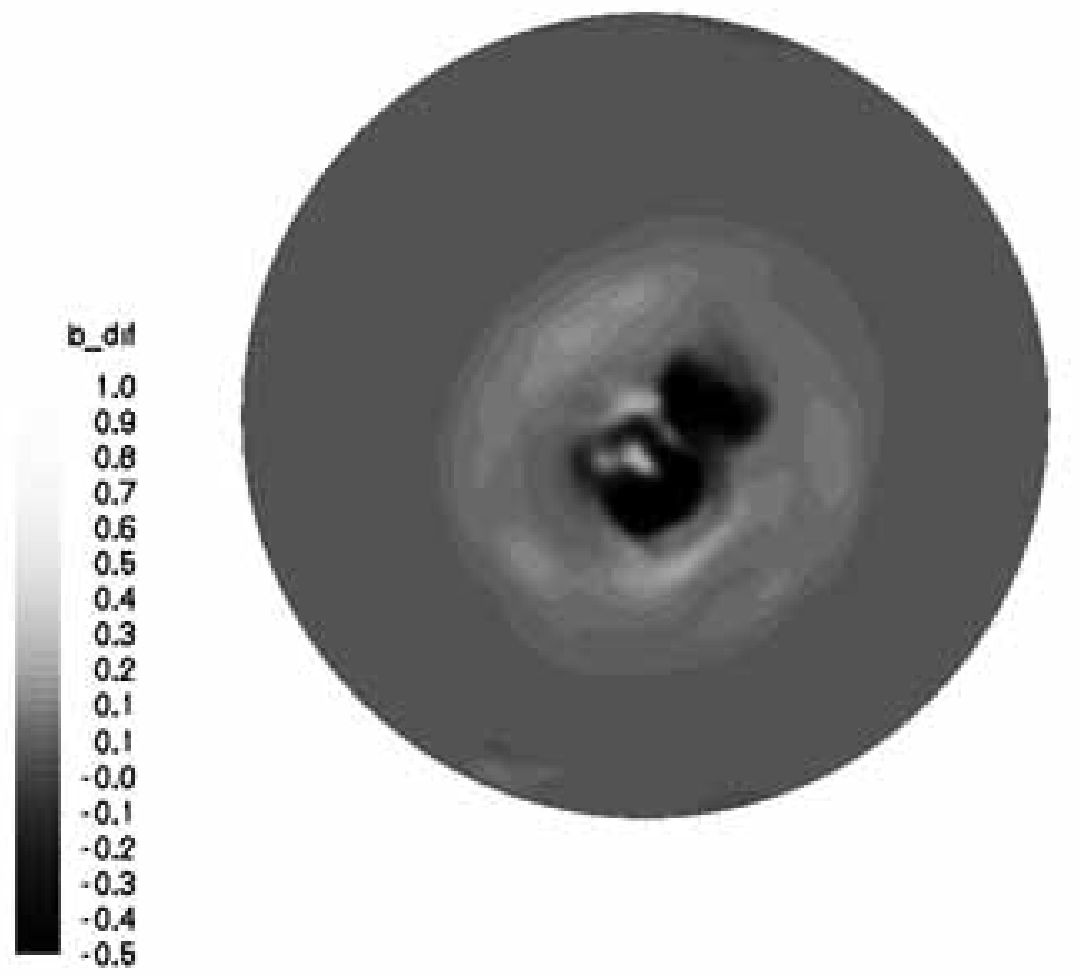}
\includegraphics[width=1.45in]{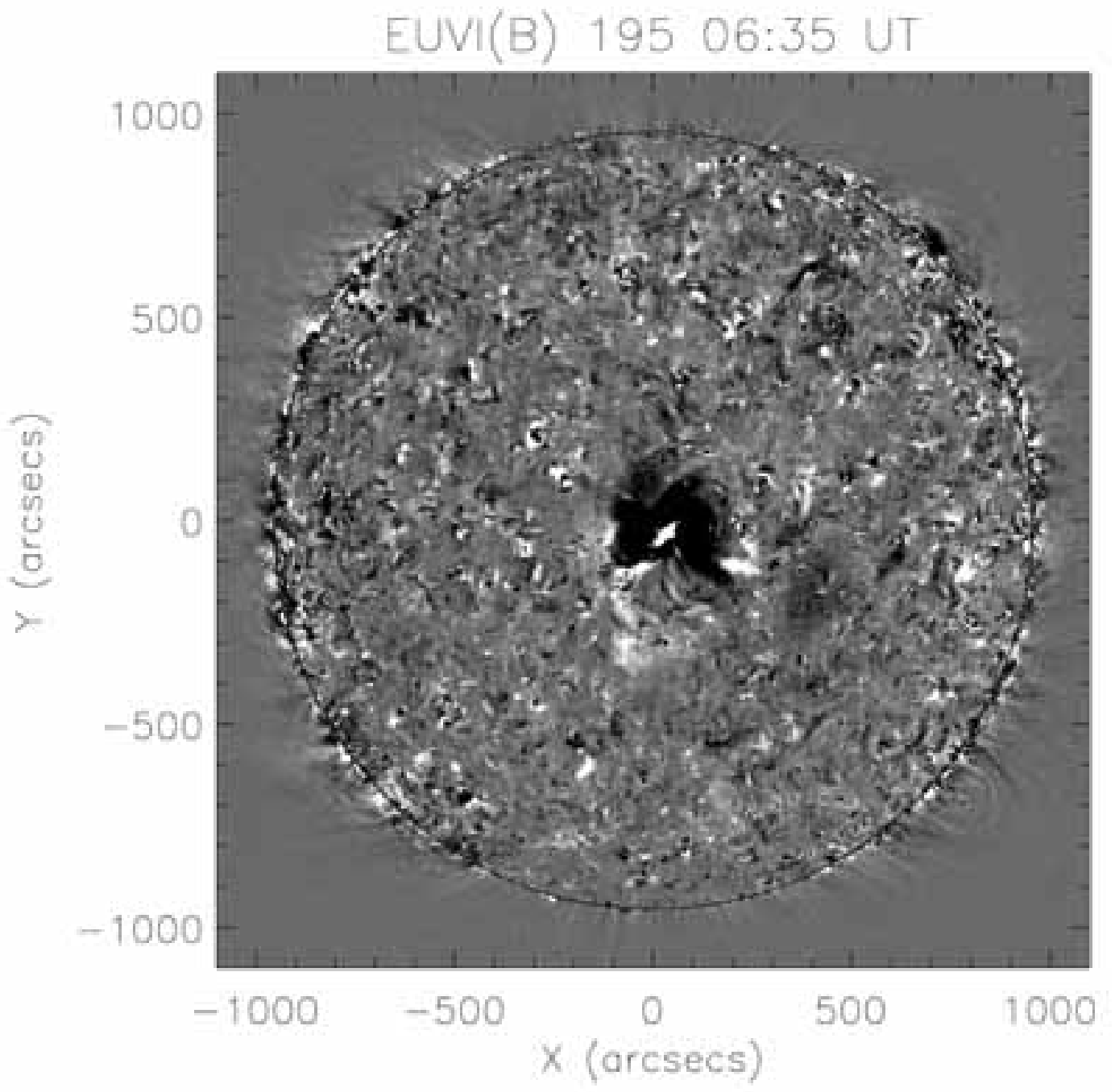}
\includegraphics[width=1.7in]{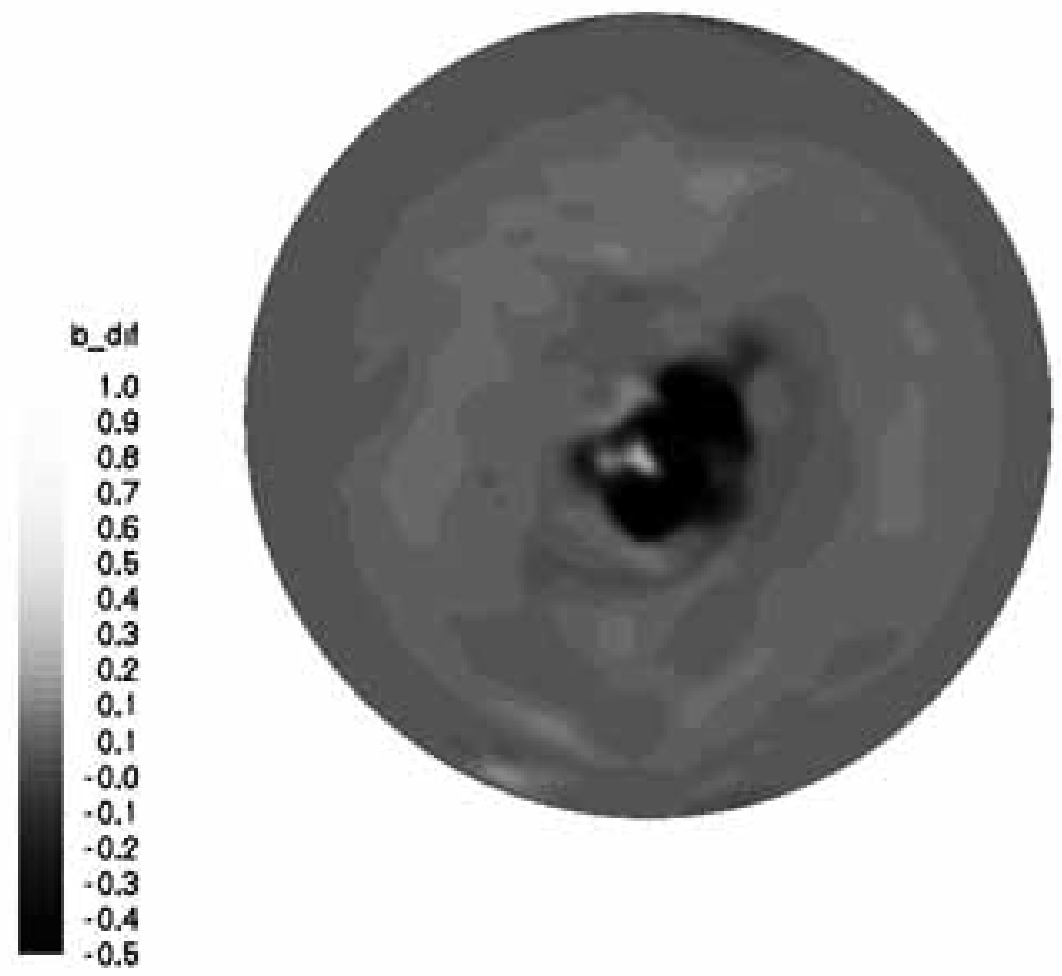}\\
\caption{A time series for the period 5:25-6:35 UT comparing STEREO-B 
EUVI Fe {\sc xii} observations and the simulation results.  Each pair compare STEREO-B EUVI base differences image (left) with base 
differences of simulated mass density at height of $1.1R_{Sun}$ (right). Overlaid on the base difference images are 
dashed white circles which act as a guide and are drawn by eye, to indicate the maximum extent of the bright front.  
The movies for this figure are: {\tt{195b\_diff.mov}} and {\tt{densityfrontdif.mov}.}  The corresponding data from the viewpoint of STEREO-A
are included as movies: {\tt{195a\_diff.mov}} and {\tt{densitysidedif.mov}.}}
\label{fig:f3}
\end{figure*}

Figure~\ref{fig:f3} compares the simulation results with STEREO-B EUVI data for the period 5:25-6:35 UT. The left panel 
of each pair shows an EUVI base difference image, while the right panel shows 
base difference images of the simulated mass density at a height of $1.1R_\odot$ (about 70 Mm).
We choose this height as it is consistent with calculations of the coronal wave bright front altitude from EUVI data 
\citep{Patsourakos09a}, as well as with previous height estimates from SOHO/EIT data (\textsection \ref{sec:Intro}).

We mark the leading edge of the bright front in each base difference EUVI-B image with a dashed white circle. 
We emphasize that this circle has been drawn by eye, and simply indicates the maximum extent reached by the coronal wave in 
a given image.  (The reader is 
encouraged to view the supplemental movies provided with this paper as well: {\tt{195b\_diff.mov}}, {\tt{densityfrontdif.mov}}).  
Figure~\ref{fig:f3} shows that 
the expansion of the simulated wave front is in a good agreement with the observed one. Deviations are probably due to the 
fact that the actual EUVI emissions are not simply a representation of the mass density, but a rather complicated, integrated 
function of it. 

The simulated bright front in Figure \ref{fig:f3} becomes broader as it expands further from the active region. This is 
consistent with observed properties of coronal waves from SOHO/EIT data \citep{Dere97, Klassen00, Podladchikova05a}. 
Areas of decreased mass density (corresponding to the core coronal dimming regions) can also be identified in the simulated 
data. We note that both the real and simulated bright fronts have an increasingly patchy, diffuse nature as the front 
expands away from the active region.
We measured the expansion of the leading edge of the bright front in running difference data from 05:45 - 06:15 UT, which expands
with an average velocity of $\approx$ 260 kms$^{-1}$.  For comparison, the Alfv{\'e}n velocity from the simulation (Figure \ref{fig:f5}) is $\approx$
200 kms$^{-1}$. 

\subsubsection{Two-component bright front}
\label{subsec:two_component_bf}

The simulations show a predominantly two-level bright front.  
Figure \ref{fig:f3} shows that the highest intensity is concentrated in patches located within a weaker, broadening front.  
Although these two components expand in a coupled manner during the early phase of the CME, during the later frames,
the simulated wave front is increasingly dominated by the weaker intensity component, which continues to expand ({\tt{densityfrontdif.mov}}; {\tt{densitysidedif.mov}}).

The observations become increasingly noisy as the event progresses and the coronal wave becomes more and more difficult to detect.  
In EUVI-A base difference data, 06:25 UT is the last frame where the bright front to the west of the active region is discernable ({\tt{195a\_diff.mov}}).
In the EUVI-A base difference simulation, the higher intensity patch to the west of the active region is visible until 06:35 UT ({\tt{densitysidedif.mov}}).

In the EUVI-B base difference data, the furthest reaches of the bright front (most obviously near to the south polar coronal hole)
can be identified until 06:55 UT.  The bright front is approximately stationary at this time ({\tt{195b\_diff.mov}}).  
The EUVI-A and B base difference simulations show that isolated higher intensity patches still exist at 07:15 UT.  
Near to the south polar coronal hole, the higher intensity patch is located at the same place from 06:45 UT.  
Near to the north polar coronal hole, a new higher intensity
patch develops from 07:05 UT, also remaining at the same location.

\subsection{Comparison of CME and Coronal Wave Expansion}
\label{subsec:comparison_cme_cw}

\begin{figure*}[h!]
\centering
\includegraphics[width=6.in]{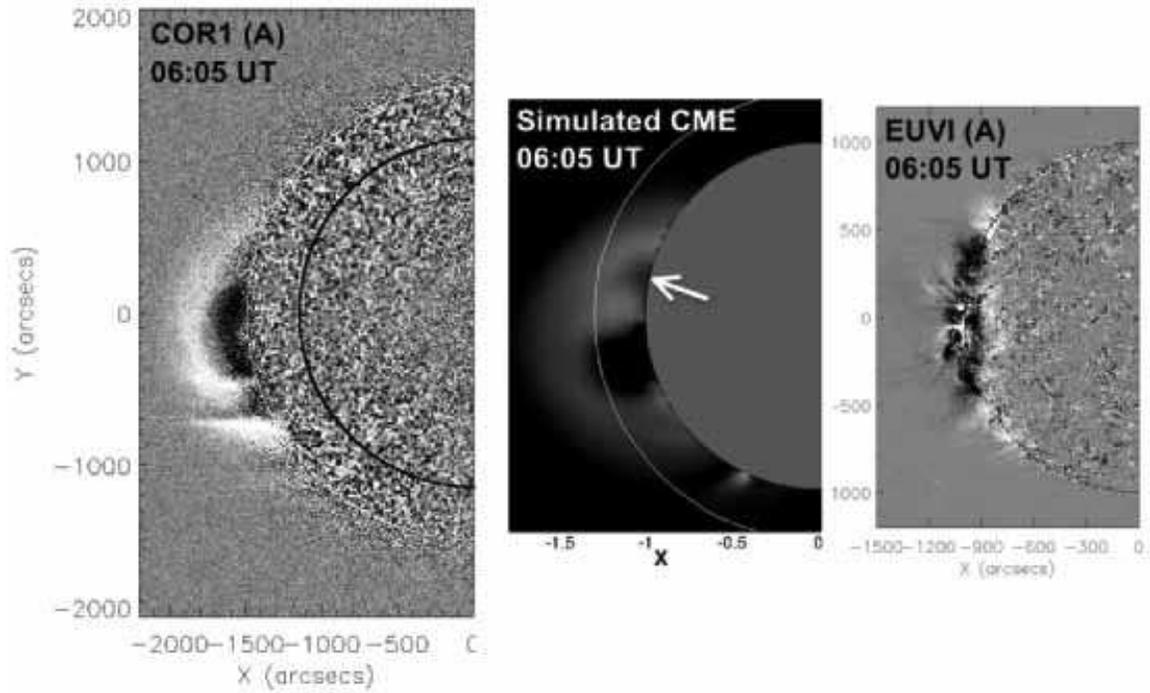}
\caption{Left panel shows the COR1-A running difference image at (06:05 - 05:55) UT.  Center panel shows the 
simulated white-light base difference image at (06:05 - 05:35) UT. Right panel shows the EUVI-A running difference 
image at (06:05 - 05:55) UT. All images are scaled to the same size.  This figure demonstrates the importance of 
the role of the simulation in developing an understanding of the true lateral extent of the CME in the low corona. 
Fitting the CME as observed in the COR1 data alone \citep[e.g. see Figure 4,][]{Patsourakos09b}, a large part of the 
CME in the low corona is missed. When proper consideration of the CME extent in the low corona is made,
the lateral extent of the CME maps to the coronal wave observed in EUVI data. The white 
arrow in the center panel indicates a region which is also part of the CME cavity (c.f. bottom panels, Figure \ref{fig:f2} at 
06:35 UT).} 
\label{fig:f6}
\end{figure*}

Figure \ref{fig:f6} compares snapshots at 06:05 UT from COR1-A, the simulated CME, and EUVI-A, all scaled to the
same size. The COR1-A and EUVI images are running 
difference images, the simulation is a base difference
image.  \cite{Patsourakos09b} show a fit of the 3D CME model of 
\cite{Thernisien06,Thernisien09} to the COR1-A data for this event at 06:05 UT. This leads them to determine an extent for the CME in 
the low corona that is much too small to match the coronal wave in the corresponding EUVI data, since the 3D model is 
essentially made up of a spherical bubble attached to a conical leg. We note that a similar approach was also used in 
\cite{Patsourakos09a}. In both papers, the authors interpret the apparent misfit between the CME model extension in the 
low corona and the EUVI coronal wave as evidence that the coronal wave and CME are different structures and conclude that the 
coronal wave is a fast-mode MHD wave.

Our simulation results at 06:05 UT are shown in the middle panel of Figure \ref{fig:f6}. The simulation gives information about 
the low corona below 1.3 R$_{\odot}$ (200 Mm), the region obscured by the occulting disk in the COR1 data.  Comparison 
of the middle and right panels of Figure \ref{fig:f6} show that the extension of the CME in the low corona maps 
very well to the coronal wave in the EUVI base difference data.  

In particular, the simulation results show a secondary cavity located to the north of the main CME cavity (marked by 
the white arrow in the middle panel of Figure \ref{fig:f6}). Comparison with the corresponding EUVI base
difference data shows secondary coronal dimmings
developing at the same location (right panels, EUVI (A), Figure \ref{fig:f4}). 
Despite the lack of spectral 
diagnostics for secondary dimmings, this correlation between the secondary CME cavity and the secondary coronal 
dimmings is consistent with plasma evacuation.

A time-series movie of the simulated COR1-A white-light data ({\tt{COR1joint1.mov}}) shows that the secondary cavity expands and merges with 
the main CME cavity, so that the low corona is really ``opened'' to a large lateral extent. This analysis demonstrates 
the important role of the simulation in developing an understanding of the true lateral extent of the CME in the low corona.  
 
In \textsection \ref{subsec:two_component_bf} we noted that the higher intensity patches of the coronal wave front no longer
expand as of $\sim$ 06:55 UT.
Correspondingly, the simulation results show that the CME has stopped expanding significantly in a lateral direction by this time.
(The reader is referred to movie {\tt{COR1joint1.mov}}).
  
\subsection{Three Dimensional Magnetic Field Topology}
\label{subsec:3DMFT}

\begin{figure*}[h!]
\centering
\includegraphics[width=2.0in]{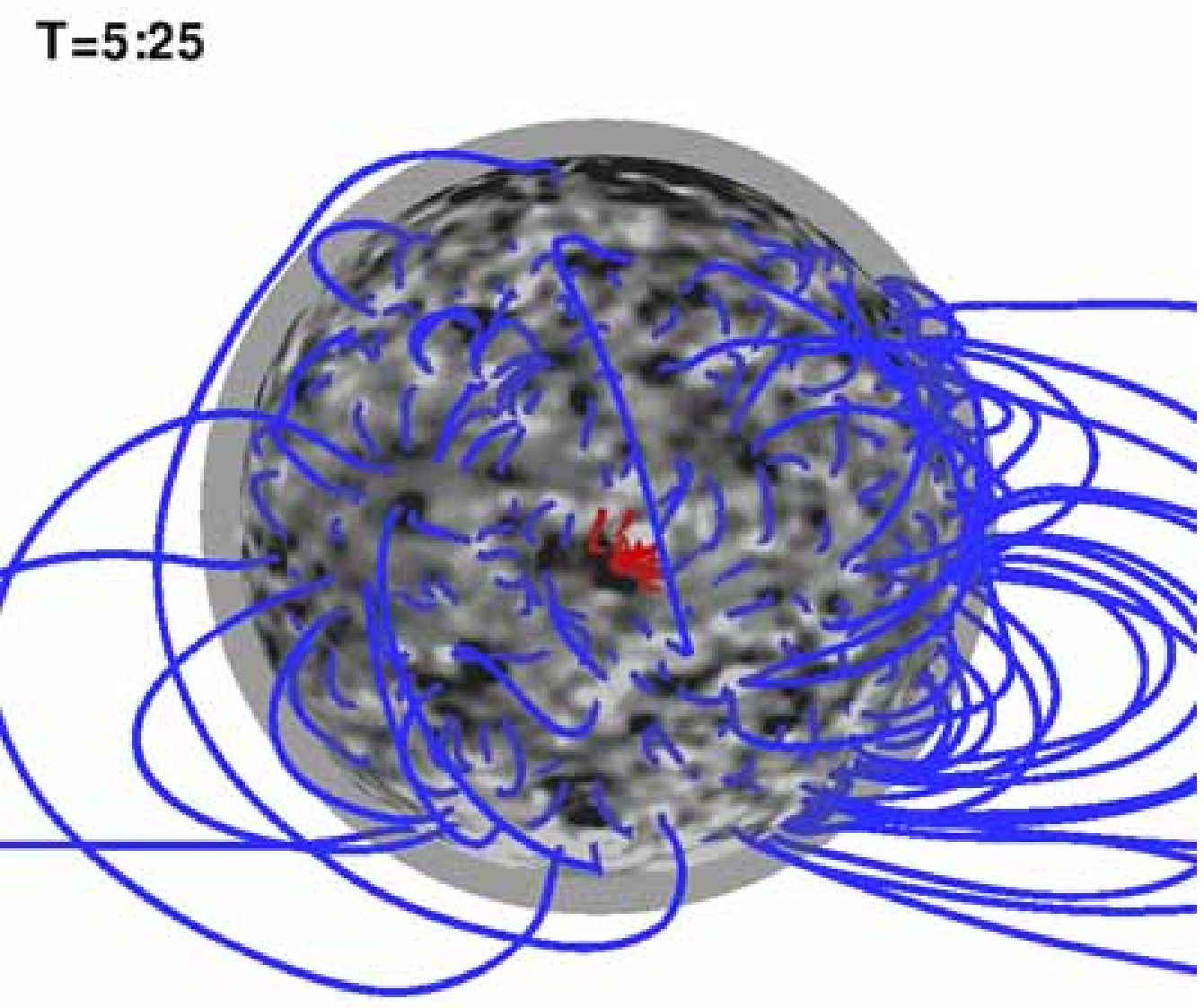} 
\includegraphics[width=2.0in]{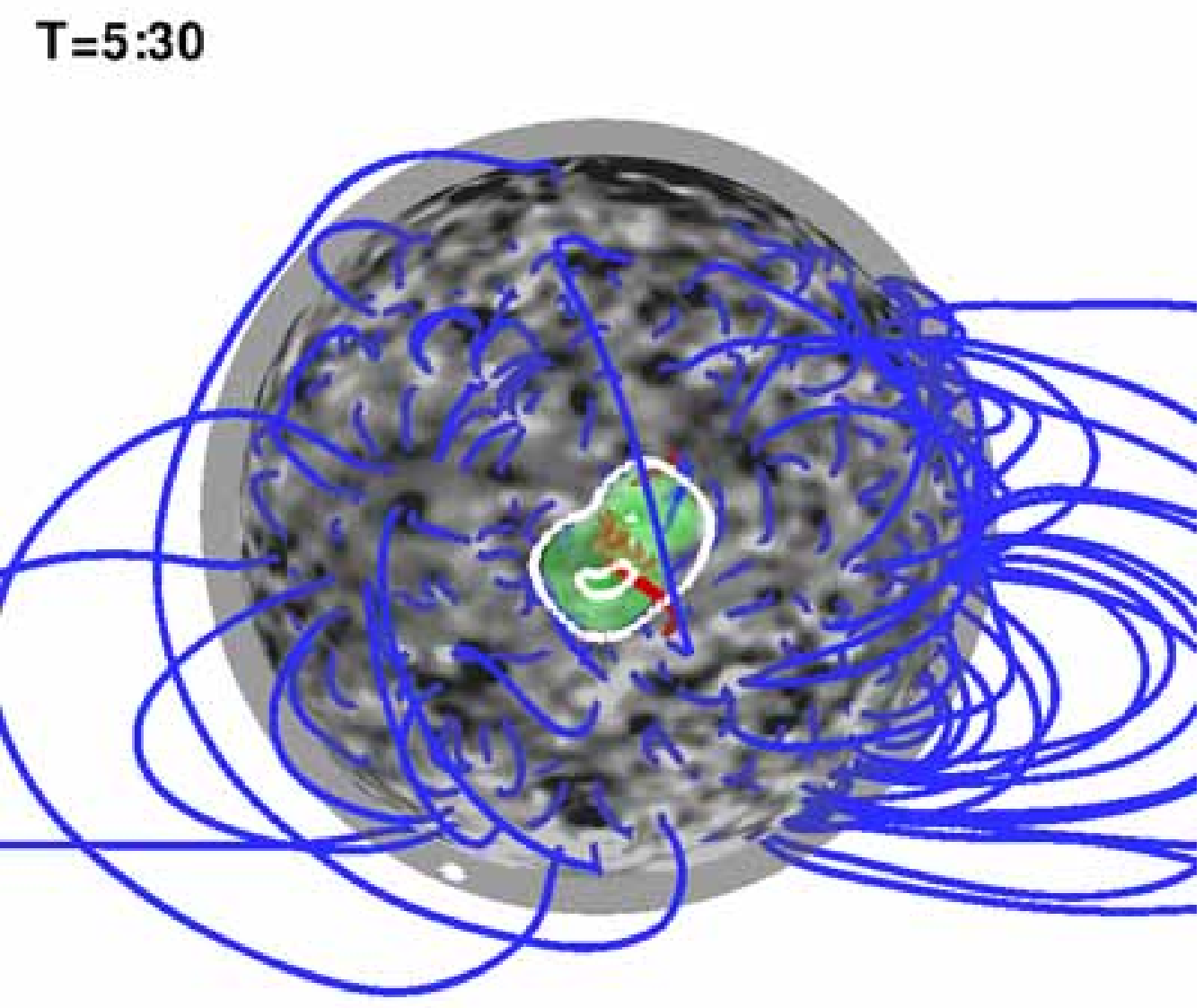}  
\includegraphics[width=2.0in]{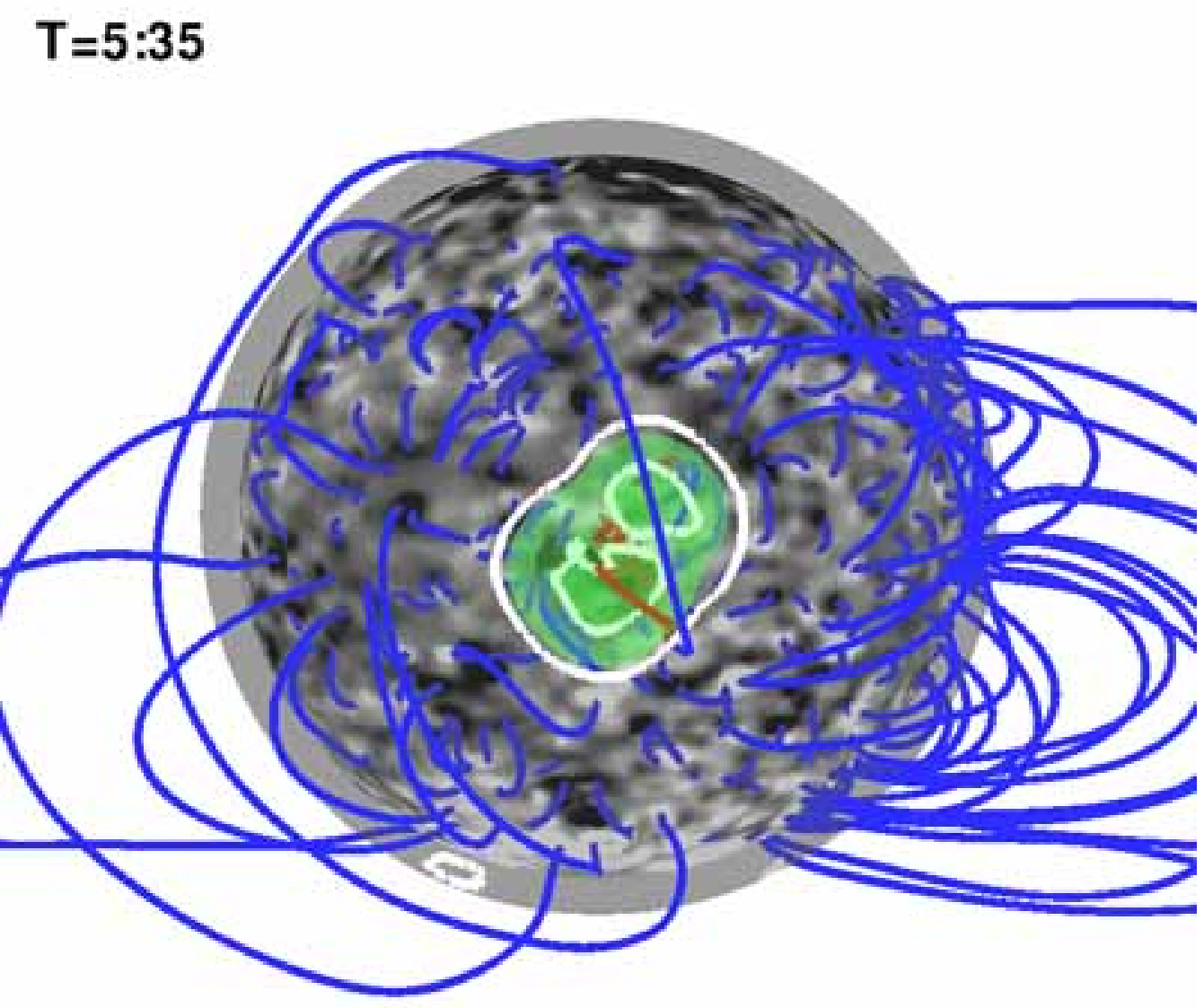} \\
\includegraphics[width=2.0in]{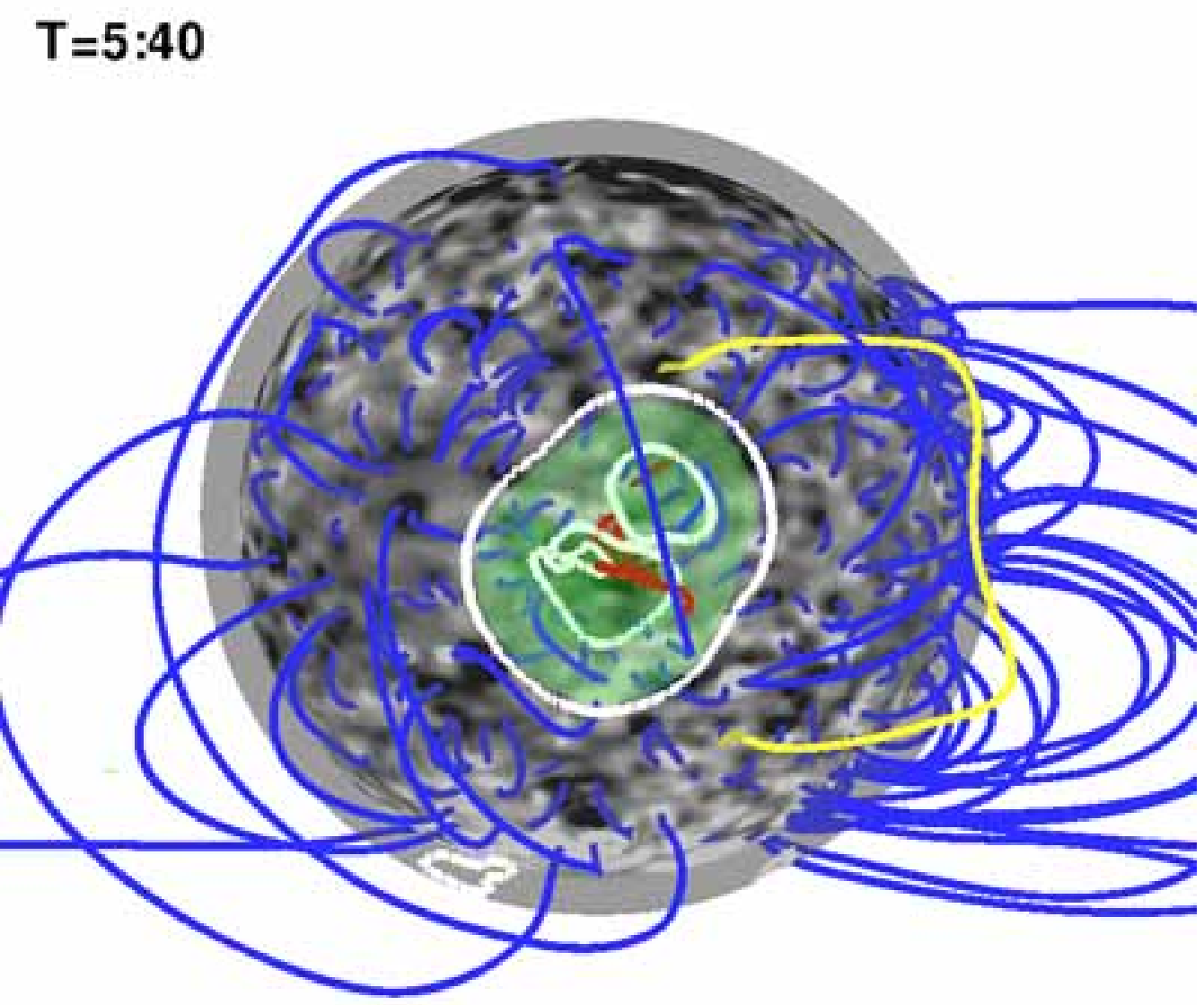}  
\includegraphics[width=2.0in]{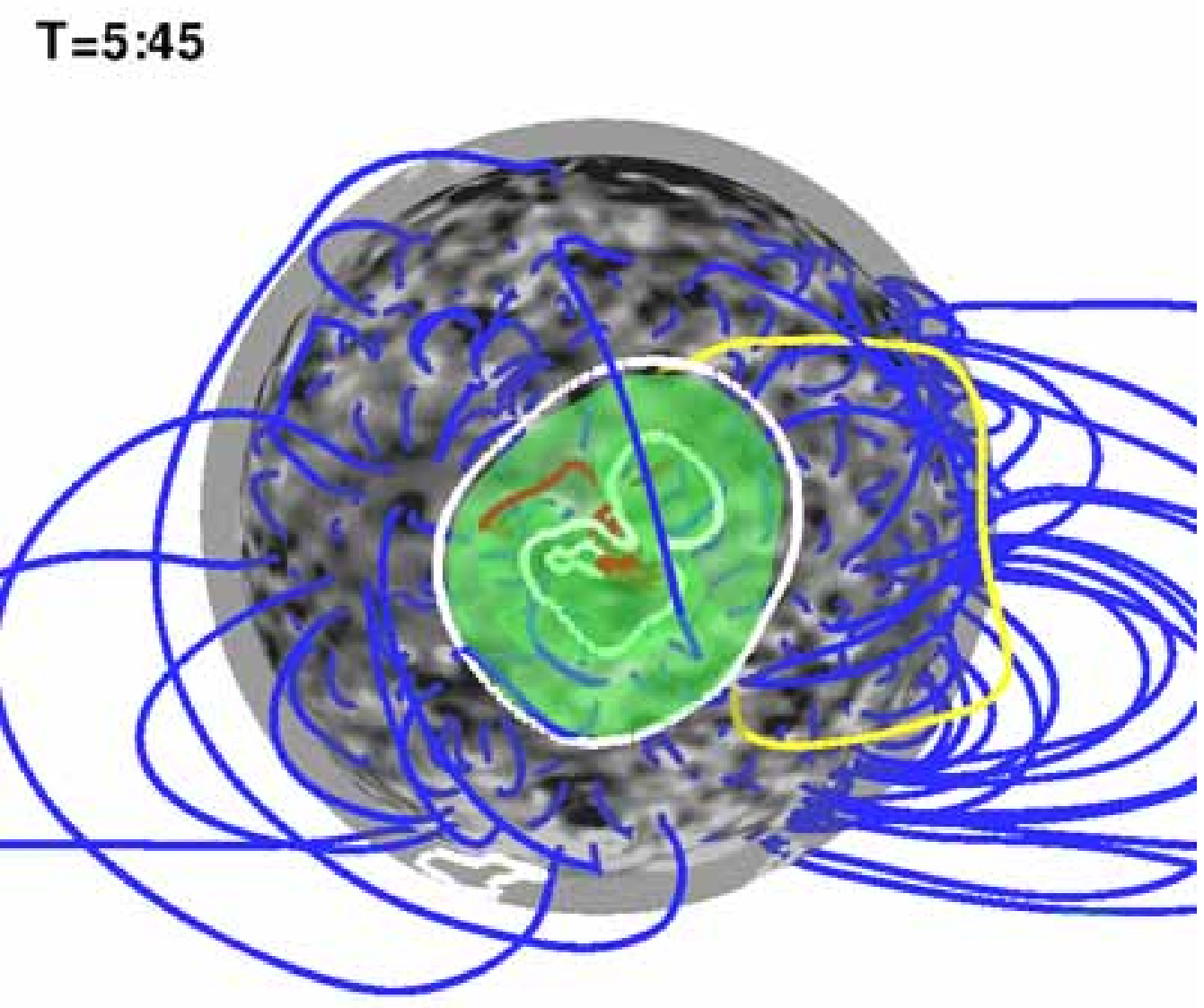} 
\includegraphics[width=2.0in]{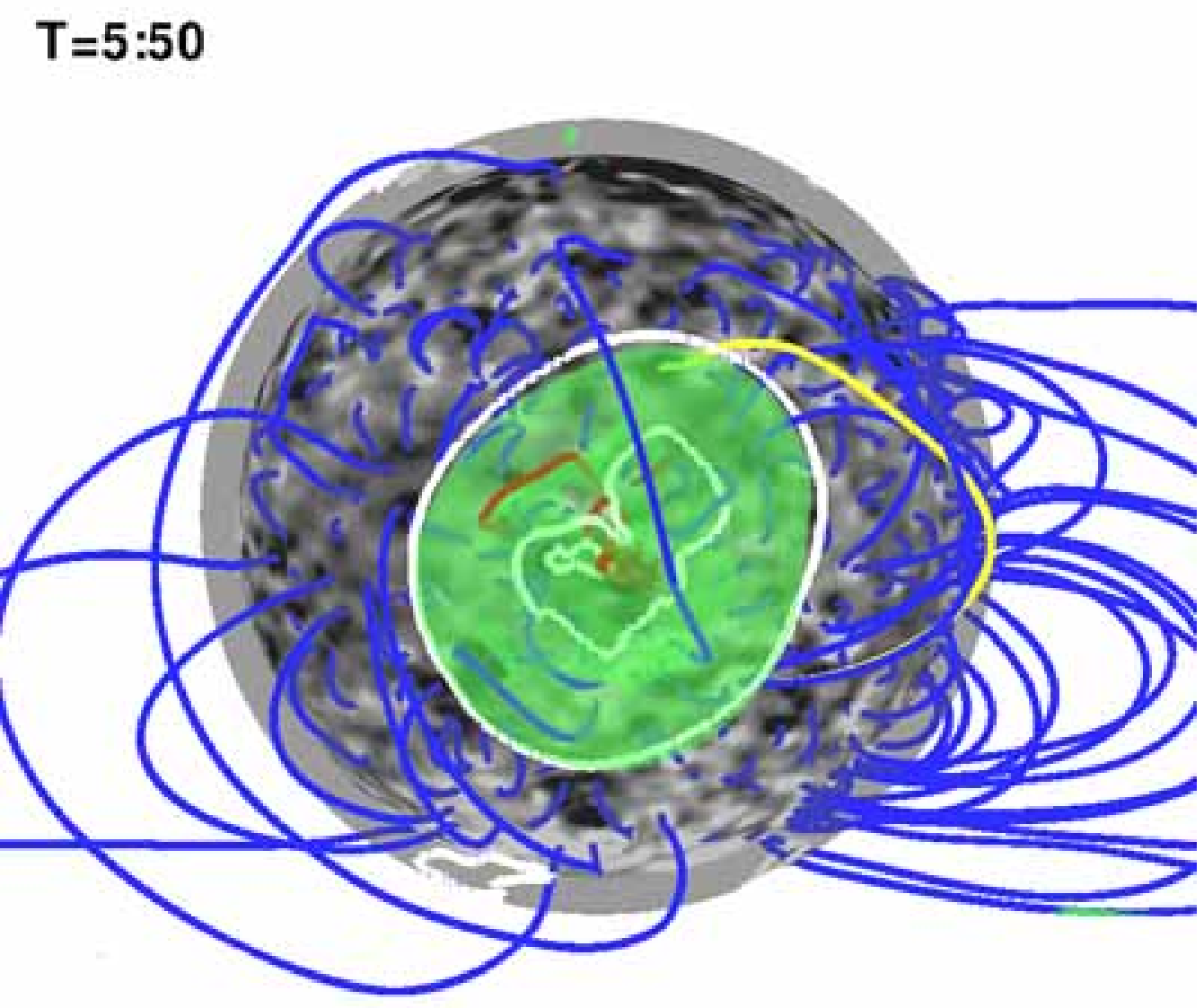} \\
\includegraphics[width=2.0in]{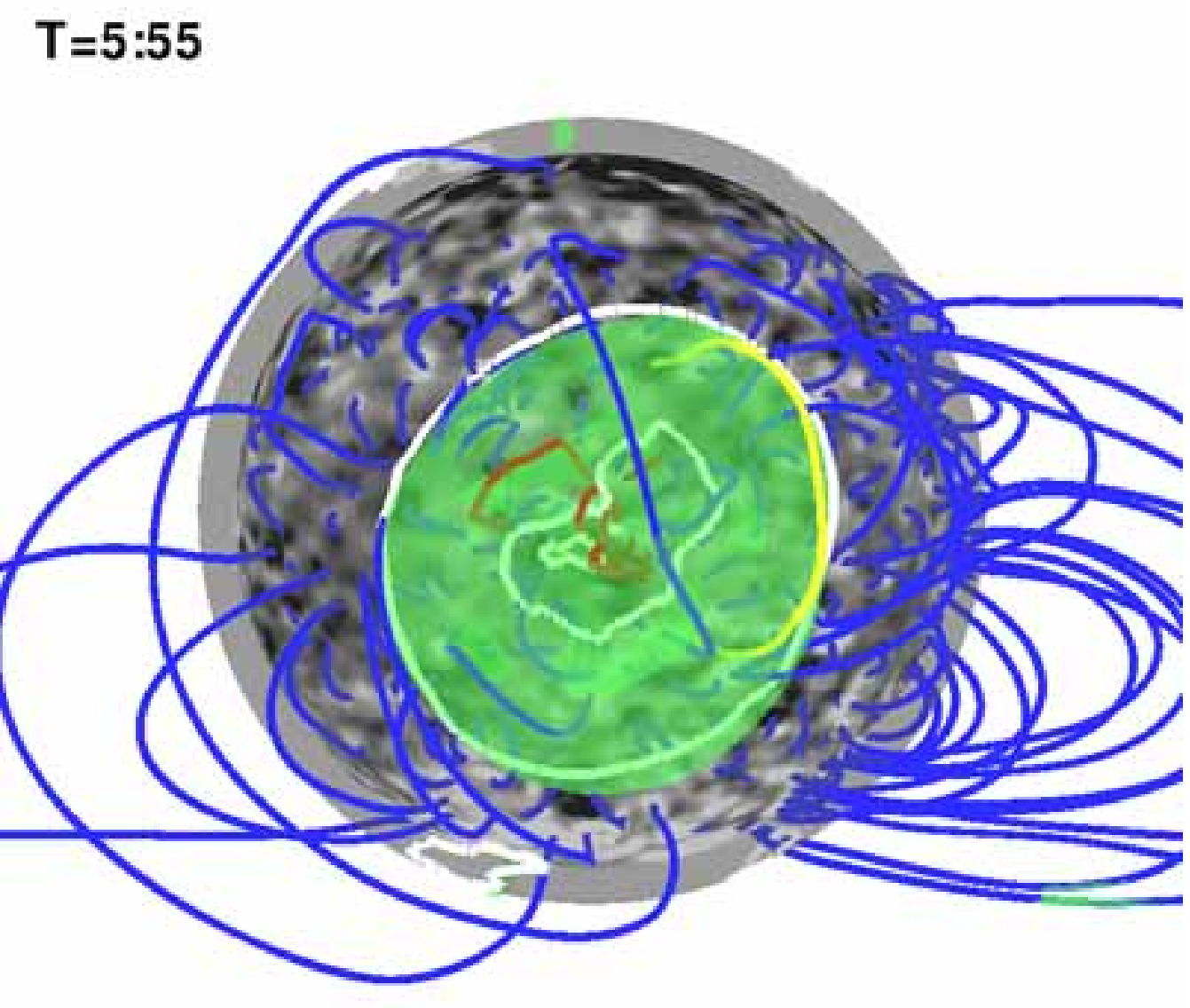} 
\includegraphics[width=2.0in]{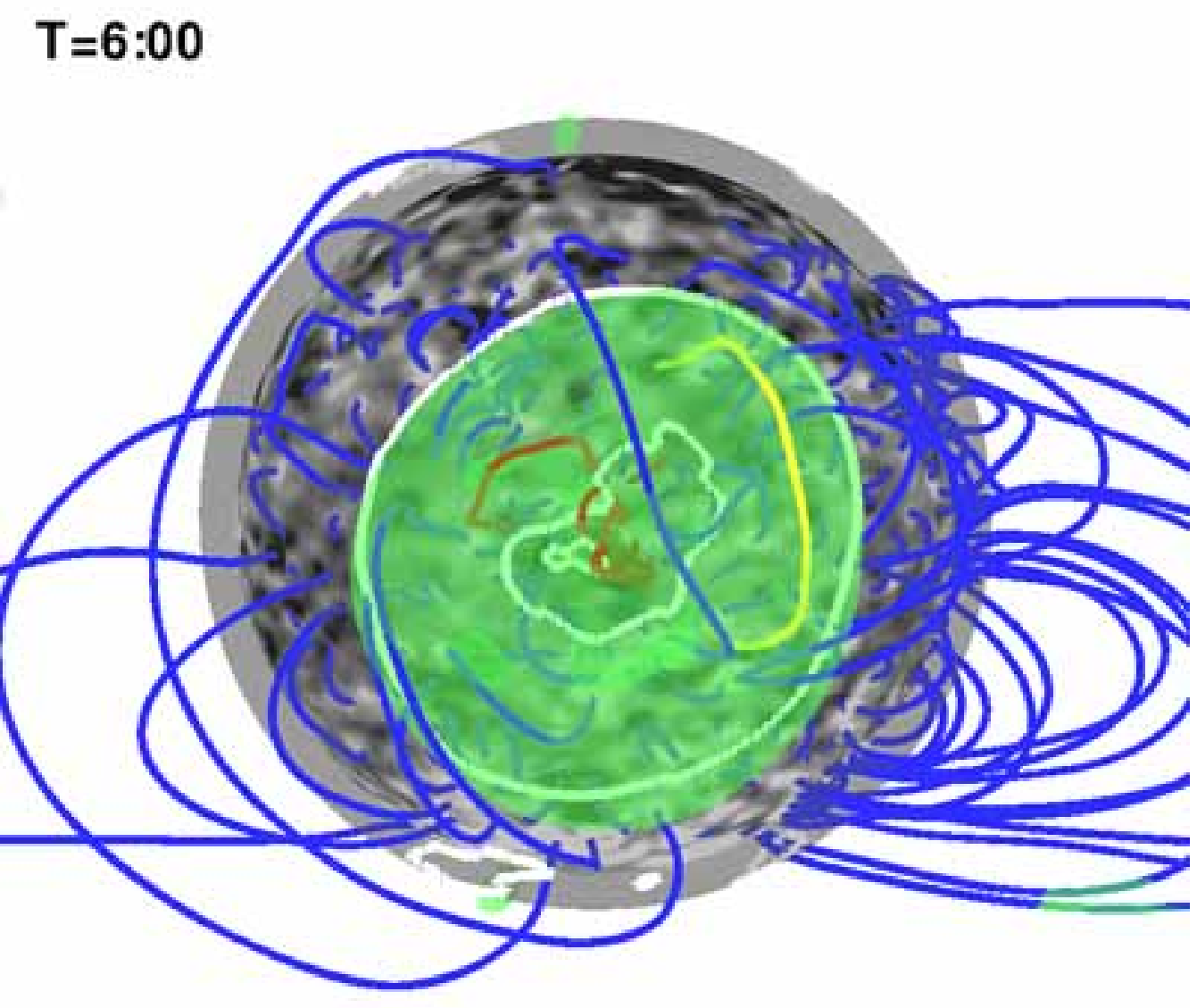} 
\includegraphics[width=2.0in]{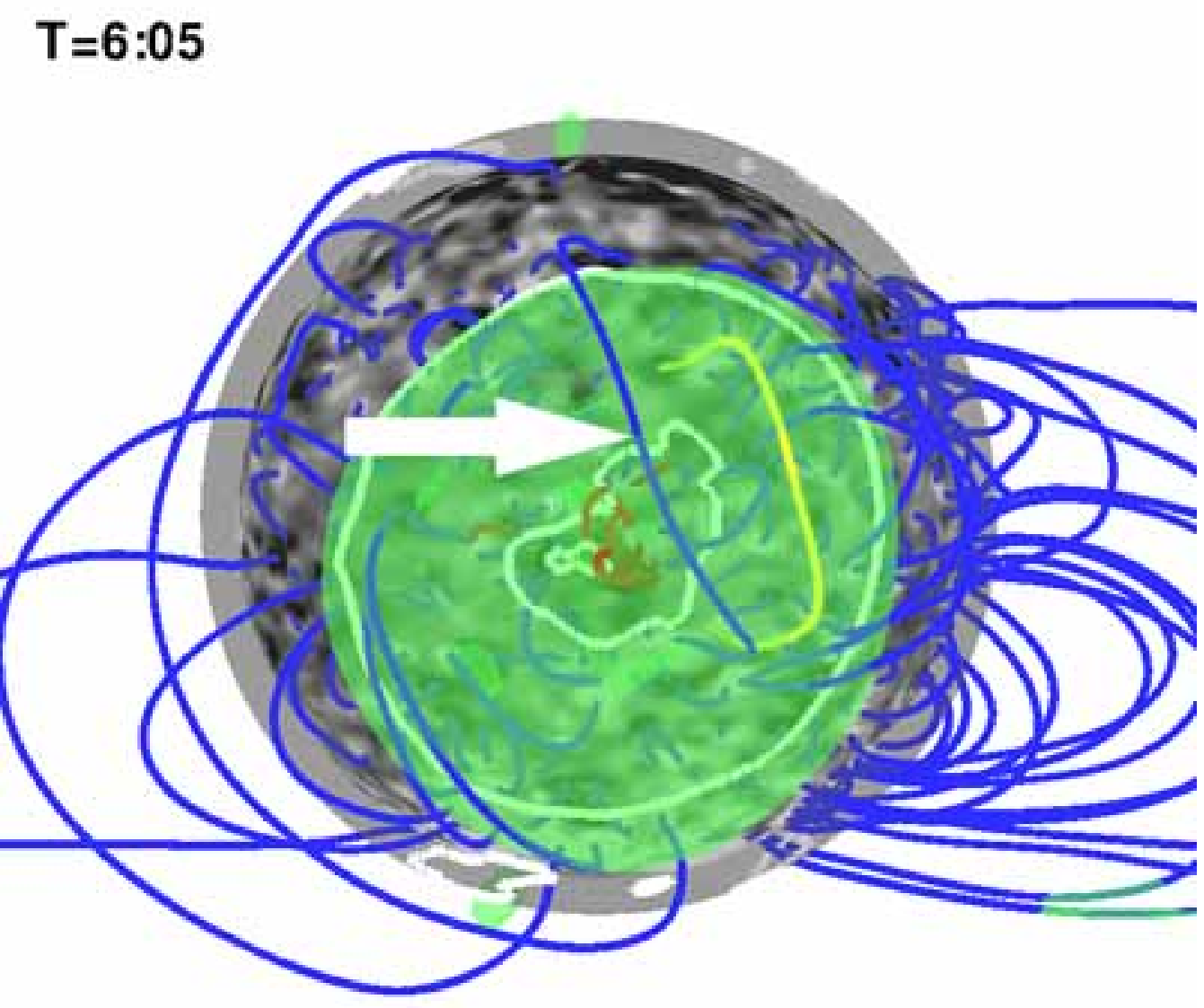} \\
\includegraphics[width=2.0in]{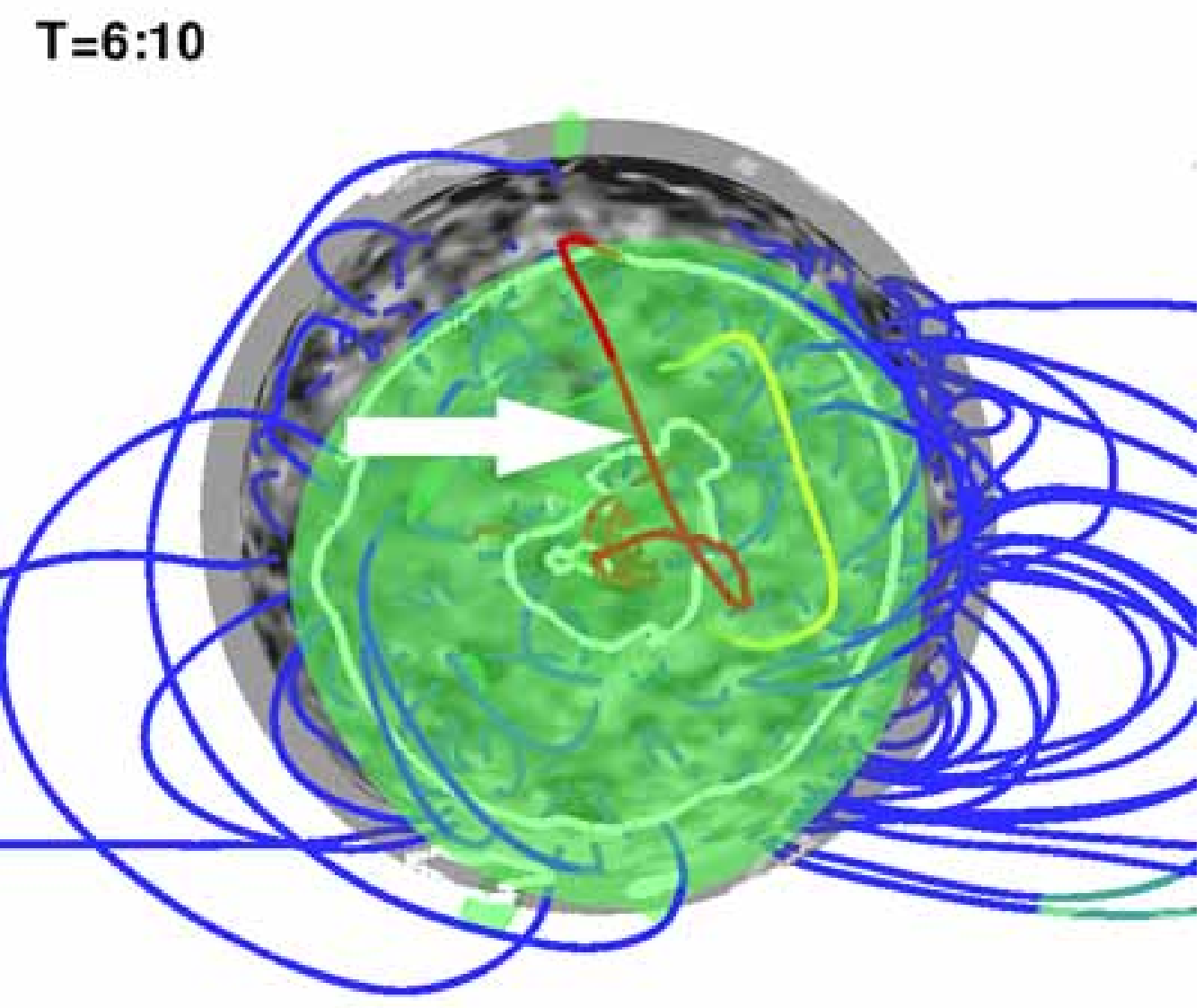} 
\includegraphics[width=2.0in]{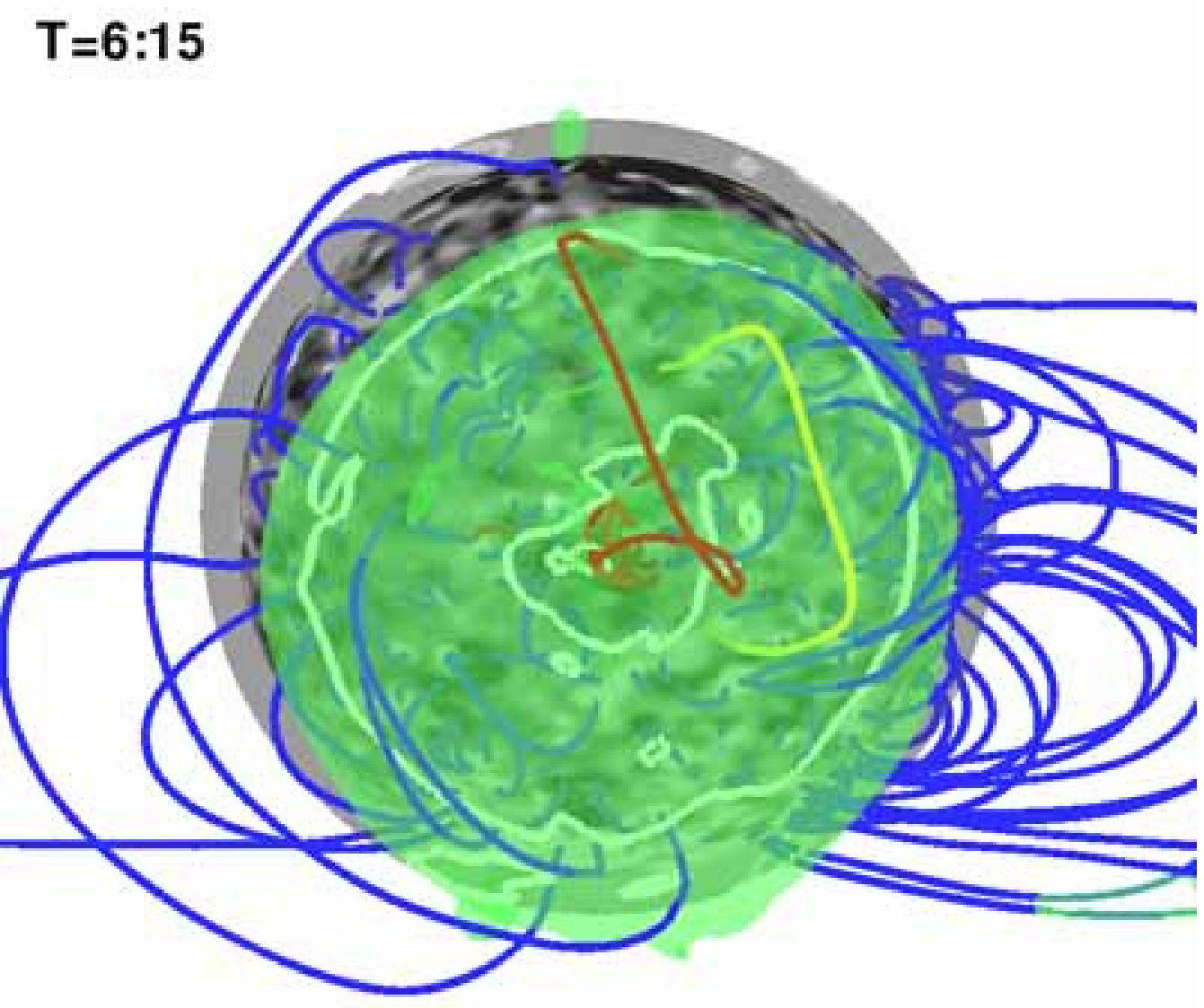} 
\includegraphics[width=2.0in]{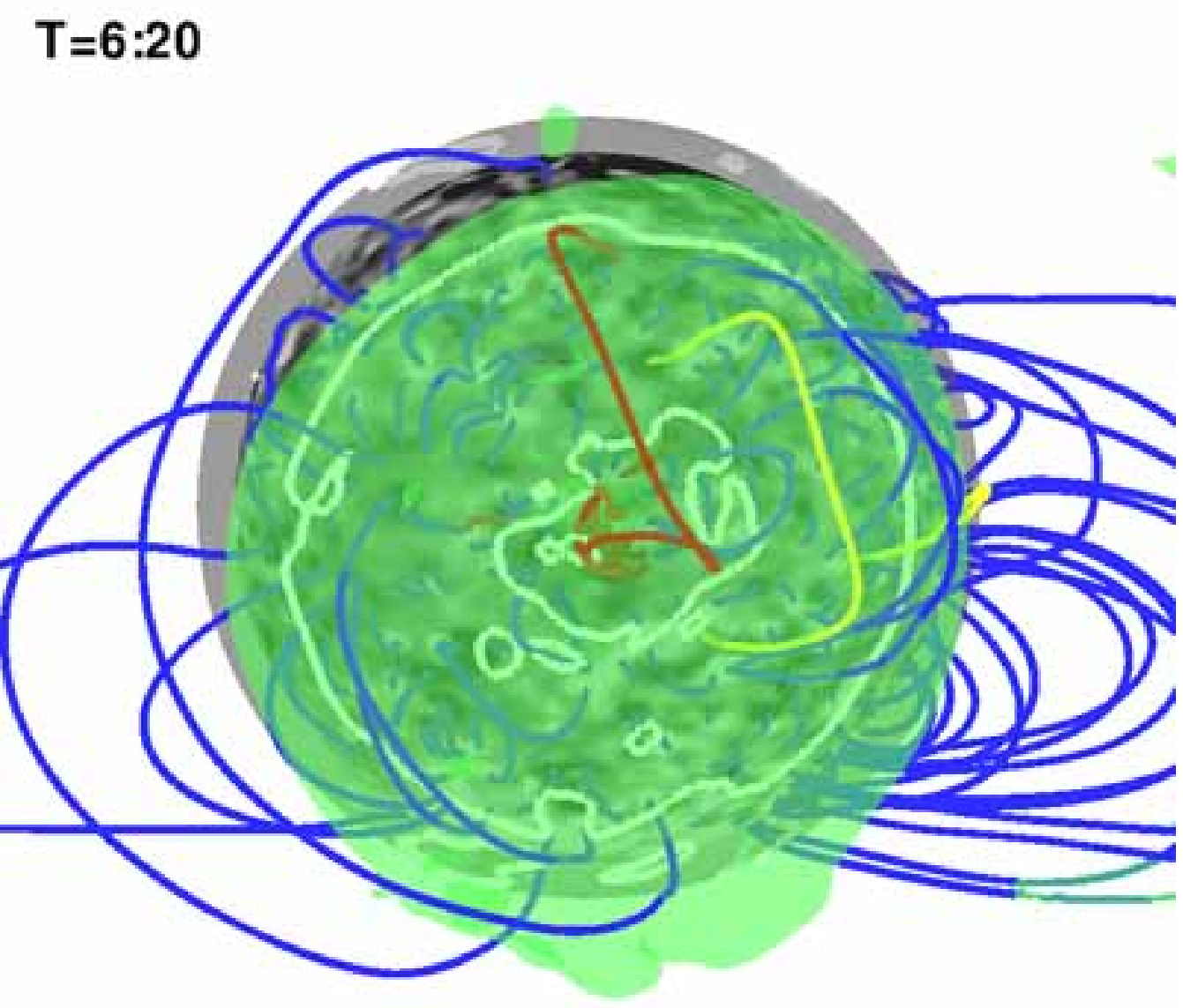} 
\caption{A time series (from 05:25 - 06:20 UT, for frames until 07:15 UT see movie {\tt{SB.mov}}) of the simulation results matched to STEREO-B viewing angle. Inner sphere 
shows the photosphere with the radial magnetic field strength (magnetogram data). Outer sphere 
is at a height of $1.1R_\odot$ colored with white line contours of mass density 
base difference (same as Figure \ref{fig:f3}). The light Green shade 
represents an iso-surface of mass density with a base ratio of 1.1. The flux rope core field lines are drawn 
in Red and surrounding field lines are drawn in Blue. A Blue field line is changed to Red if it reconnects 
with a flux rope core field line, and to Yellow if it reconnects with a surrounding field line.
This figure is also included as a movie: {\tt{SB.mov}.} }
\label{fig:f7}
\end{figure*}

\begin{figure*}[h!]
\centering
\includegraphics[width=2.0in]{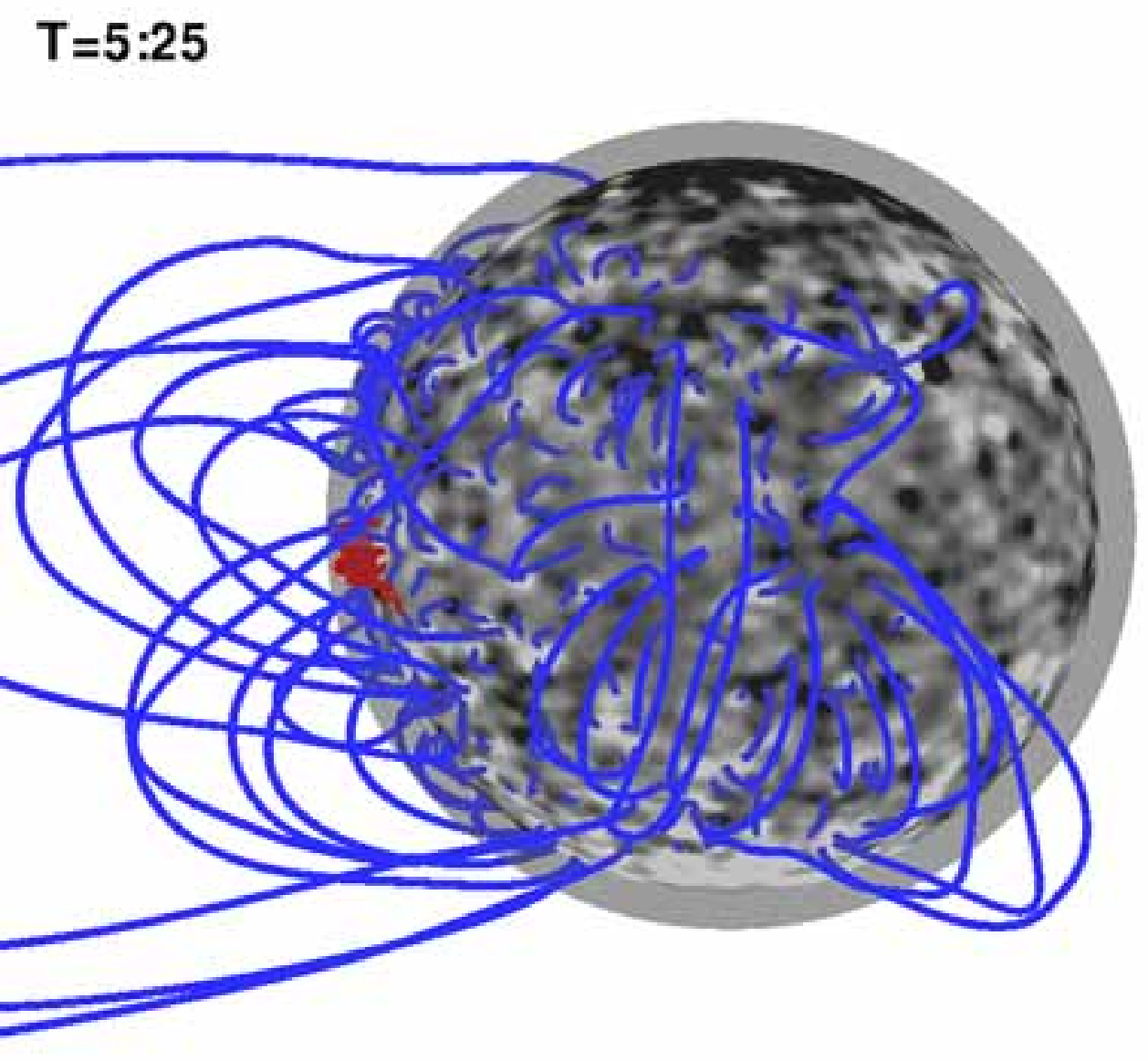} 
\includegraphics[width=2.0in]{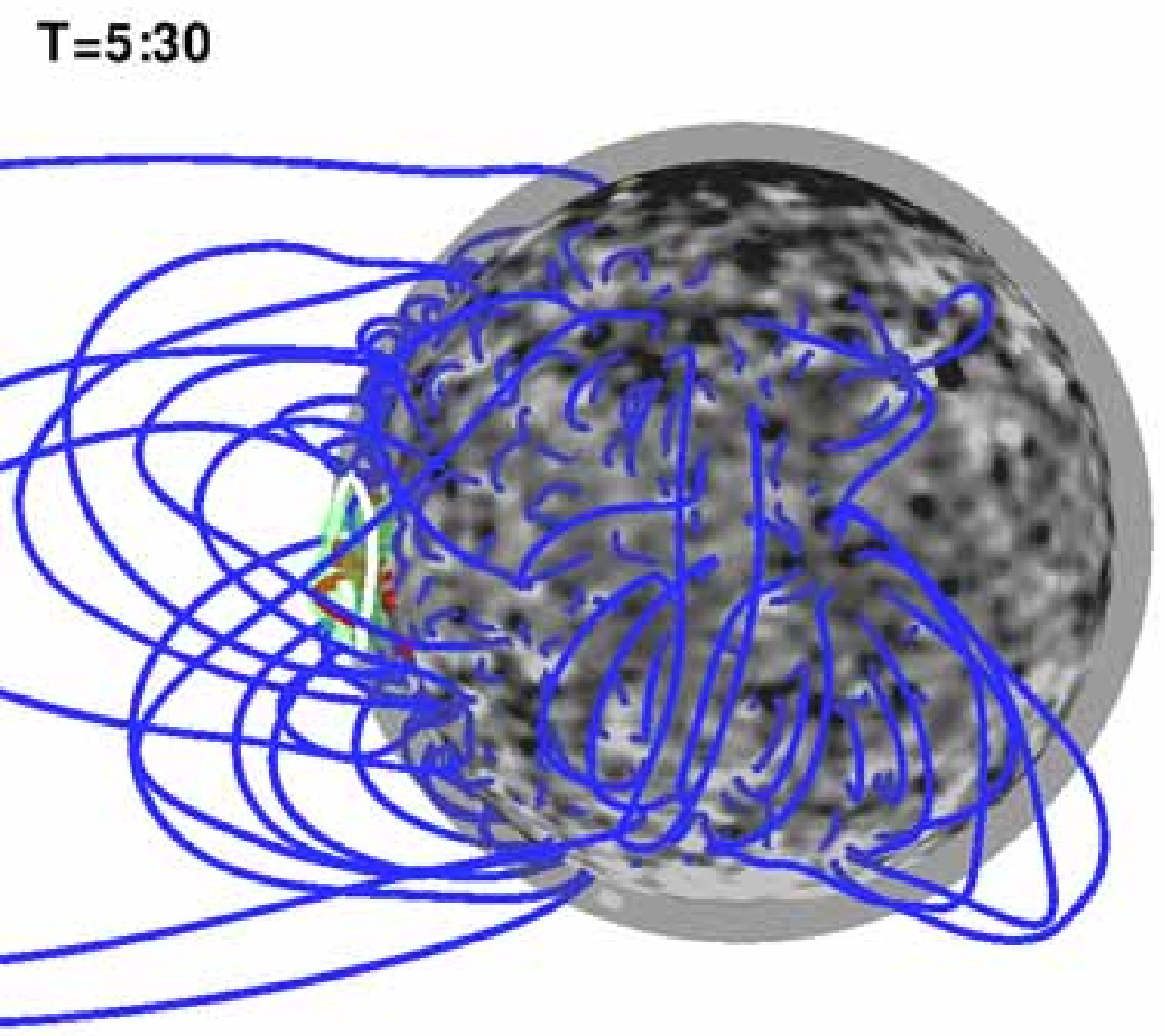}  
\includegraphics[width=2.0in]{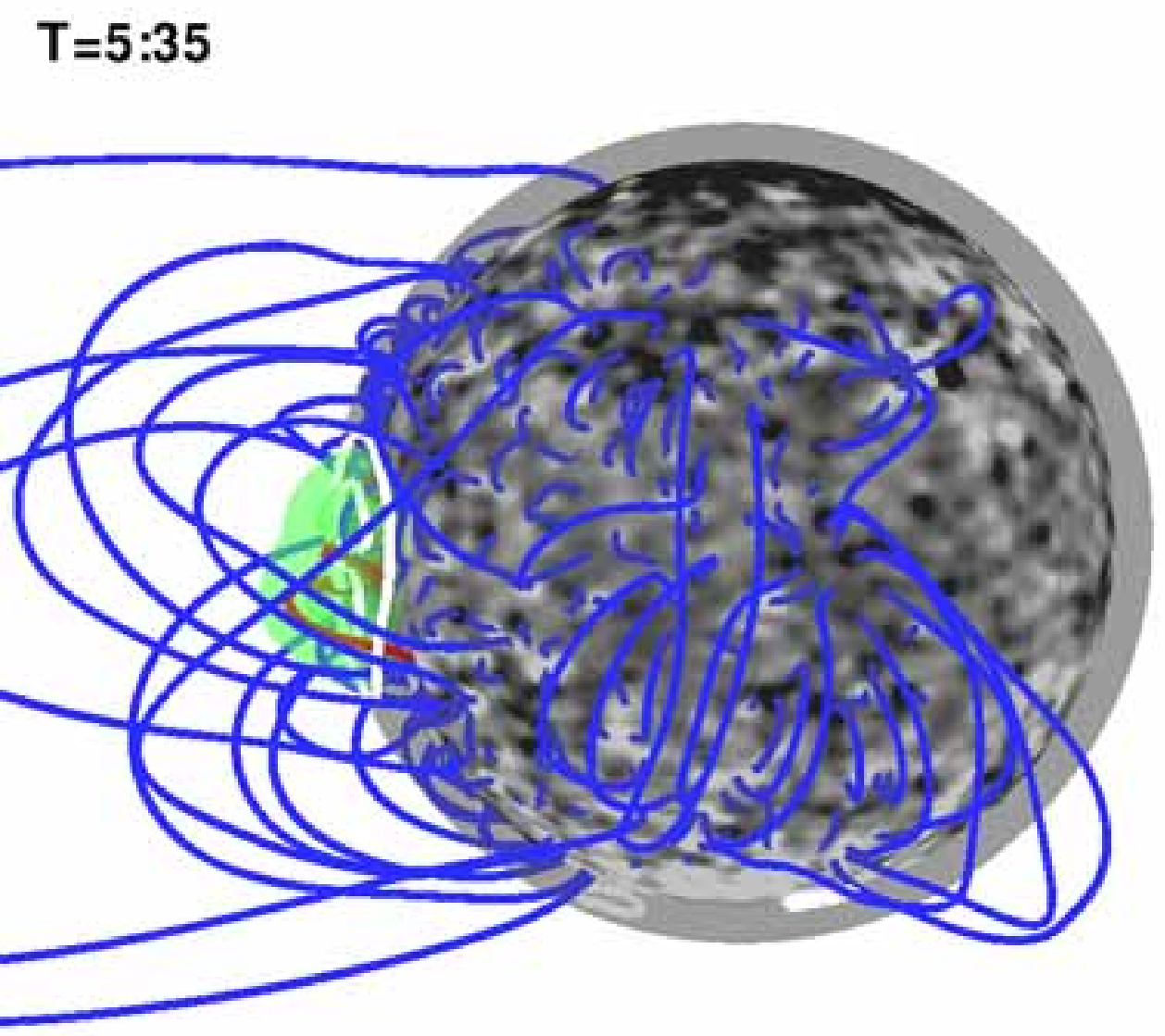} \\
\includegraphics[width=2.0in]{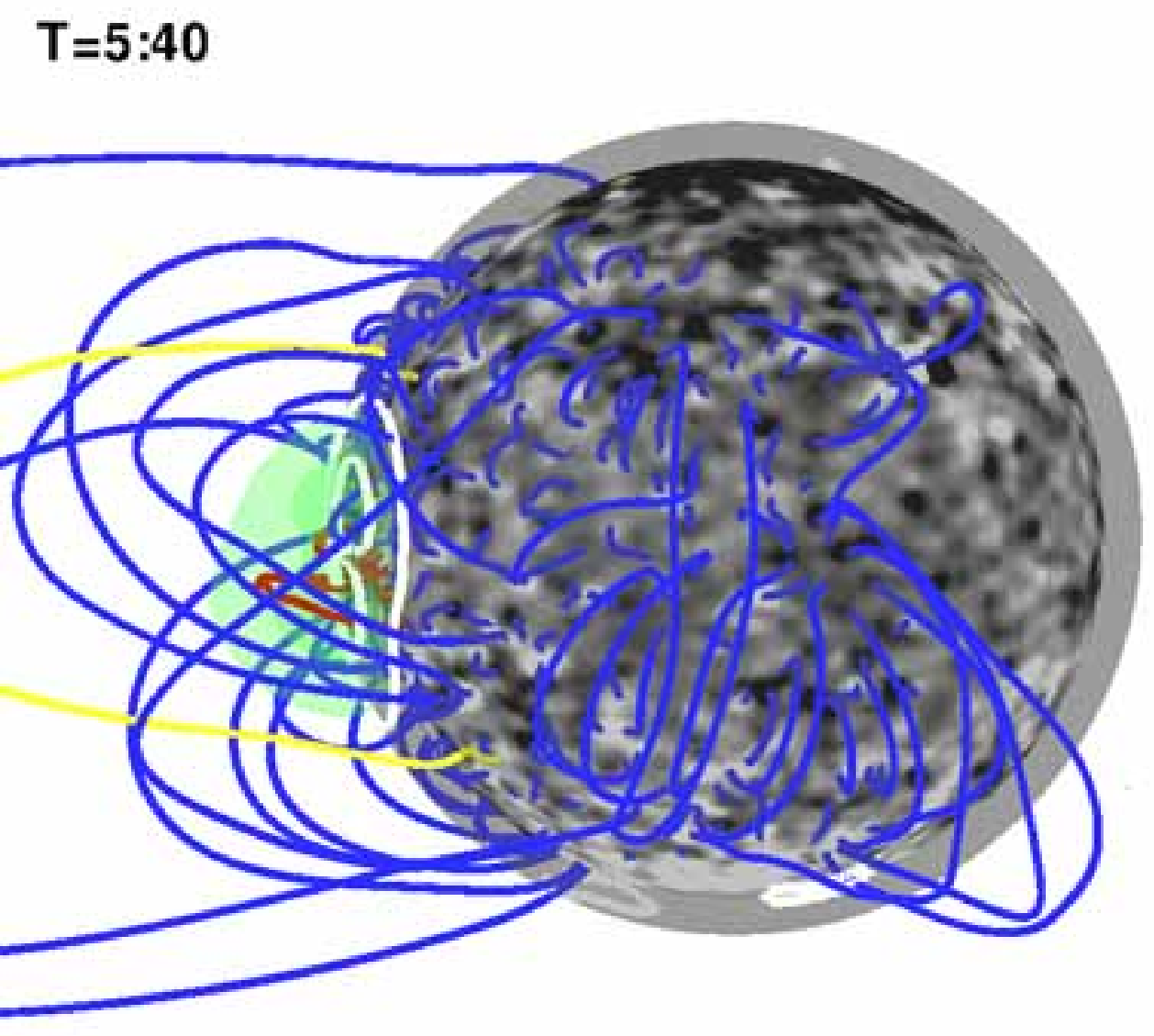}  
\includegraphics[width=2.0in]{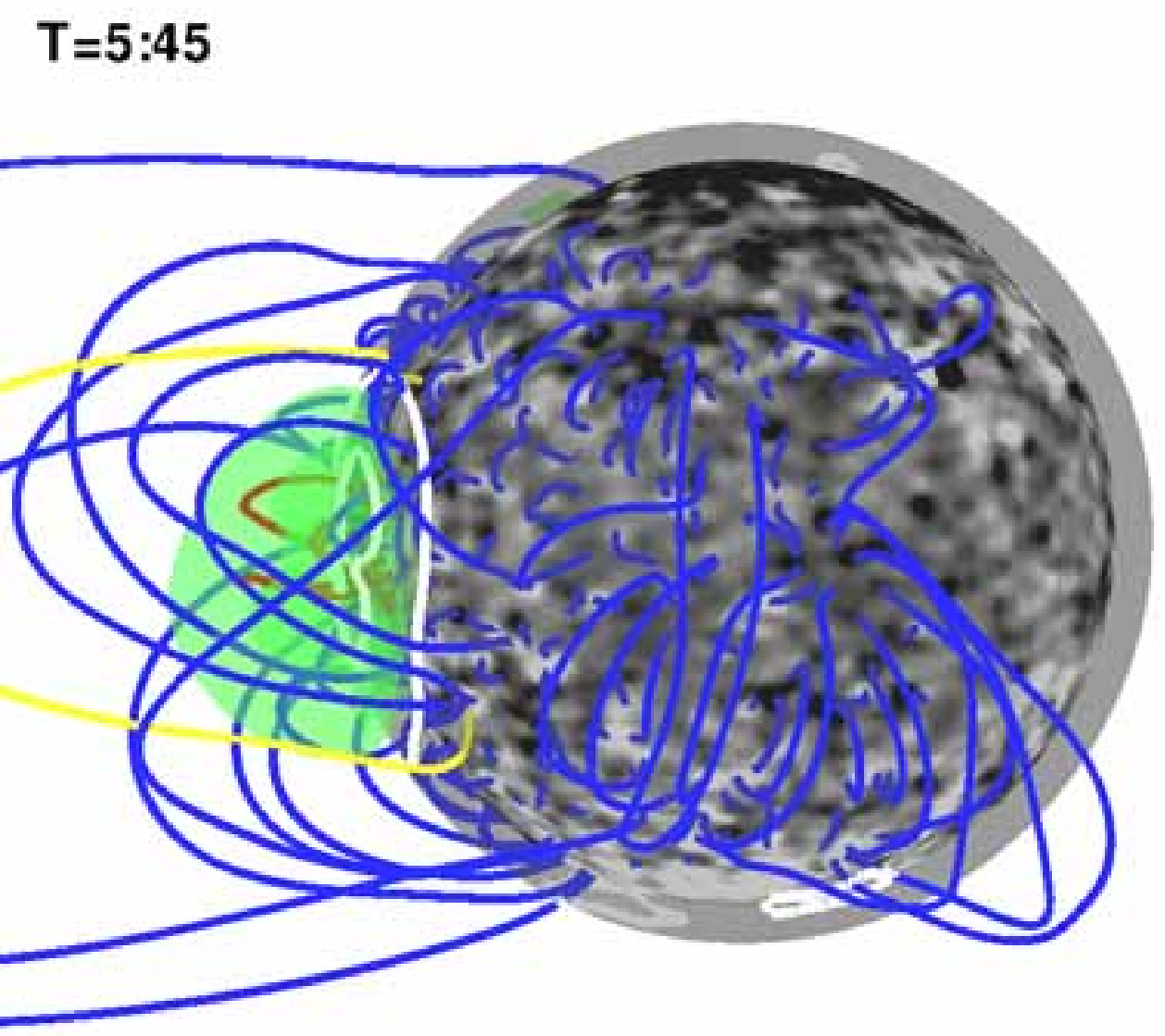} 
\includegraphics[width=2.0in]{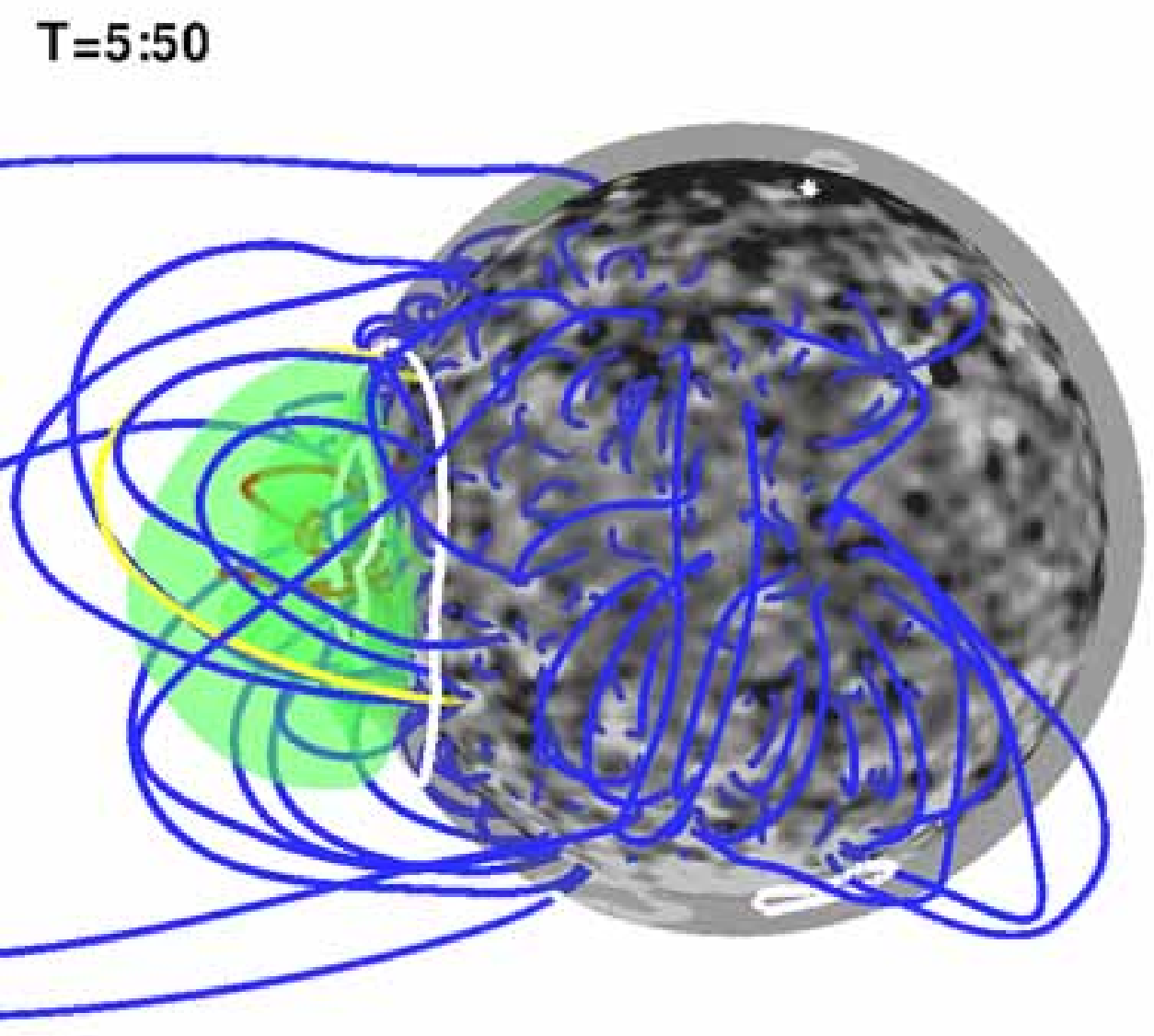} \\
\includegraphics[width=2.0in]{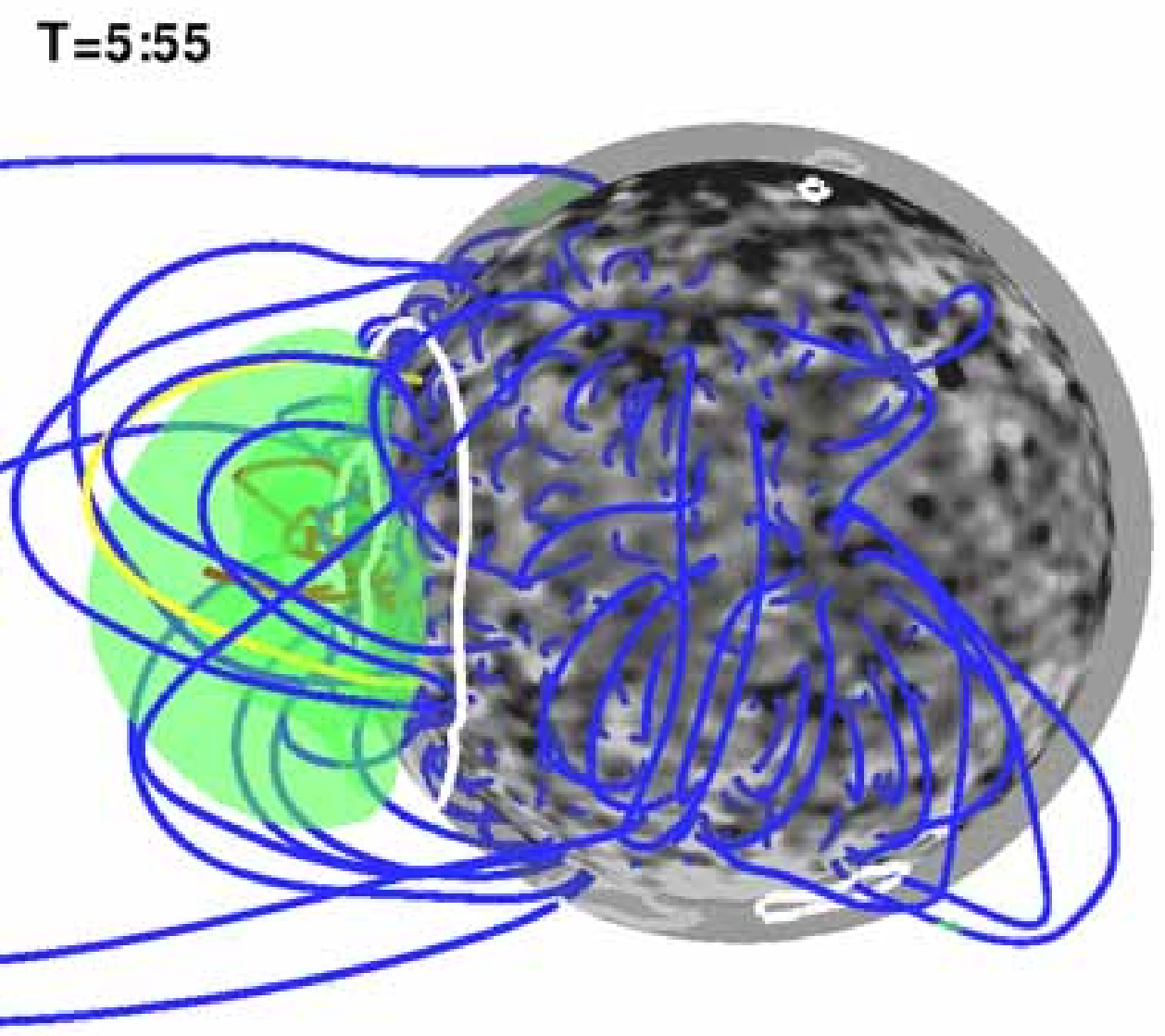} 
\includegraphics[width=2.0in]{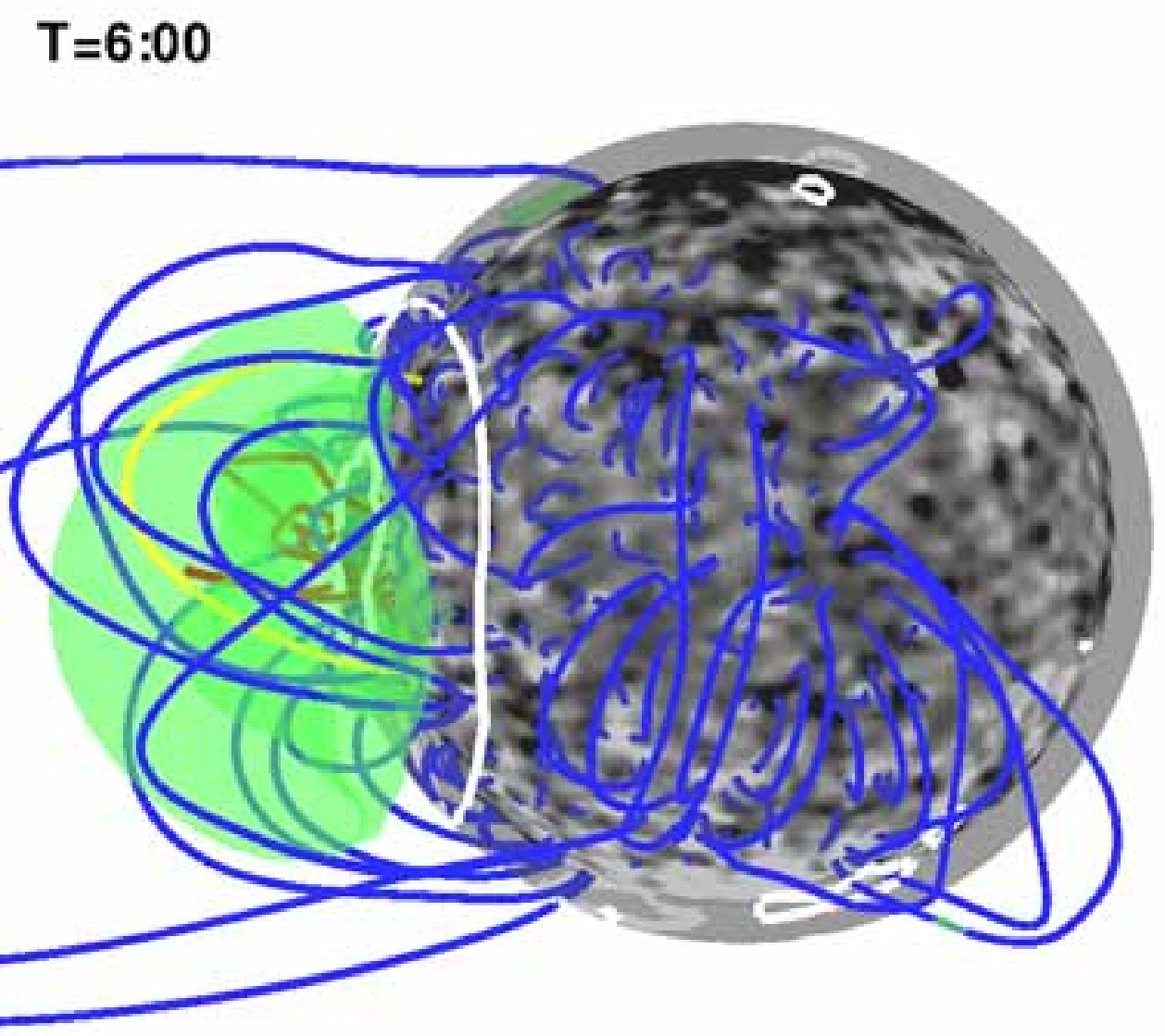} 
\includegraphics[width=2.0in]{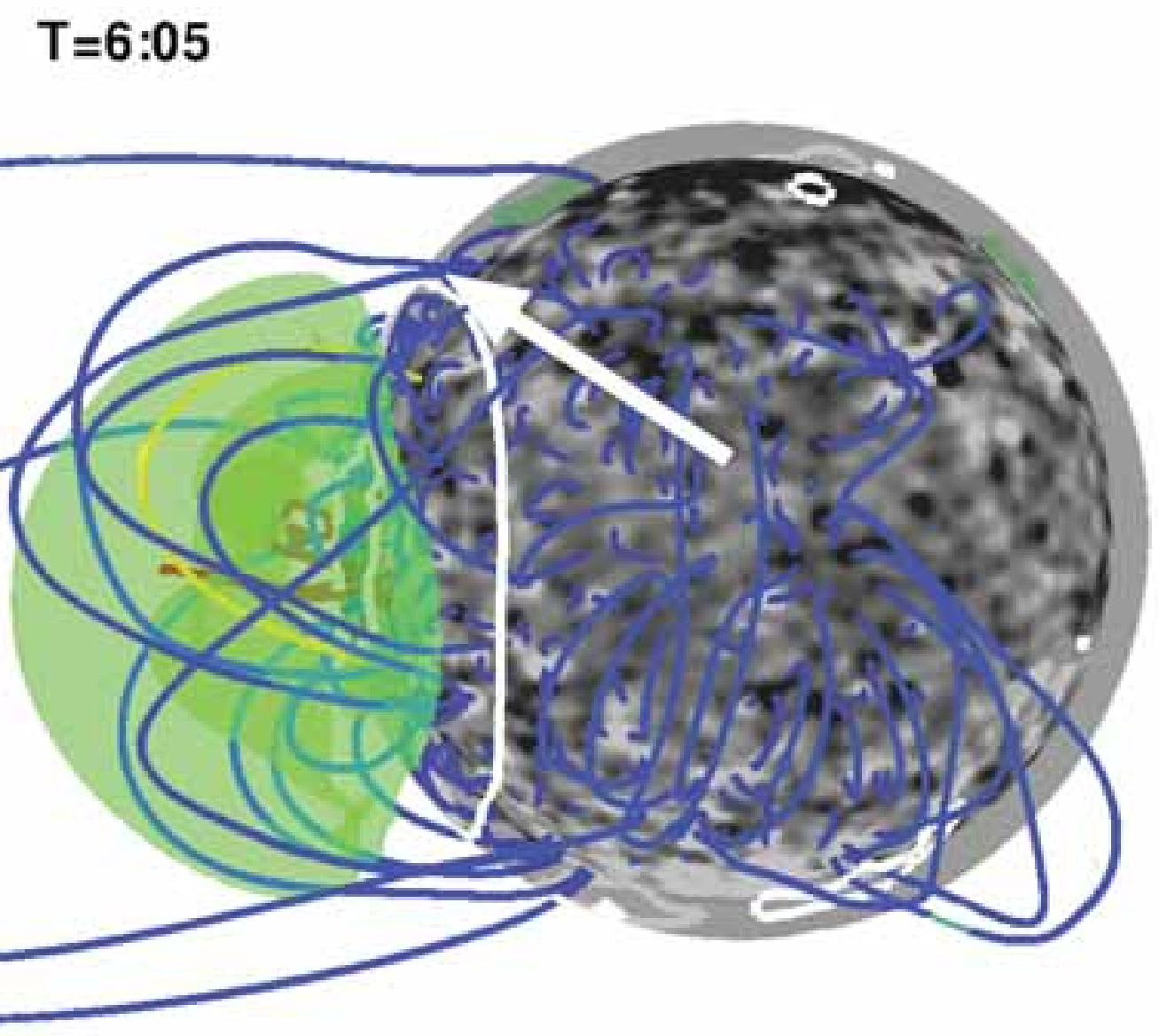} \\
\includegraphics[width=2.0in]{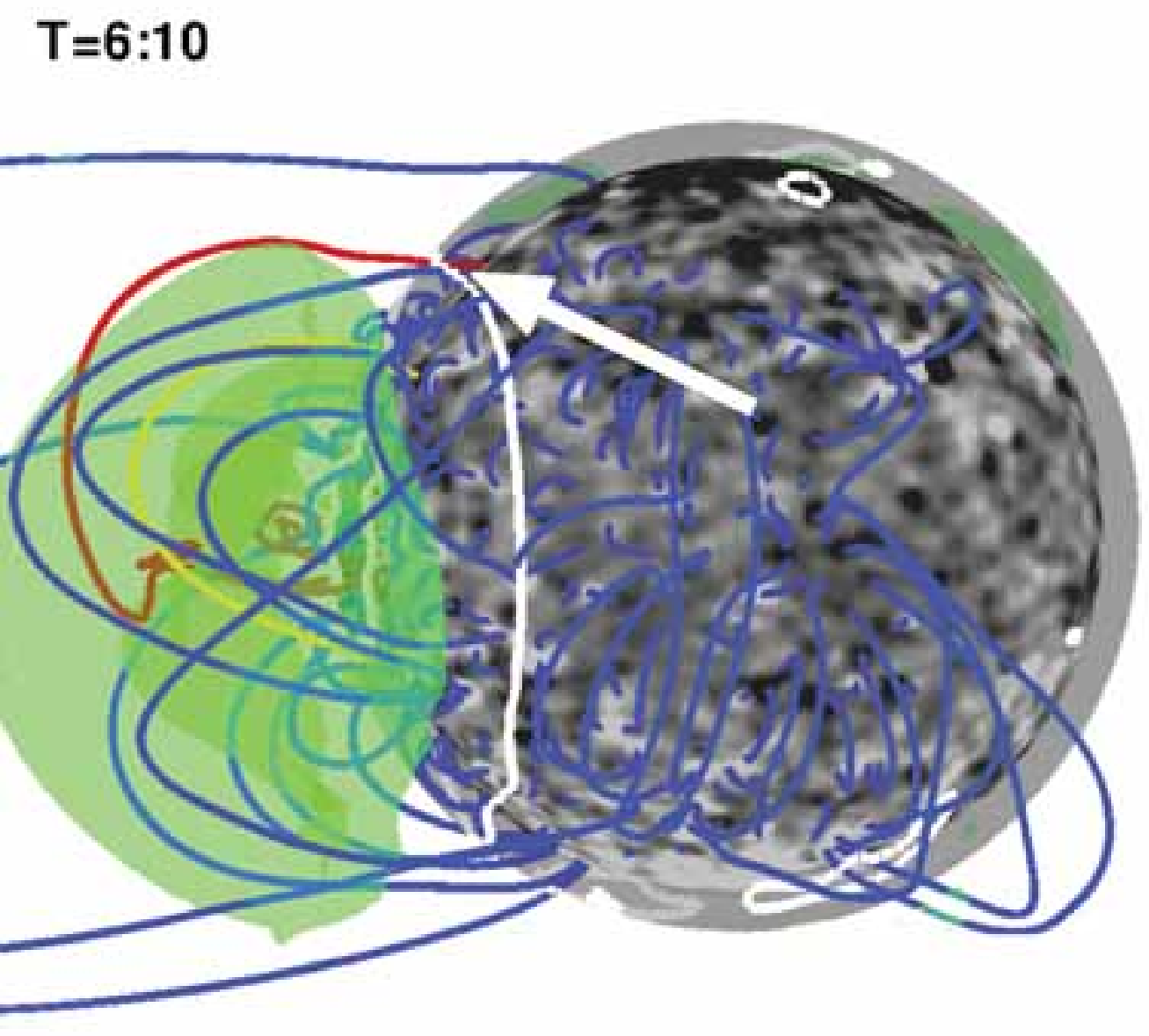} 
\includegraphics[width=2.0in]{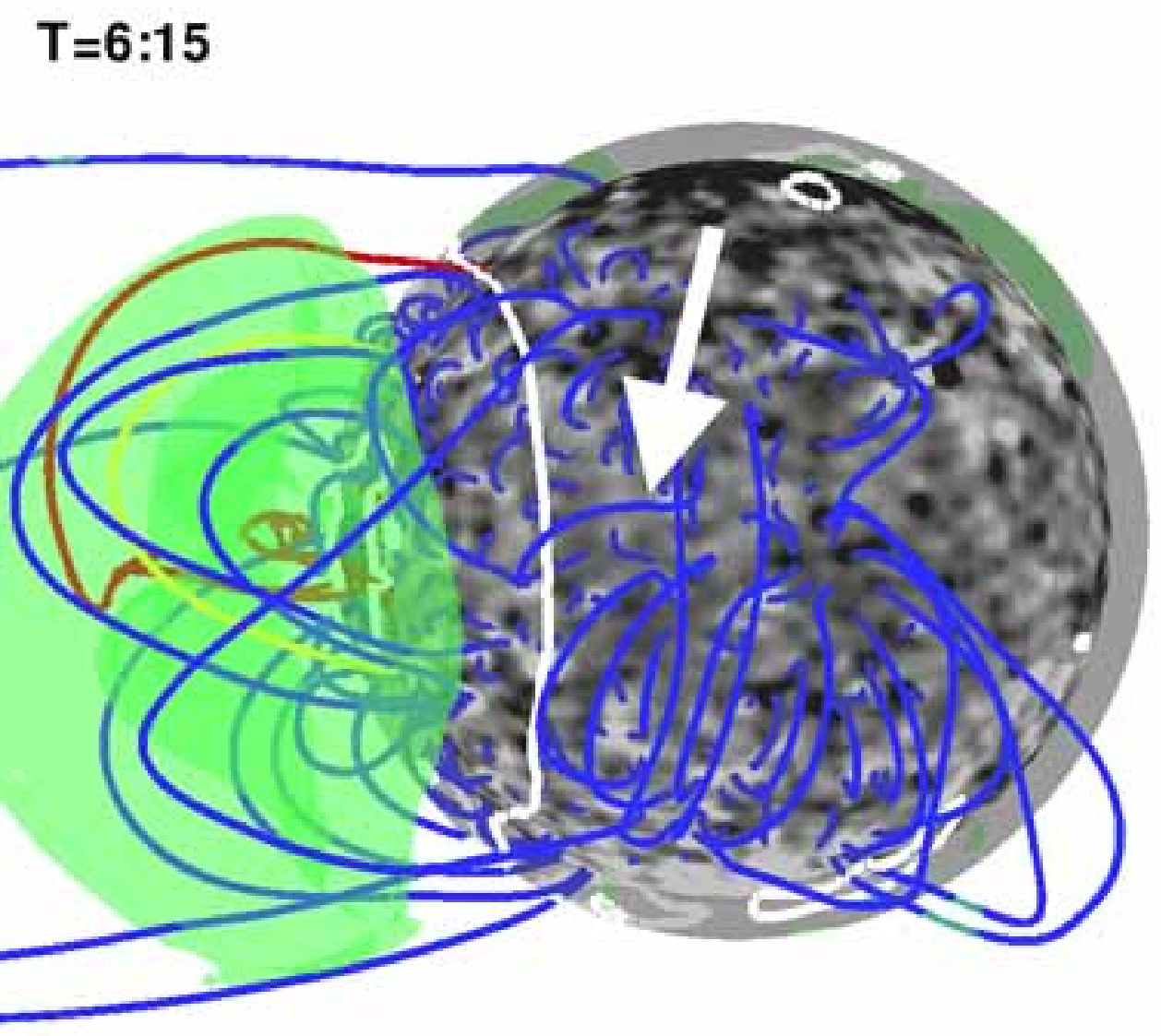} 
\includegraphics[width=2.0in]{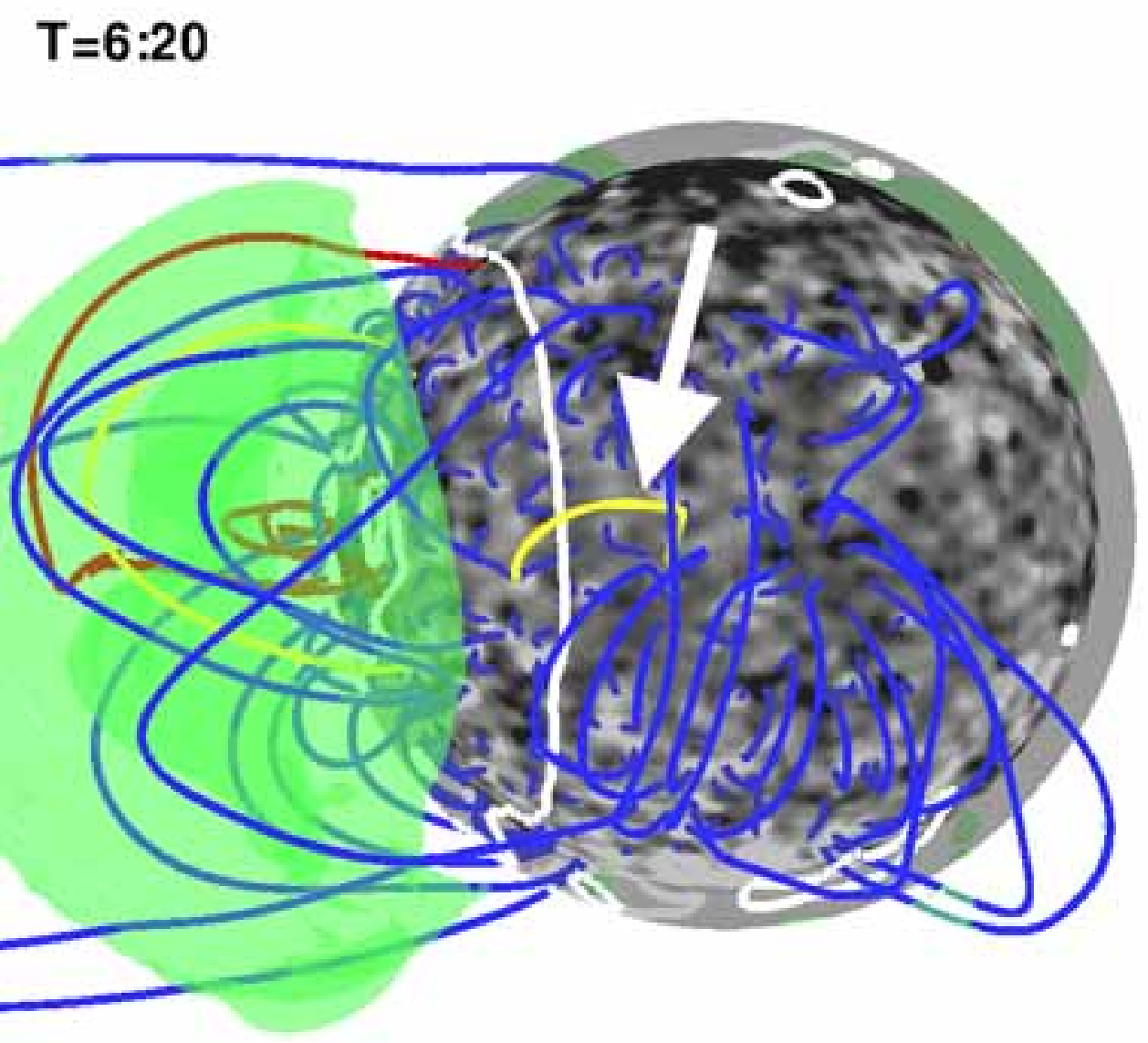} 
\caption{Same as Figure~\ref{fig:f7}, but matched to STEREO-A viewing angle.  (See movie {\tt{SA.mov}}).}
\label{fig:f8}
\end{figure*}

Figures \ref{fig:f7} and \ref{fig:f8} show time-series plots of the simulation results matched to the STEREO-B (on-disk) and 
STEREO-A (limb) viewing angles, respectively. (The reader is encouraged to view the movies that correspond to these 
figures, {\tt{SA.mov}} and {\tt{SB.mov}}).  

The inner sphere shows the photosphere with the radial magnetic field strength from the magnetogram data. The outer 
sphere (light grey) is at height of $1.1R_\odot$ (70 Mm) and represents the altitude at which coronal waves are observed. 
The white contours represent the density-enhanced front (same as displayed in Figure \ref{fig:f3}). The green shade 
represents an iso-surface of mass density with a base ratio (ratio between the current image and pre-event image) of 1.1. 

Selected core field lines of the magnetic flux rope are drawn in Red and some surrounding field lines of a range of sizes 
are drawn in Blue. Where 
the core flux rope field (Red) reconnects with a surrounding field line (Blue), the Blue field line changes to Red, 
indicating the new extended connectivity of the core flux rope field. Reconnections between surrounding magnetic 
field lines (i.e. Blue and Blue), are shown by the newly reconnected field line changing to Yellow. The same 
magnetic field lines have been plotted in both Figures \ref{fig:f7} and \ref{fig:f8}, so that we can study the 
same development from the two different perspectives.  

\begin{figure*}[h!]
\centering
\includegraphics[width=3.0in]{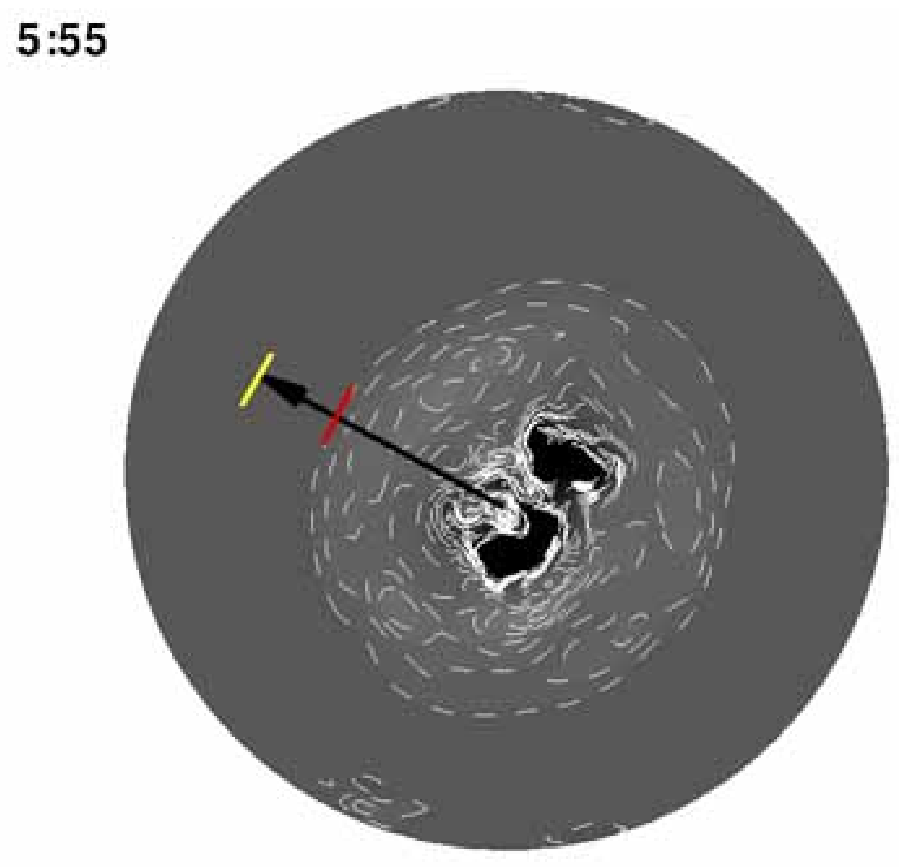} \\
\includegraphics[width=6.0in]{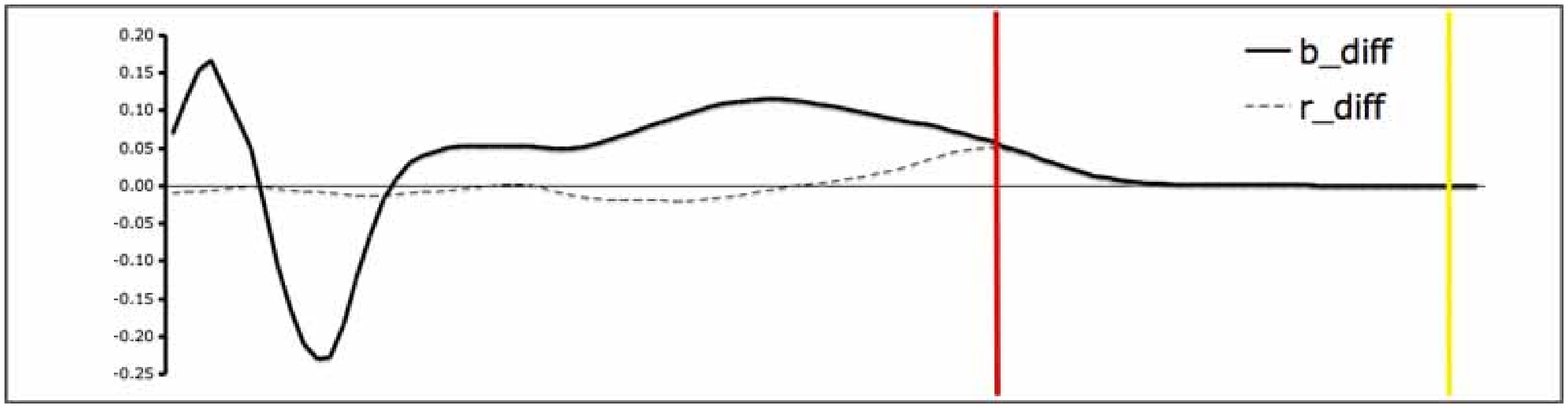} \\
\includegraphics[width=6.0in]{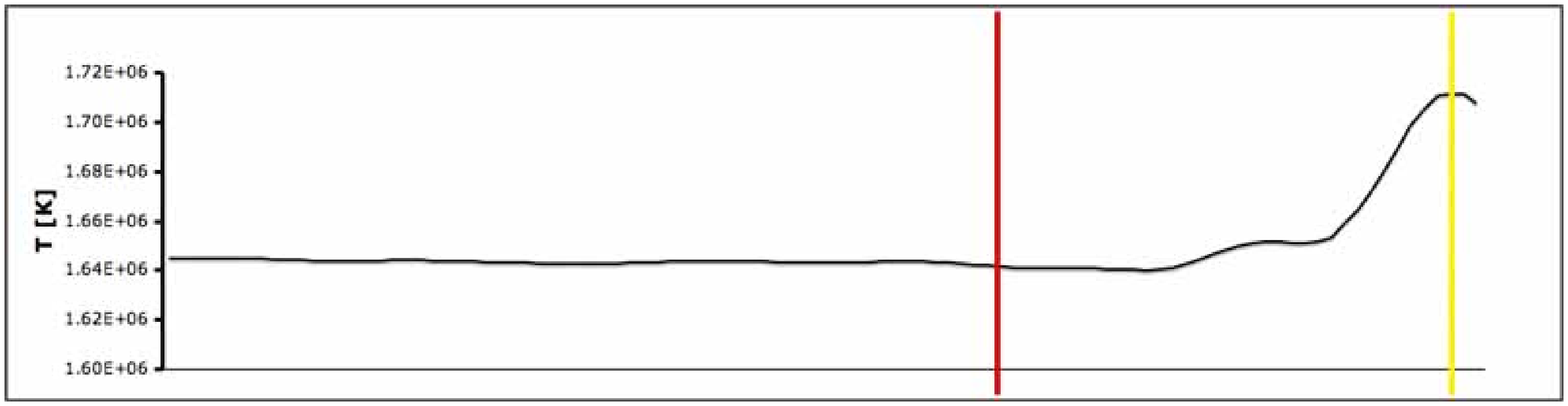} 
\caption{An extraction of density base and running differences (second panel), and 
temperature (bottom panel), at $r=1.1R_\odot$ along the line shown in the top panel. Red line marks 
the location of the coronal wave front, while yellow line marks the location of the CME shock.}
\label{fig:f9}
\end{figure*}

Referring to Figure \ref{fig:f7}, we see that the green iso-surface of increased mass density approximately maps to 
the white contour at each time frame. Figure~\ref{fig:f9} shows a line profile of the density base 
and running differences, as well as the temperature along the path of the coronal wave at $r=1.1R_\odot$ (shown by the black arrow in
the top panel). It can be seen that the temperature jump (indicating the shock front) is ahead of the density increase associated 
with the coronal wave. This means that the green shade represents the CME front and not the shock. Indeed, comparison of 
the green iso-surface with the white-light simulation and observational data (Figure \ref{fig:f6}) further suggests 
that it represents the outer-most shell of the expanding CME.  The existence of a major reconnection 
(discussed below), further suggests that the green iso-surface represents the actual CME rather than a shock, since 
there is a close matching between the reconnected magnetic field line (Red) and the iso-surface.    

Reconnection of a given field line first occurs when the white contour reaches the 
vicinity of one of the reconnecting field lines. This applies both for reconnections between the flux rope and surrounding magnetic field 
(Red), as well as for reconnections between surrounding fields (Yellow). We therefore conclude that 
the reconnections are directly driven by the expanding CME.

Further, we observe a major reconnection between the core flux rope field (Red) and the overlying field (Blue) at 
06:05 - 06:10 UT (marked by narrow white arrows in Figures \ref{fig:f7} and \ref{fig:f8}). This 
reconnection transfers the connectivity of the core magnetic field from near the equator to a latitude of $\sim$ 60$^{\circ}$.
On comparison with Figure \ref{fig:f6}, this reconnection explains the development of the 
secondary cavity (\textsection \ref{subsec:comparison_cme_cw}), and dramatic lateral expansion of the CME in 
the low corona at this time.

These results show that where the core magnetic flux rope is able to reconnect with overlying or surrounding favorably orientated 
field, the CME ``opens'' up the low corona to a very large lateral extent.  A clear example was reported in an observational
study by \cite{Manoharan96} concerning reconnection with a trans-equatorial loop and the subsequent 
formation of two dimmings in quiet Sun regions.  Figures \ref{fig:f7} and \ref{fig:f8} show that quiet Sun loops across the 
whole range of sizes are deflected by the CME expansion.  Where the orientation is favorable, reconnection occurs.
Whether CMEs are large-scale
by nature or become large-scale through nurture was considered by \cite{vanDrielGesztelyi08}.  From our results, 
we conclude that CMEs become large-scale by nurture, through stretching of the magnetic field and reconnection in the low corona with the 
surrounding magnetic environment \citep{Pick98}.


\section{DISCUSSION}
\label{sec:Discussion}

\subsection{Physical cause of the bright front emission}
\label{subsec:temp_density}

It is difficult to separate temperature and density effects within single narrow bandpass observations 
(such as 195 \AA\ used by EUVI) because line-of-sight effects make the optically-thin EUV data complex to interpret, 
and because intensity changes can theoretically be caused by alterations in temperature and/or density.  
Indeed, observations have contributed evidence in favor of both temperature 
\citep[e.g.][]{WillsDavey99,Gopalswamy00b} and density \citep[e.g.][]{Warmuth05,White05,WillsDavey06} 
enhancements. 

\cite{Delannee08} find that the plasma can be brightened from both a current shell around the 
expanding flux rope via Joule heating, as well as from an increase in density due to compression.  
They find that the emission from the current shell generates a more conspicuous brightening than 
that from the plasma compression; however, they note that ``the dissipation of the current densities 
at low altitude would not be responsible for the observed structure''.  This work is consistent with our 
simulation results presented in Figure \ref{fig:f3}, which show that the coronal wave bright front is well 
described by the mass density enhancement at a height of $1.1R_\odot$.  We note that this does not 
exclude some additional contribution from temperature effects.  Indeed, some heating component is 
expected as the plasma is compressed.

\subsection{What causes the enhanced mass density?}
\label{subsec:enhanced_mass_density}

\cite{Delannee99} conjectured that fast expansion of the magnetic field should compress the plasma 
at the boundaries between expanding stable flux domains, naturally leading to the enhanced emission. 
\cite{Chen02, Chen05a} showed that stretching of the overlying magnetic field leads to compression of 
plasma in the legs of the CME, producing an intensity enhancement.  \cite{Attrill07a} suggested that magnetic reconnections between the 
expanding CME and favorably oriented surrounding quiet Sun magnetic field drive weak flare-like brightenings
making up the bright front.  
The simulation demonstrates that the plasma is indeed compressed as a result of the expansion of the CME
(Figure \ref{fig:f3}) and that stretching of the overlying magnetic field and magnetic reconnection with
surrounding field do occur (see Figures \ref{fig:f7} and \ref{fig:f8}).  
Although we only display selected field lines from the simulation (thus probably missing some stretching and 
magnetic reconnection events), these mechanisms are still constrained 
to the footpoints of overlying magnetic field lines \citep[see][]{Chen02}, and locations for favorable reconnection.
They are not responsible for the emission of the entire bright front, only to the higher concentrations
of intensity within it (Figure \ref{fig:f3}).

In order to directly drive compression between the expanding CME and surrounding magnetic field to the 
spatial extent covered by the coronal wave, it is necessary that the CME really expands to a massive, 
global extent low down ($<$ 200 Mm) in the corona.  Our simulation results confirm (Figure~\ref{fig:f6} and \textsection \ref{subsec:3DMFT})
that this is indeed the case. 

\subsection{Persistent Brightenings}
\label{subsec:brights_dims}

\begin{figure*}[h!]
\centering
\includegraphics[width=6.in]{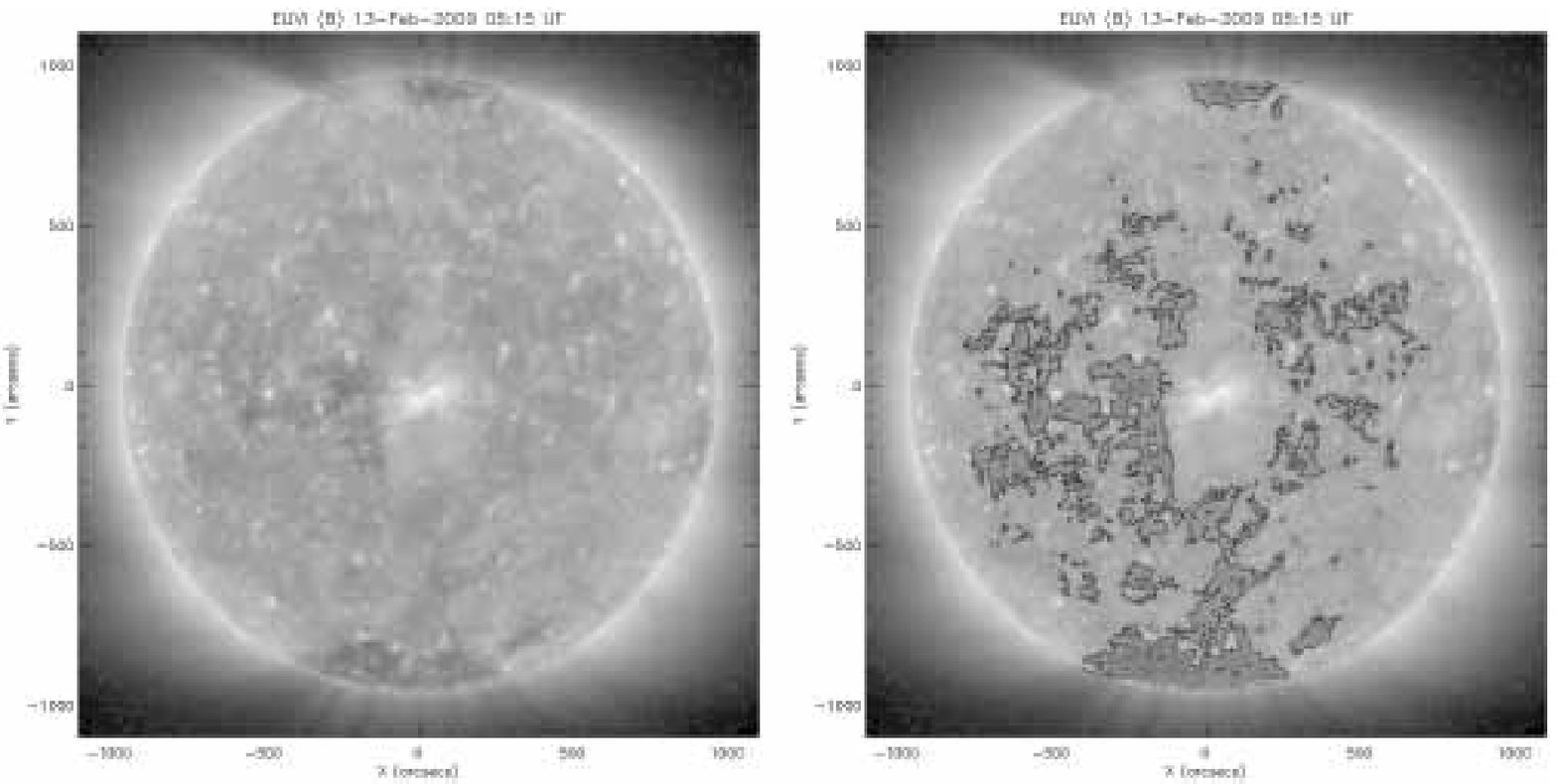}
\includegraphics[width=6.in]{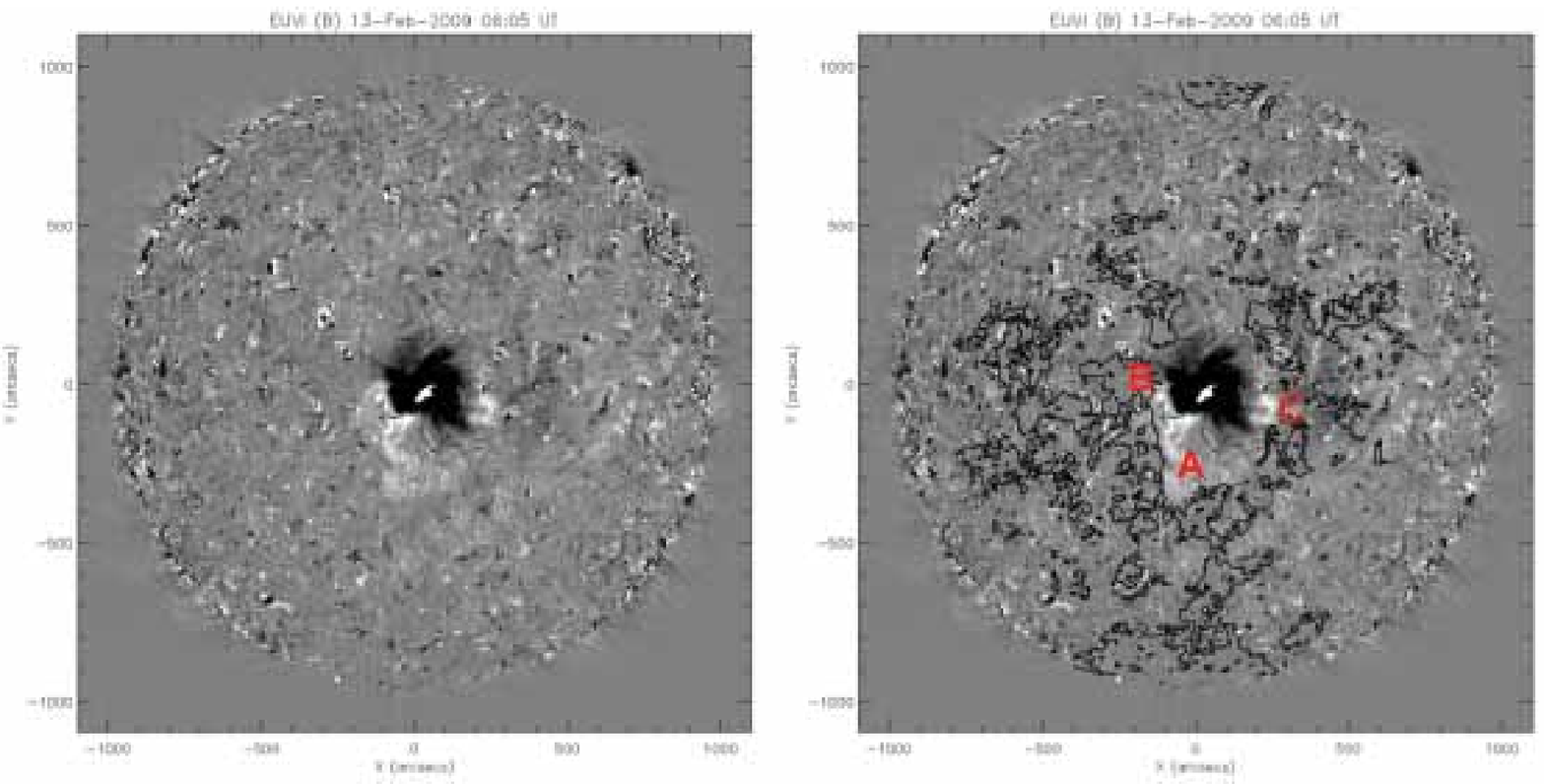}
\caption{Top panels show pre-event EUVI-B 195 \AA\ Fe {\sc xii} images at 05:15 UT.  Black contours are overlaid, outlining 
regions of low intensity.
Bottom panels show base difference EUVI-B images at 06:05 UT. 
The same contours as in the top right panel are overlaid on the base difference data. 
The persistent brightenings (white regions) lie at the boundaries of the deep, core dimming regions.  A low-latitude 
coronal hole lies to the east of brightenings ``A'' and ``B''. Brightenings marked ``A'' are 
evident in the simulation results as well (see Figure \ref{fig:f3}), and hence are due primarily to a density increase. 
Brightenings ``B'', and ``C'' on the other hand, are much more closely linked to regions of ongoing magnetic reconnection 
between the core flux rope magnetic field and surrounding, favorably orientated magnetic field (see Figure \ref{fig:f7}).}
\label{fig:f1}
\end{figure*}

\begin{figure*}[h!]
\centering
\includegraphics[width=3.0in]{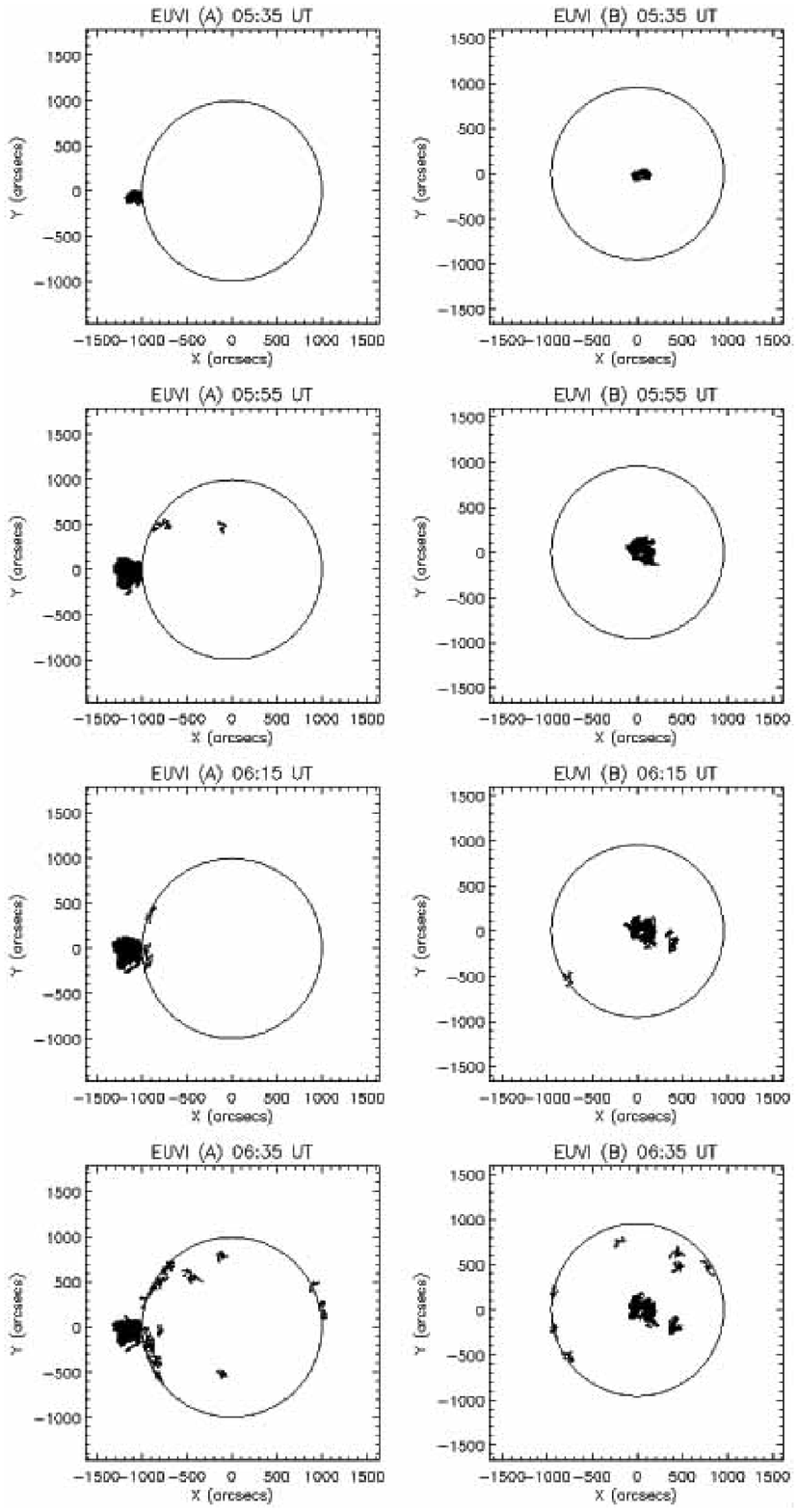}
\hspace{1cm}
\includegraphics[width=3.0in]{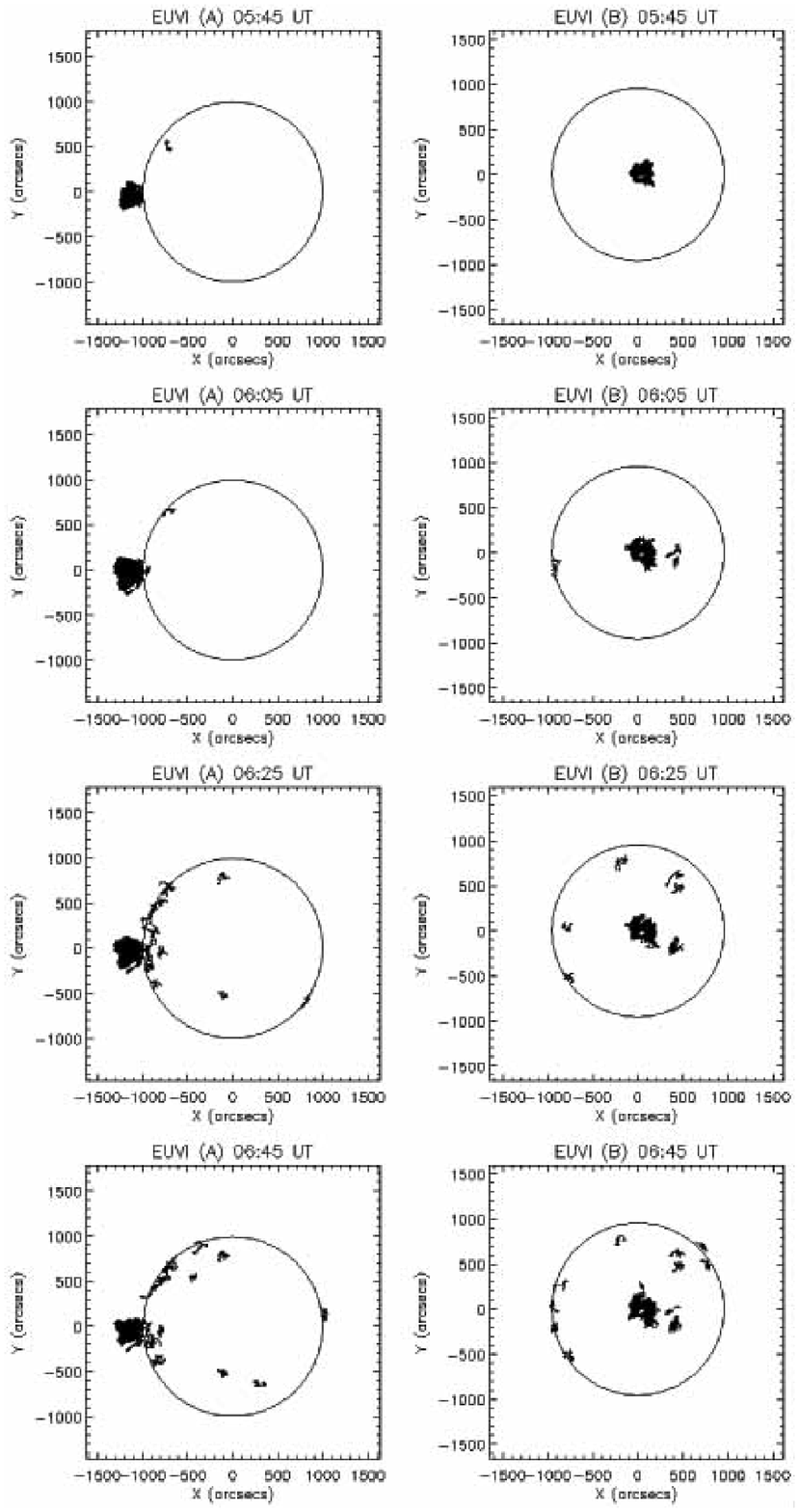}
\caption{Output from the automatic dimmings extraction algorithm (Attrill \& Wills-Davey, 2009). The deep, core dimming 
is extracted in the immediate vicinity of the active region (c.f. bottom panels, Figure \ref{fig:f1}). In addition, widespread, secondary dimmings are also 
detected, spread across a large fraction of the solar disk (c.f. Figure \ref{fig:f3}).} 
\label{fig:f4}
\end{figure*}

\begin{figure*}[h!]
\centering
\includegraphics[width=3.4in]{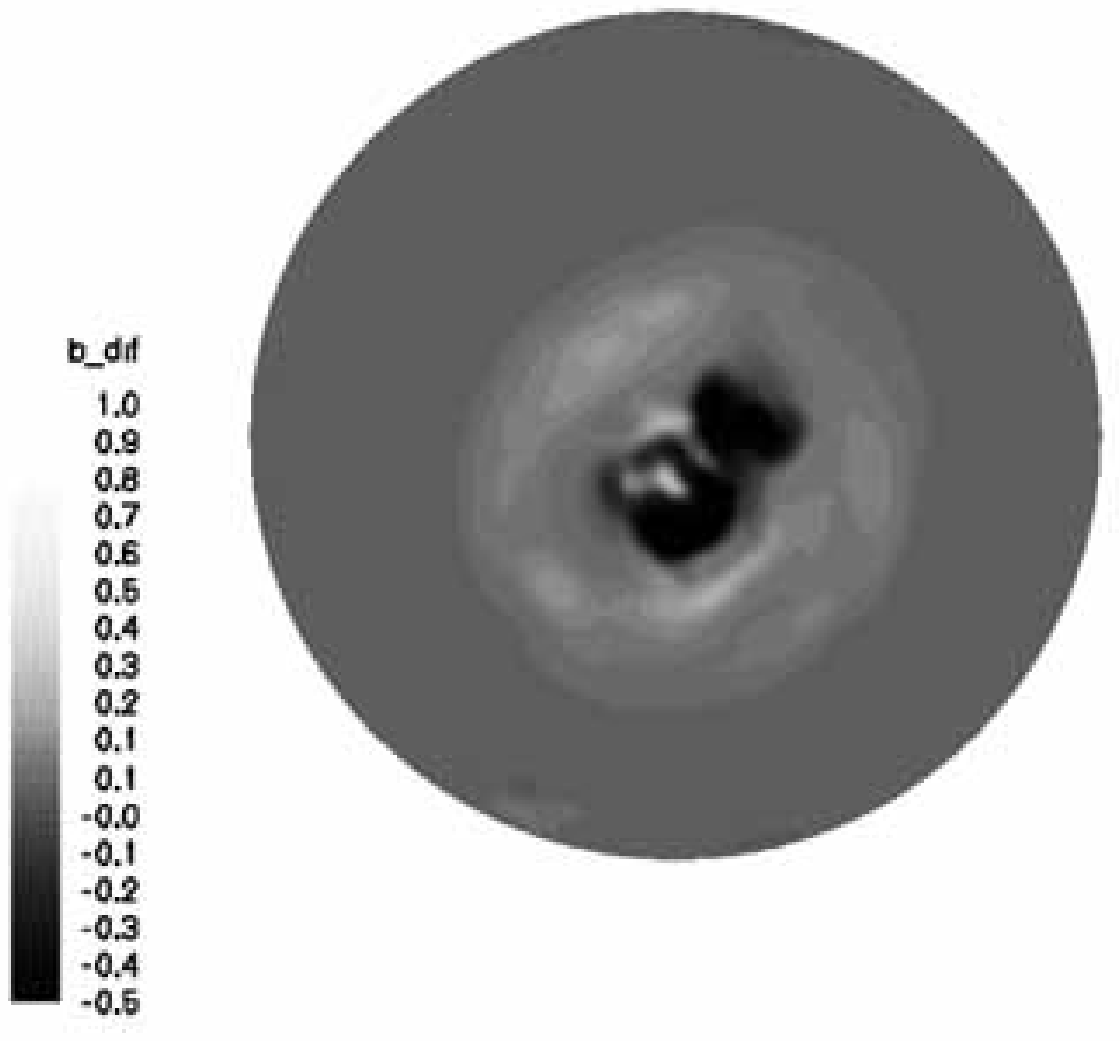}
\includegraphics[width=3.0in]{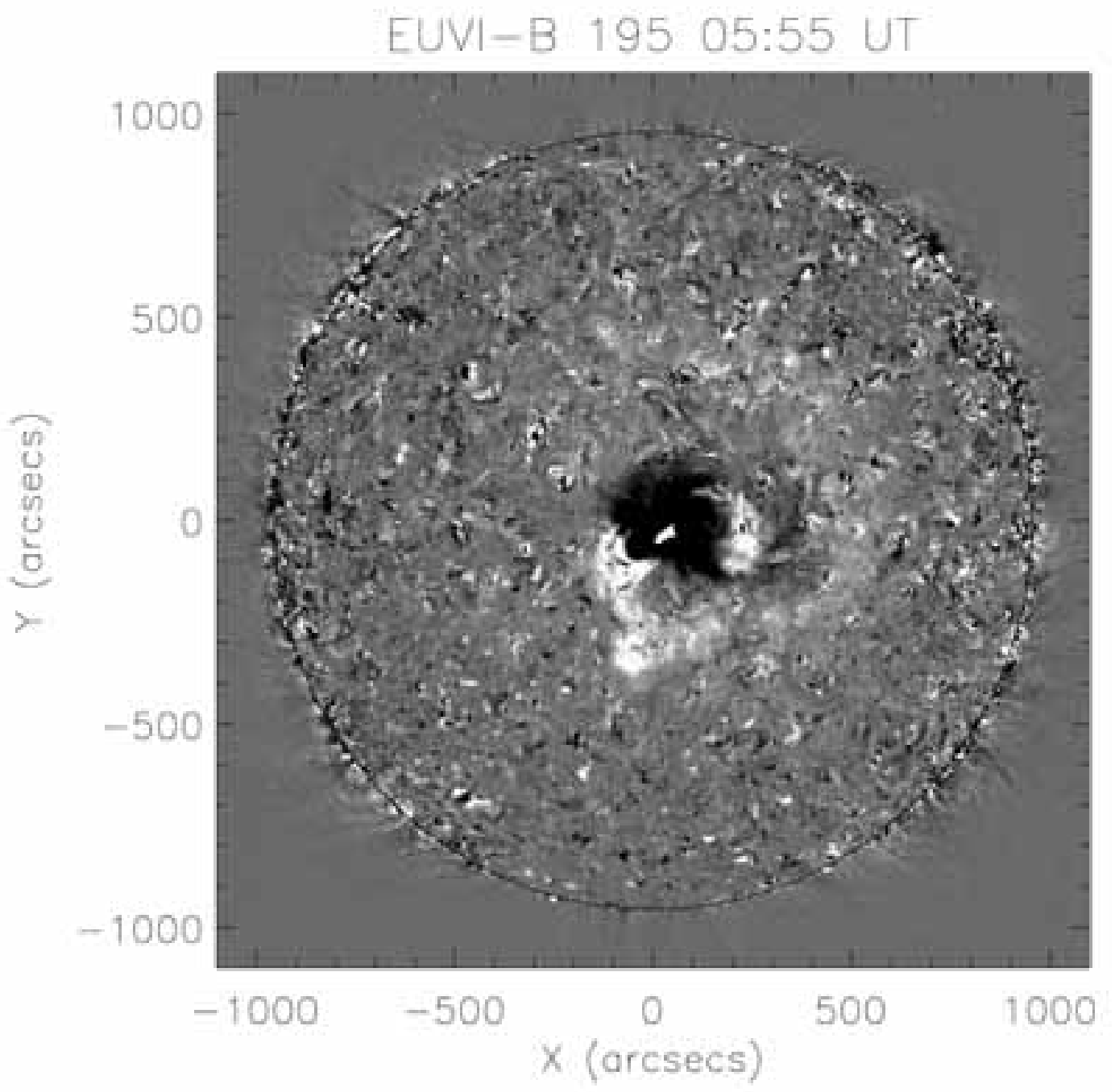} \\
\includegraphics[width=6.8in]{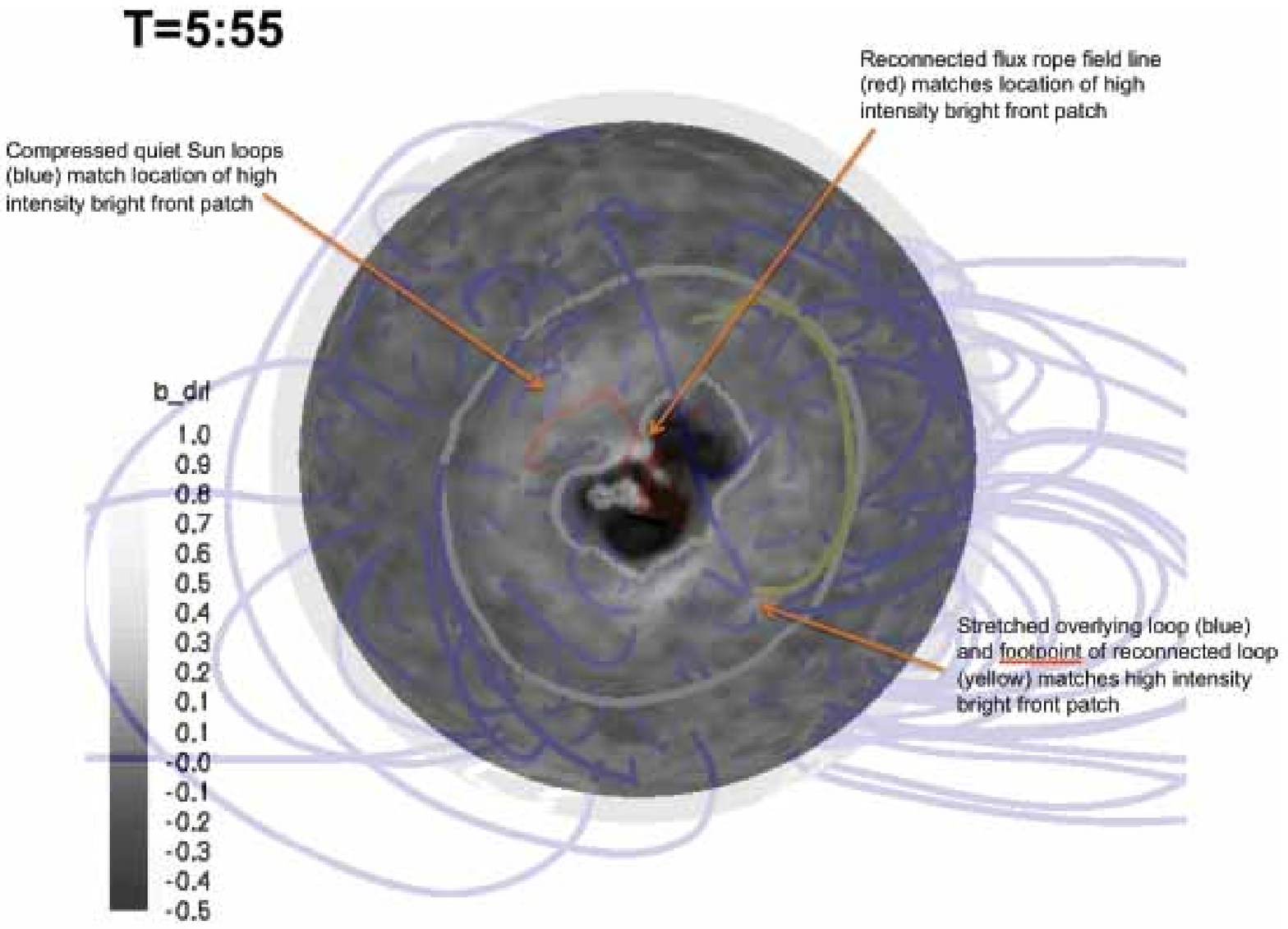}\\  
\caption{Top left panel shows a base difference image of mass density at 1.1 R${\odot}$ from the simulation. 
Top right panel shows the EUVI-B base difference data at 05:55 UT.  Bottom panel shows an overlay of the mass density 
and magnetic field from the simulation.  Orange arrows indicate the different mechanisms that contribute to 
the bright front observed in the base difference data at 05:55 UT.
}
\label{fig:f10}
\end{figure*}

Stationary brightenings have previously been reported at coronal hole boundaries by 
e.g. \cite{Thompson98, Veronig06, Attrill07b} and at separatrices formed in the large-scale magnetic topology 
\citep[e.g.][]{Delannee99, Delannee07}.  Brightening ``A'' in Figure \ref{fig:f1} is likely due to compression between the expanding CME and 
``open'' magnetic field of the coronal hole to the East of AR 1012, since this brightening 
is also seen in the simulation results (Figure \ref{fig:f3}), which show enhanced mass density.  
Brightenings ``B''  and ``C'' in Figure \ref{fig:f1} are located at the same place as ongoing magnetic 
reconnections seen in the simulation (Figure \ref{fig:f7} and Figure \ref{fig:f10}).
\cite{Attrill07a} reported persistent brightenings similar to these at the outer edge of deep, core dimming regions
for two events during solar minimum of cycle 23.  They suggested that these brightenings may be due to ongoing 
reconnection between the expanding flux rope and surrounding, 
favorably orientated magnetic field.  Our results for this event are consistent with such an interpretation.

\subsection{Secondary Dimmings}
\label{subsec:sec_dims}

Behind the expanding bright front, we detect localized regions of secondary dimming (Figure \ref{fig:f4}).  
Secondary dimmings were originally reported by \cite{Thompson00b}. Such dimmings 
may be understood in a wave context \citep[e.g. as in][]{Wu01}, since a rarefaction shock develops
trailing a large-amplitude perturbation \citep{Landau87}.  However, such dimmings 
would be short-lived, with a duration on Alfv{\'e}n timescales ($\sim$ few minutes) contrary 
to observations \citep{Cliver05,Delannee07}.  Although secondary dimmings have a much lower 
average intensity level (e.g. $\sim$ 50~counts/pixel) before the event, compared to the core dimming ($\sim$~100 counts/pixel), 
the relative magnitudes of both the core and secondary dimmings are substantial (e.g. $\sim$~50\% and $\sim$~20\%, respectively).  
Further, like the core dimmings, the secondary dimmings remain at a reduced intensity level for an extended period ($>$ 1 hour).  

\begin{figure*}[h!]
\centering
\includegraphics[width=3.4in]{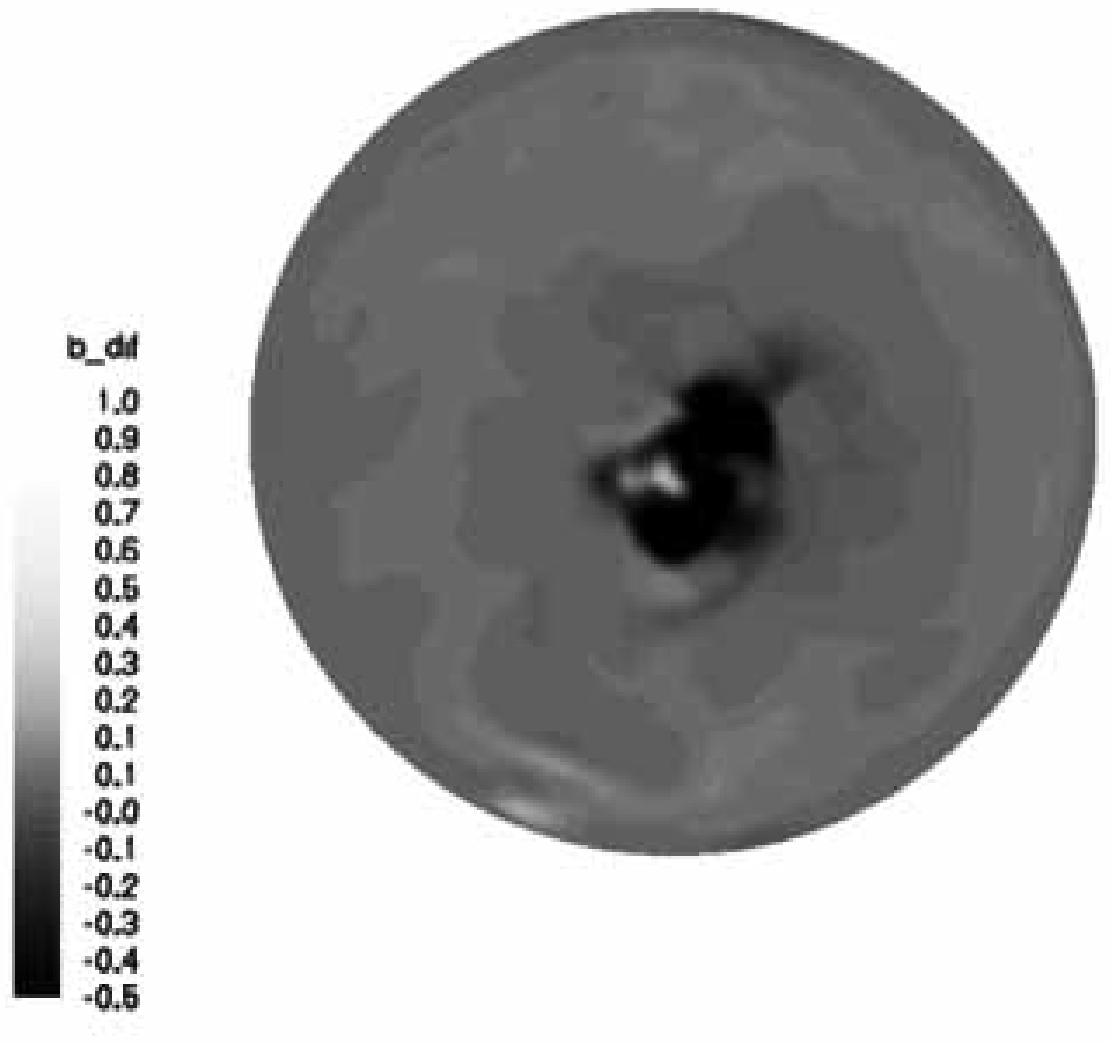}
\includegraphics[width=3.0in]{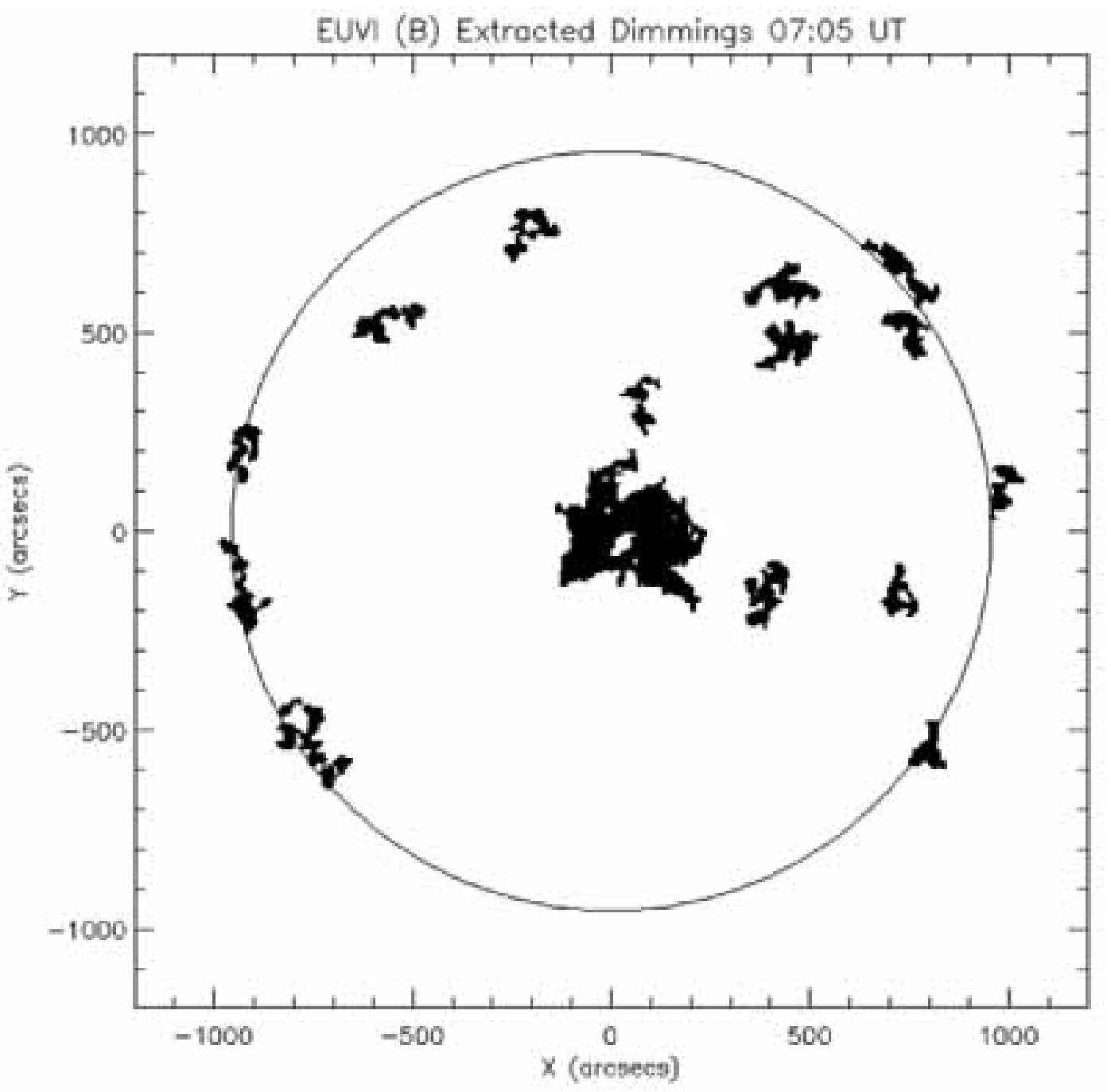} \\
\includegraphics[width=6.1in]{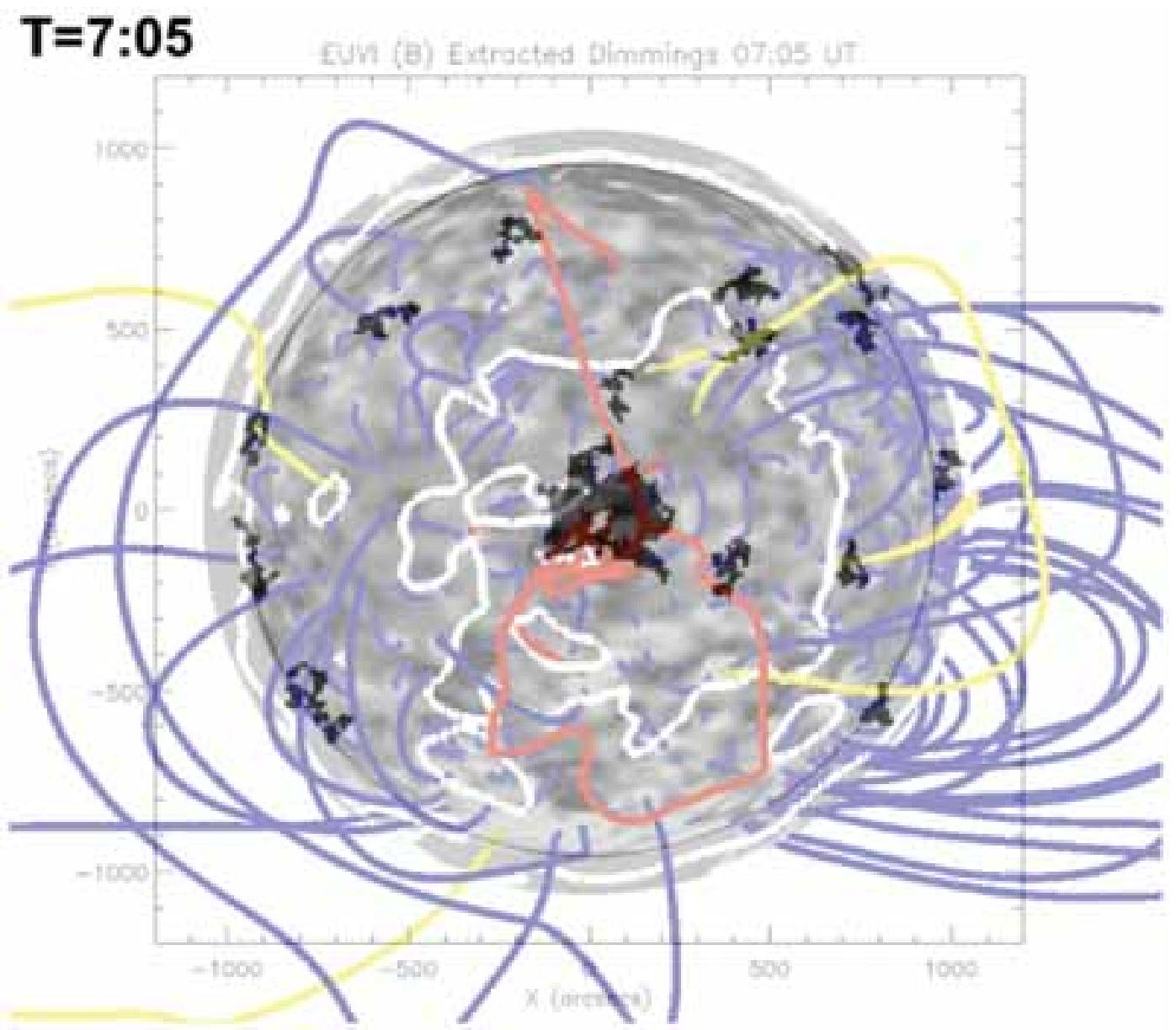}\\  
\caption{Top left panel shows a base difference image of mass density at 1.1 R${\odot}$ from the simulation. 
Top right panel shows the core and secondary dimmings at 07:05 UT, extracted using the automatic dimmings algorithm. 
Bottom panel is an overlay of the extracted dimmings and selected magnetic field lines showing the correspondence 
between reconnected magnetic field (red and yellow lines) with the locations of secondary dimmings at 07:05 UT.  
}
\label{fig:f11}
\end{figure*}

The locations of these secondary dimmings also appear to be closely associated with the magnetic
fields reconnected through CME expansion. For example, Figure \ref{fig:f6} shows a secondary dimming to the north of the source region
that corresponds to the secondary cavity in the simulation seen at 06:05 UT.  This dimming extends the CME cavity northward (see {\tt{COR1join1.mov}}). 
Further evidence of this can be found in 
Figure \ref{fig:f11} \citep[also see][]{Mandrini07}.  These results show that overlying and neighboring magnetic 
field is ``opened'' through magnetic reconnection, extending the CME footprint in the low corona. 
Therefore, we also interpret the secondary dimmings in this event as being due to density depletion, although spectral 
diagnostics have yet to confirm or refute this interpretation.

\subsection{Understanding the two-component bright front}
\label{subsec:understanding_2bf}

Our results show that the bright front observed in \emph{base} difference EUV data and the CME 
are strongly coupled (e.g. Figure \ref{fig:f6}).  These higher-intensity brightenings are due to the CME compressing 
the plasma (against both surrounding and overlying magnetic field).  The brightest 
concentrations in the data and simulations in Figure \ref{fig:f3} show correspondence with the regions of reconnection in 
Figure \ref{fig:f7}, both red and yellow field lines.  Figure \ref{fig:f10} shows a direct comparison.

When the CME has expanded to its maximum lateral extent, the brightest parts of the coronal wave either 
disappear or become stationary before fading (\textsection \ref{subsec:two_component_bf}).  This is the result
of multiple factors: the CME is no longer directly compressing plasma, the overlying field has already stretched, and magnetic 
reconnections with surrounding favorably-orientated field have had time to occur.
  
What remains is a weaker, more uniform component that is consistent with an MHD wave interpretation.  The dynamic 
expansion of the CME is a highly energetic, impulsive event, therefore wave(s) are expected to be generated.  
The simulation results show that this weaker component exists 
throughout the expansion of the CME.  The later frames of the simulation show that it continues to expand even 
after the considerable CME lateral expansion has finished.  In this later stage, the coronal wave is 
freely propagating \citep[e.g.][]{Veronig08}.  
In the \emph{running} difference EUVI-A data ({\tt{195a\_rdiff.mov}}), which highlights the \emph{leading edge} 
of the disturbance, the Western expansion can be followed considerably later than in the 
base difference images, until at least 06:55 UT (c.f. 06:25 UT, the reader is encouraged to compare the running and base difference movies for EUVI-A: {\tt{195a\_rdiff.mov}} and {\tt{195a\_diff.mov}}).  It is more likely that the weaker component can be detected in running difference images, which better show subtle changes.

With this analysis, we believe we are able to reconcile the different (wave and non-wave) interpretations of coronal waves.  
When EIT waves were first discovered, they were studied using \emph{running} difference data.  
This method highlights the (often faint) leading edge of the disturbance, making it useful for identifying waves; this technique was
probably perpetuated because researchers \citep[e.g.][]{Thompson99} originally identified their observations as a strong candidate 
for the predicted coronal counterpart to the chromospheric Moreton wave \citep{Moreton60a,Uchida68}.  

In the late 1990s, Delann{\'e}e et al., (and later Chen et al., and Attrill et al.), preferentially used \emph{base} 
difference images because they show real brightenings and dimmings \citep[e.g.][]{Chertok05}.  The motivation for using base 
difference images is due to the focus of these authors on coronal dimmings, which are strongly 
connected with coronal waves and CME events.  It is not possible to study coronal dimmings with running 
difference images.  However, base difference images do not show faint features so well.  We have shown that the 
\emph{base} difference brightenings are closely linked to the CME and magnetic field evolution;
hence, the development of non-wave models.  

For some time, both of these methods have been applied,
with different studies producing disparate conclusions.  In most 
cases, however, researchers have attempted to find a single 
solution--either wave or non-wave--applicable to all aspects
of diffuse coronal wave events.  Over time, this has led to seemingly
contradictory evidence, selectively supporting wave or non-wave
models, depending of the focus of the study.
\cite{Zhukov04} first introduced 
the concept of a coupled coronal wave, consisting of an eruptive mode and a wave mode, based on comparative 
analysis of EIT running and base difference data.  Our results are consistent with such a picture.
The combined observational and simulation results presented here allow us to firmly establish and understand the 
contribution from each of the various mechanisms.  In retrospect, it is not surprising that wave and non-wave interpretations 
have failed to be reconciled when the different data sets highlight different things!  

\subsection{Implications for impulsive electron events}
\label{subsec:impulsive_e}
In interplanetary space, field lines following the Parker spiral connect the western longitudes of the 
Sun to spacecraft at 1 AU.  When an impulsive electron event occurs in the corona, energetic particles 
travel along these field lines to 1 AU.  However, sometimes impulsive electron events are clearly related 
to flares that occur on the eastern half of the solar disk, up to $\sim$ $1R_\odot$ from the Parker spiral footpoint.
\cite{Krucker99} suggested that EIT waves might explain how the flare site and Sun-spacecraft magnetic 
field lines are connected.  They concluded that at the time of electron release, the EIT wave had 
not expanded far enough to reach the footpoints of the Parker spiral.
However, by considering the EIT wave as a fast-mode MHD wave (which moves faster at higher altitudes due to 
decreasing density), they showed that the calculated wave front at higher altitude ($\sim$ $1.5R_\odot$) 
is fast enough to connect the flare site with the Sun-spacecraft field line.

We would like to suggest an alternative possibility.  Our simulation results 
(Figures \ref{fig:f7} and \ref{fig:f8}) show that reconnection between the core magnetic flux rope
(intimately connected with the flare region) and the overlying or surrounding magnetic field occurs at a similar height range.
The reconnection(s) subsequently displace the flux rope connectivity out of the flare region to a distance of $\sim$ $1R_\odot$,
consistent with observations.


\section{CONCLUSIONS}
\label{sec:Conclusions}

Using a combination of multi-wavelength observational analysis and a global MHD simulation driven by real data, 
we study the 13 February 2009 coronal wave -- CME -- dimmings event observed by STEREO, in quadrature.
We find that the diffuse bright front emission is primarily due to mass density enhancement.
This is caused by a 
combination of both wave and non-wave mechanisms, both of which are directly-driven by the CME, which expands to a considerable lateral extent in the low 
($<$ 200 Mm) corona.
The reorganization of the magnetic field through reconnection facilitates lateral expansion of the CME, leading to far-reaching 
compression and ``opening'' of the surrounding magnetic field.  
This is further evidenced by secondary dimmings that form across the solar disk.

We find the diffuse coronal wave front displays a dual structure, consisting of
a brighter and a weaker component.
The brighter component, observed primarily in base difference EUVI data, 
is due to plasma being compressed by the expanding CME (against both 
surrounding and overlying magnetic field).  Some of the bright patches correspond to regions of reconnection 
where the field orientation is favourable for this to occur.
This non-wave component maps directly to the CME footprint at every stage of the evolution.
Thus, there is a strong coupling between the development of the coronal ``wave'' bright front, CME and associated dimmings.  

The weaker component, which is most likely to be detected in running difference EUVI data is present throughout the event, and
ultimately decouples from the bright component after the CME ceases lateral expansion late in the event.
This suggests that the weaker component is an MHD
wave, initially driven by the expanding CME, later becoming freely propagating.

This work demonstrates the considerable insight gained from advanced numerical simulations well constrained by observations.  
We hope that this work can progress the coronal wave community away 
from divisive, separatist theories toward a more cohesive, holistic approach to understanding the complexity 
of EUV coronal waves.  Future work should focus on the combined analysis of other well-observed events and 
on what coronal ``waves'' can tell us both about their driving CMEs, and the structure and dynamics of the surrounding corona.  
The potential for coronal seismology can now be pursued with the confident identification of the true wave component.

As this study demonstrates, detailed global MHD simulations are essential for furthering development of 
comprehensive physical models.  We must now include rigorous quantitative data 
analysis for comparison with such models.  This goal will be forwarded both by the continuing development of automated measurement techniques and by 
the upcoming launch of the Atmospheric Imaging Assembly aboard the Solar Dynamics Observatory.


\acknowledgments

We would like to thank an unknown referee for his/her comments and suggestions. We thank Yang Liu and Hao Thai for 
providing the magnetogram data, and Michelle Murray for suggestions about displaying the simulation data. 
We also thank Pascal D{\'e}moulin, Nariaki Nitta-san, Tibor T\"{o}r\"{o}k and and Veronica Ontiveros for helpful discussions.  This work has been supported by the SHINE through NSF grant ATM-0823592 and the NASA grant NNX09AB11G-R. 
Simulation results were obtained using the Space Weather Modeling 
Framework, developed by the Center for Space Environment Modeling, at the University of Michigan with funding 
support from NASA ESS, NASA ESTO-CT, NSF KDI, and DoD MURI.  


\clearpage


\clearpage

\clearpage

\clearpage

\clearpage

\clearpage

\clearpage

\clearpage

\end{document}